\documentclass[12pt,a4paper,twoside]{book}
\usepackage{t1enc}
\usepackage[utf8]{inputenc}
\usepackage{graphicx,color,amsmath,amssymb,cite,float}
\usepackage{anysize, epigraph, epsfig, subfig, wrapfig, url}
\usepackage{fancyhdr}
\usepackage[english]{babel}

\marginsize{3cm}{4cm}{2cm}{5cm}
\let\tmp\oddsidemargin
\let\oddsidemargin\evensidemargin
\let\evensidemargin\tmp
\reversemarginpar

\usepackage[bottom]{footmisc}     

\makeatletter
\def\cleardoublepage{\clearpage\if@twoside \ifodd\c@page\else
  \hbox{}
  \thispagestyle{empty}
  \newpage
  \if@twocolumn\hbox{}\newpage\fi\fi\fi}
\makeatother

\fancypagestyle{plain}{\pagestyle{fancy}}

\newcommand{\HRule}{\rule{\linewidth}{0.3 mm}}

\def\CPP{{C\kern-.05em\raise.23ex\hbox{+\kern-.05em+}}}
\def\midtilde{\raise-0.5ex\hbox{\textasciitilde}}

\begin{document}

\begin{titlepage}
\begin{center}

  \HRule \\[0.5cm]
    {\Large  \textsc{\textbf{Experimental and theoretical studies}}}   \\[0.5cm]
    {\Large \textbf{Analysis of Low-mass Dilepton Enhancement \\[0.3cm]
                  in 200 GeV Au+Au Collisions at RHIC}} 
  \HRule \\[1.3 cm]

  \LARGE{\textbf{M.Sc.\ Thesis \\[2cm]}}

  \large{\it Author: \\}
  \Large{\textbf{Vargyas, M\'arton \\ [0.1cm]}}
  \large{ELTE TTK\\
  m.vargyas@gmail.com\\[1.5cm]}

  \large{\it Supervisor: \\}
  \Large{\textbf{Prof.\ Cs\"org\H o, Tam\'as \\ [0.1cm]}}
  \large{ELTE TTK \\
  csorgo.tamas@wigner.mta.hu\\[0.6cm]}

  \begin{figure}[h!]
    \center
      \includegraphics[scale=0.22]{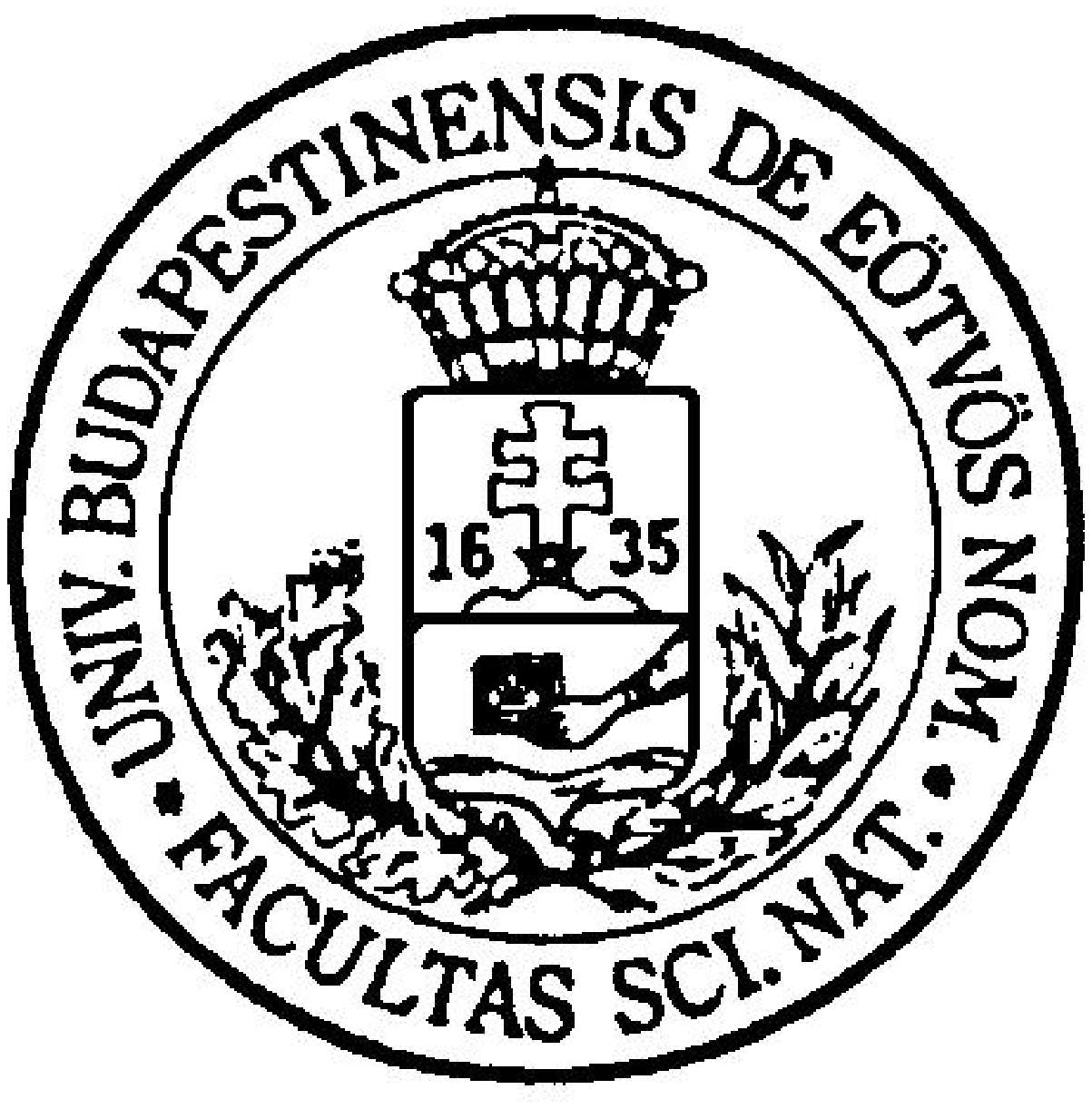}
  \end{figure}

  \large{May 29, 2012}

  \thispagestyle{empty}
  \pagebreak

\end{center}
\end{titlepage}

{\color{white} disappear please from this page}
\thispagestyle{empty}
\pagebreak

\color{black}
\thispagestyle{empty}
\vspace{4cm}
\epigraph{\large Palma sub pondere crescit}
\pagebreak


\chapter*{\centering \begin{normalsize}Abstract\end{normalsize}}
\vspace{-10pt}
\noindent 
In case of an $U_{A}(1)$ symmetry restoration in a hot and dense hadron matter, the mass of the produced hadrons and mesons can significantly change, and their production cross-section can also be modified. According to recent indirect measurement of Ref.~\cite{Vertesi:2009io} the $\eta'$ meson can suffer an at least 200 MeV mass drop. The decrease of the $\eta'$ mass results in an enhanced production, thus we can search for this effect via the decays of $\eta'$ even if these decays happen in vacuum, outside of the hot and dense hadronic matter, where the mass modification happens. These decays of $\eta'$ change the two-pion correlation functions, and the dilepton spectrum. 

In this M.Sc.\ Thesis I search for the signature of an $\eta'$ enhancement in the PHENIX dilepton~\cite{PPG088} spectrum in  $\sqrt{s_\textrm{NN}} = 200$ GeV Au+Au collisions, which has a significant enhancement in the low-mass region as well, e.g.\ in the ($0.1 - 1.0$) GeV invariant electron-positron mass range. This is also the region of the $\eta'$ meson's Dalitz-decay ($\eta'\rightarrow \textrm{e}^+ \textrm{e}^- \gamma$), so the $\eta'$ enhancement might be responsible for at least a part of the excess. Other mesons' (other) properties can also be changed in the hot, dense medium, but in this thesis I focus on the mass modification of the $\eta'$ meson. To explore the role of $\eta'$, I used EXODUS~\cite{EXODUS} simulations to generate different dilepton spectra, corresponding to different $\eta'$ properties. The conclusion here was that the excess can not be described with just the $\eta'$ mass modification, but the agreement with data has been improved significantly.

Another idea which might brings us closer to understand the excess is to examine the radial flow of the mesons, which was not included in the original PHENIX analysis~\cite{PPG088}. Radial flow is important in the low $p_\textrm{T}$ range, where it describes the particles' spectra well, just in the region where the dilepton spectrum has the most contributions from. Thus examining the effect of the radial flow seems to be inevitable, as it might be responsible for certain parts of the excess.

The results summarized here are work in progress, internal results obtained with the framework of the PHENIX Collaboration at RHIC. I have presented them to the Collaboration in 3 talks during 2011-2012.

\thispagestyle{empty}
\pagebreak

\tableofcontents
\thispagestyle{empty}
\pagebreak

 
\begin{table}
\caption*{\Large \bf List of common abbreviations and notations}
\begin{center}
\begin{tabular}[H]{ l l }
\hline \hline \\
BNL   & Brookhaven National Laboratory \\
RHIC  & Relativistic Heavy-Ion Collider \\
BBC   & Beam Beam Counter \\
ZDC   & Zero Degree Calorimeter \\
DC    & Drift Chamber \\
PC    & Pad Chamber (PC1, PC2, PC3 correspond to the different layers of PC)  \\
RICH  & Ring Imaging Cherenkov Counter \\
EMCal  & Electromagnetic Calorimeters, stand for both PbGl, and PbSc detectors \\
& \\
$p_\textrm{T}$ & transverse momentum, $p_\textrm{T} = p_x^2 + p_y^2$ \\
$m_\textrm{T}$ & transverse mass, $m_\textrm{T} = \sqrt{p_\textrm{T}^2 + m^2}$ \\
$\sqrt{s_\textrm{NN}}$ & center of mass energy \\
${\langle u_\textrm{T} \rangle}^2$ & the average transverse flow \\
$T_0$ & freeze-out temperature of the hadron gas \\
$T_\textrm{eff}$ & effective slope of the of hydro spectra, e.g.\ $e^{-m_\textrm{T}/T_\textrm{eff}}$, where $T_\textrm{eff} = T_0 + m {\langle u_\textrm{T} \rangle}^2$\\
\end{tabular}
\end{center}
\end{table}
\thispagestyle{empty}
\pagebreak

\chapter{Introduction}
The results presented here are reports on a work in progress that proceeds in the framework of the PHENIX Collaboration. Published PHENIX data are presented from Ref.~\cite{PPG088} for Au+Au collisions.

\noindent Note, that this M.Sc.\ Thesis deals with one of the most difficult, unresolved and unsolved puzzles of relativistic heavy-ion physics, namely the low-mass dilepton enhancement in 200 GeV Au+Au collisions at RHIC. To our best knowledge, no published theoretical model has yet been able to describe these measurements.

\chapter{The experiment}

\section{RHIC (Relativistic Heavy Ion Collider) }

The experimental scene, the Relativistic Heavy Ion Collider (RHIC) is located at Brookhaven, in the center of Long Island, near New York in the USA. This collider is an intersecting storage ring of 3.8 km circumference, accelerating particles with alternating electromagnetic field. It is able to collide protons, and heavy-ions (Au+Au, Pb+Pb, Cu-Cu, U+U), and even particles with different mass and charge, e.g.\ d+Au and Cu+Au. The acceleration energy can also be varied in a broad range, from $\sqrt{s_\textrm{NN}} = 7.7 - 200$ GeV for Au+Au, and up to 510 GeV for p+p collisions, which makes it the most versatile collider ever built. It sets another world record by being the highest energy collider for polarized proton-proton collisions, but the corresponding spin physics will not be discussed in the current manuscript.

\begin{figure}[h!]
  \centering
  {\includegraphics[scale=0.6]{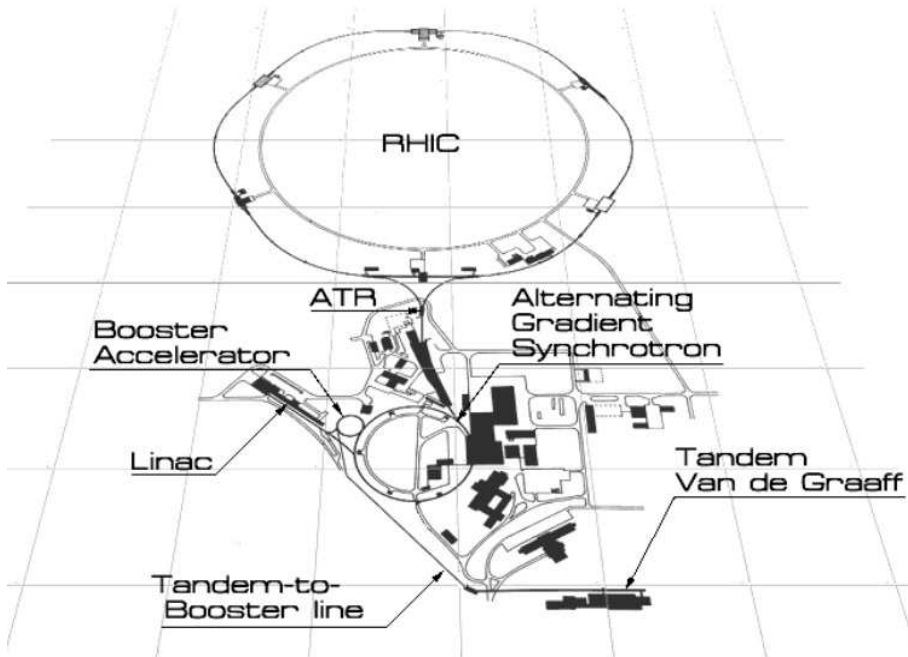} }
\caption{Schematic picture of the RHIC accelerator with its pre-accelerators}
  \label{f: rhic_introduction}
\end{figure}

The acceleration process is completed in steps, before the beam reaches the main ring, it goes through pre-accelerators (the Linac, Booster and AGS). An overview of the experiment and the acceleration setup is indicated on Fig.\ \ref{f: rhic_introduction}\footnote{
Figures \ref{f: rhic_introduction}-\ref{f: PHENIX2012} are courtesy of the PHENIX Experiment at Brookhaven National Laboratory's Relativistic Heavy Ion Collider.}. \\

\noindent {Heavy-ions}, using gold ions as an example, first go through the Tandem Van de Graaf linear accelerator, which uses static electricity to accelerate atoms removing some of their electrons. These ions leave Tandem with approximately $5\%$ of the speed of light (equivalent with 1 MeV energy/nucleon). The ionization here results $Q = +32$ $e$ electric charge by the end of this stage. Later these ions enter two smaller synchrotons, the Booster and the AGS (Alternating Gradient Synchroton)
, two circular accelerators, which accelerate them up to $0.37$ $c$ (95 MeV/nucleon) and $0.997$ $c$ (8.86 GeV/nucleon) and ionize them to $Q = +77$ $e$ and $+79$ $e$ electric charge, respectively. So actually before entering the RHIC ring, the gold atoms loose all of their electrons.  \\ 
 
\noindent {Protons} start in a different linear accelerator, the Linac, but after that they follow the same path as the heavy-ions. They leave the Linac with 200 MeV, the BOOSTER with 2 GeV and the AGS with 23 GeV energy. RHIC is also the only spin-polarized proton accelerator ever built, so its recent world record of $\sqrt{s} = 510$ GeV for polarized p+p collision outdated an earlier record (500 GeV) that was also reached at RHIC. \\

\noindent After leaving AGS, the beam is passed into the AGS-to-RHIC transfer line, where a switching magnet splits the ions and protons to a so-called "blue" and "yellow" beams. At RHIC, two accelerator rings are located inside the tunnel which inherit the name of the beams injected into them. The blue one circulates ions clockwise, while the yellow ring accelerates ions counter-clockwise. RHIC's two concentric rings are made up of 1,740 superconducting magnets.
Due to further acceleration by RHIC, protons can reach $\sqrt{s_{\textrm{NN}}} = 510$ GeV, the heavy-ions gain typically $\sqrt{s_{\textrm{NN}}} = 200$ GeV energy/nucleon. To operate effectively, the particles (both protons and ions) are collected into highly concentrated "bunches", each containing billions of ions, and 55-110 bunches circulate in a ring at the same time. \\

\noindent The blue and yellow ring intercept each other in 6 points, 4 of which a detector-complex has been built on, the PHOBOS, BRAHMS, PHENIX and STAR experiments. PHOBOS and BRAHMS were designed as special purpose experiments, while STAR and PHENIX were foreseen as major user facilities with strong upgrade programs. PHOBOS was a tricky setup of silicon detectors, measuring charged particles from the collision, BRAHMS (Broad RAnge Hadron Magnetic Spectometer) was looking for a new state of matter, the color glass condensate (a detailed description of this state of matter can be found in Ref.\ \cite{CGC}). Both detector-complexes were designed to detect particles with high pseudorapidities (close to the beam direction)%
\footnote{One definition of pseudorapidity is: $\eta = \frac{1}{2}\ln{\frac{|p|+p_{z}}{|p|-p_{z} }}$, so high $p_z$ is equivalent with high pseudorapidity. The second one is directly related to the angle between the particle's momentum and the beam direction (here the z-axis): $\eta~=~-\ln \left[\tan{\frac{\theta}{2}}\right]$. Comparing the two definition it is clear, that particles with high pseudorapidity fly out with small $\theta$, cor\-res\-pon\-ding to particles that propagate close to the beam direction.}, 
but their data taking period has already been completed: PHOBOS has been closed at 2005, BRAHMS a year later. The remaining two detector-complexes are STAR (Solenoid Tracker at RHIC), which covers the whole solid angle, and is able to reconstruct the path of charged particles with high accuracy, and PHENIX, which will be detailed subsequently, given that PHENIX research is the subject of this thesis. All the four RHIC experiments published the so-called "RHIC White Papers"  in 2005, announcing the discovery of a new state of matter. More information about this discovery can be found in Refs.~\cite{Adcox:2004mh,Adams:2005dq,Back:2004je,Arsene:2004fa}. \\

\noindent In 2010 PHENIX published its direct photon measurement in 200 GeV Au+Au collision, which revealed, for the first time experimentally, that the initial temperature of the hot and dense fireball and its nearly perfectly flowing fluid is hot enough to melt hadrons to a plasma of quarks (and supposedly gluons too). This measurement was based on virtual photon decays to lepton pairs, and the effective initial temperature was found to be $T_\textrm{init} > 300$ MeV. The Hagedorn spectrum of hadrons suggests that hadrons as we know them can not exist at temperatures $> T_\textrm{Hagedorn} \sim 180$ MeV.

\section{PHENIX experiment}
\subsection{Introduction of PHENIX}
The PHENIX ({\bf P}ioneering {\bf H}igh {\bf E}nergy {\bf N}uclear {\bf I}nteraction e{\bf X}periment) detector-complex was designed to detect rare events of charged particles, and it is specialized to measure particles coming directly from the collision, the so-called direct probes of the quark-gluon plasma. These particles are not participating in the strong interaction, thus they can pass through the dense and hot medium without final state interactions, hence they carry information about the initial state and time evolution of the quark-gluon plasma. In particular, PHENIX specialises on photons, electrons, muons as well as hadrons. 

\begin{figure}[h!]
  \centering
   \vspace{-1cm}
  \includegraphics[scale=0.5]{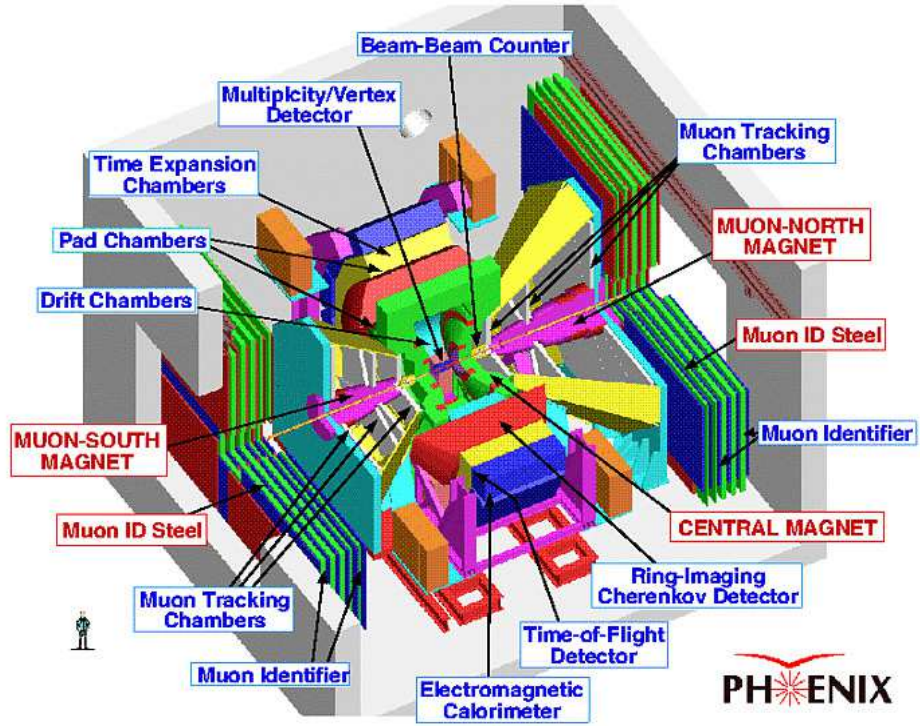}
  \caption{A shematic drawing of the PHENIX experiment as of 2012} \label{f: PHENIX2012_3D}
\end{figure}

\begin{figure}[H]
  \centering
  \vspace{50pt}
  \includegraphics[scale=0.85]{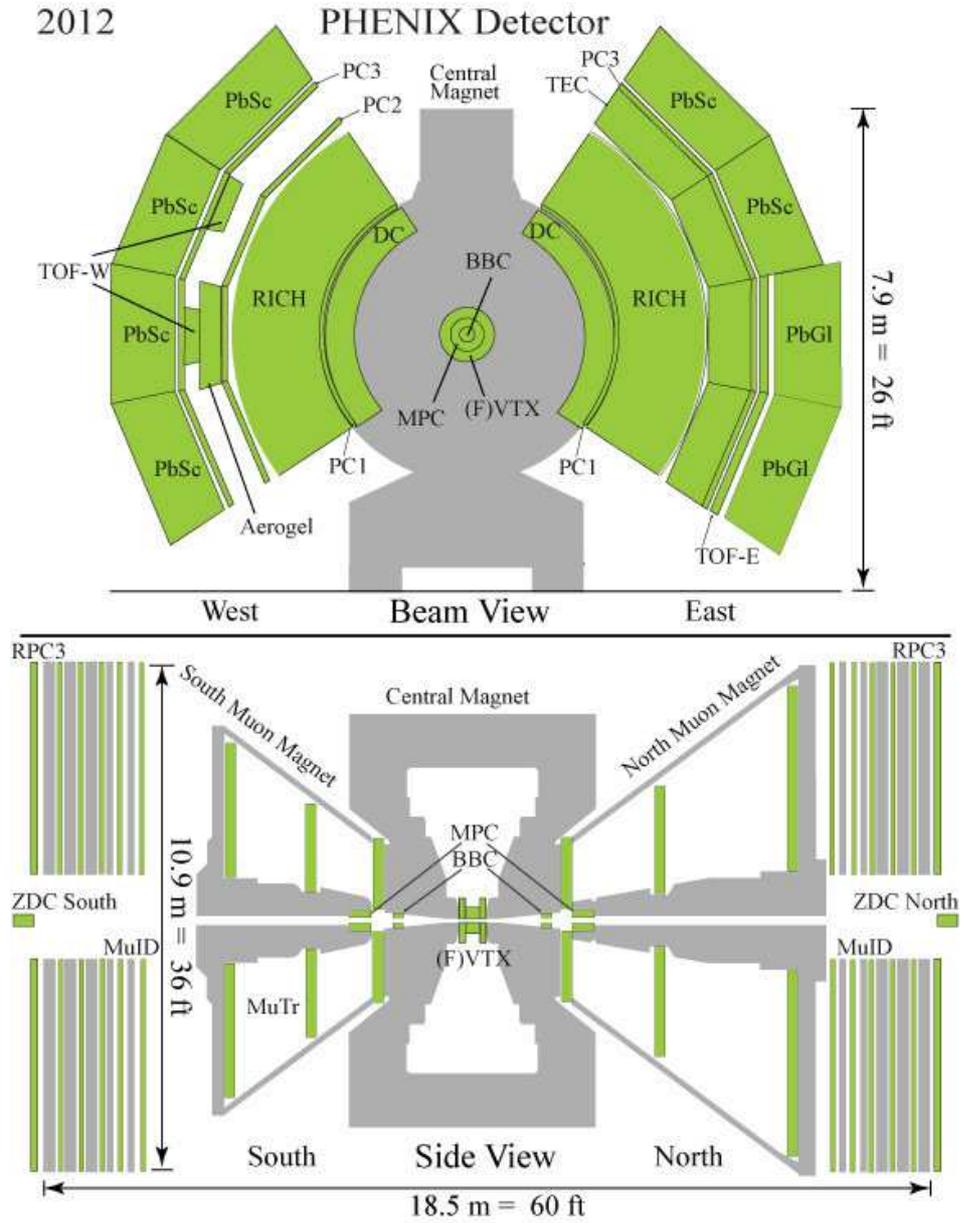}
  \caption{The cross-section drawing of PHENIX from different views} \label{f: PHENIX2012}
\end{figure}

PHENIX is a huge international collaboration, it involves $\sim$ 500 scientists from 13 countries. It consists of several subdetectors, modules controlling these detectors and electronics responsible for communication and data-taking. The total weight of all three of the PHENIX magnets and steel are 1,657 tons, the whole weight of PHENIX is around 3,000 tons. Its dimensions are also impressive, it is 12 meters wide and four stories high. A schematic drawing of the PHENIX experiment as of 2012 is illustrated on Fig.~\ref{f: PHENIX2012_3D}, while a detailed cross-section drawing of PHENIX, showing the Central Arm and the Muon Arm can be seen on Fig.~\ref{f: PHENIX2012}. 
  
By location, its detectors can be grouped into the two central arms, where most of them are placed transverse to the beam, and the two muon arms which focus on the measurement of muons, that are located closer to the RHIC rings (north and south arms). By functionality, the detectors broadly fall into 3 categories: the tracking detectors, the calorimeters and the event characterization detectors. PHENIX also includes three huge magnets that bend the trajectories of charged particles.  \\

\noindent {\bf Tracking detectors} measure the tracks of charged particles. The charged particles' momenta can be determined from curvature of their tracks in the magnetic field, if the PHENIX magnets are switched on. If these magnets are switched off, multiplicity and pseudorapidity density measurements are possible with tracking. At PHENIX 
the Drift Chamber (DC) and the Pad Chamber (PC) are the most relevant tracking-type detectors. \\

\noindent {\bf Calorimeters} detect the energy of the particles by absorbing them or their decay products, this is why calorimeter-type detectors are always placed on the edge of a detector-complex, they are the last station for a particle. At PHENIX the Electromagnetic Calorimeters (PbSc and PbGl) and the Zero Degree Calorimeter (ZDC) are the most important calorimeters. The PHENIX EMCals stop electromagnetic showers, but they are partially transparent for hadrons. \\

\noindent Particle identification detectors are based on various principles. For example, the Time of Flight detector (TOF) measures particle's speed obtained from their time of flight, and the path length of their tracks. Ring Imaging Cherenkov detector identifies electrons using their induced characteristic Cherenkov radiation. EMCals can also be used for particle identification, by measuring the energy of the particle with a given momentum, that is known from the tracking detectors. In the forward region special (MuID) plates were installed to identify muons. 

\begin{figure}[H]
  \centering
  \includegraphics[scale=0.4]{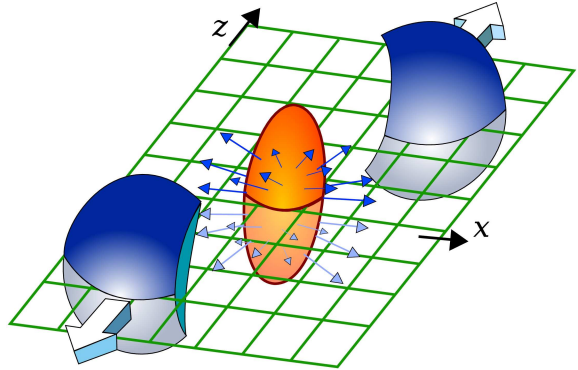}
  \caption{The reaction plane} \label{f: ReactionPlane}
\end{figure}

Beyond particle identification, to determine the reaction plane and the impact parameter of the collision is also fundamental, and this is a part of what {\bf event characterization detectors} are responsible for. They also function as triggers, selecting interesting events and controlling the data-taking. The reaction plane is determined by the beam axis and the impact parameter\footnote{The impact parameter connects the center of the two colliding particles at the time of their closest approach.}, as illustrated on Fig.~\ref{f: ReactionPlane}.

\noindent A short list of the PHENIX detectors, ordered by their location is given as follows: \\

\noindent {\bf Detectors of the Central Arm}: { the Drift Chamber (DC), the Pad Chamber (PC), the Ring Imaging Cerenkov (RICH), the Time Expansion Chamber (TEC), the Time-of-Flight detector on the East side and the West side (TOF-E, TOF-W), the Aerogel Cerenkov Counter (Aerogel), and the Electromagnetic Calorimeters (PbSc and PbGl)}. \\

\noindent {\bf The Muon Arm detectors}: { the Muon Tracker (MuTr), the Muon Identifier (MuID) and the Muon Piston Calorimeter (MPC)}. \\ 

\noindent {\bf The Event Characterization Detectors}:  { the Beam-Beam Counter (BBC), Zero Degree Calorimeters both on the north and south side (ZDC north and ZDC south, which are physically located inside the RHIC tunnel and their position is not to scale on Fig.~\ref{f: PHENIX2012}), the Forward Calorimeters, the Multiplicity Vertex Detector (MVD) and the Reaction Plane Detector. } 
    
\pagebreak

\pagebreak

\subsection{Detectors of the PHENIX dilepton measurements }

In this section an overview of the PHENIX detectors used in the dilepton spectrum measurement is presented. These are the global detectors; the Beam-Beam Counter (BBC) and the Zero-Degree Calorimeter (ZDC), two tracking detectors of the Central Arm; the Drift Chamber (DC) and Pad Chamber (PC), the Ring Imaging Cherenkov Counter (RICH) for electron identification, and the Electromagnetic Calorimeters (EMCal) for energy measurement, which also provide additional electron identification.  

The BBC and ZDC together measure the collision centrality and the vertex position along the beam direction. The centrality corresponds to the impact parameter of the collision, which is fundamental to measure. The ZDC and BBC are also used as first level triggers, so if they have hits simultaneously, data taking from all the other detectors can begin, otherwise signal in the detectors is considered as noise and data are not stored. The two BBC detectors are located 1.44 m from the collision point. 
They count the number of charged particles in pseudorapidity range 3.1 < |$\eta$| < 3.9, and deter the start-time of the collision with 20 ps resolution, which results in $\sim 2$ cm precision for p+p, and $\sim 0.6$ cm accuracy for Au+Au, respectively. 

The ZDC's are 18 m far from the interaction point in both directions almost along the z-axis with a 3.6 cm vertical difference. They are focused on measuring neutrons coming from the Coulomb dissociation of beam nuclei and the evaporation neutrons emitted by the spectators. To understand the role of evaporation neutrons, consider a non-central heavy-ion collision. The peripheral part is not involved in the collision, it breaks off the nucleus and continues its movement along the z-axis. The neutrons and protons are almost uniformly placed inside the nucleus, so this splitting does not change the neutron/proton ratio. But bigger nuclei prefer to have more neutrons than the smaller ones, so this newly formed nucleus is not stable and emits a neutron with transverse momentum low compared to the typical 100 GeV momentum of the beam. These are called evaporation neutrons and due to the large beam energy used at RHIC, they move towards the ZDC. To be able to separate these neutrons from the spectators, other particles moving towards the beam direction and specially from the beam, the ZDC's were placed just after the dipole (DX) magnets, which bend the beam back to RHIC's two separate rings (the "yellow" and "blue" ring), thus keep charged particles off the ZDC's.

\begin{figure}[H]
  \center
  \vspace{20pt}
  \subfloat{\includegraphics[scale=0.45]{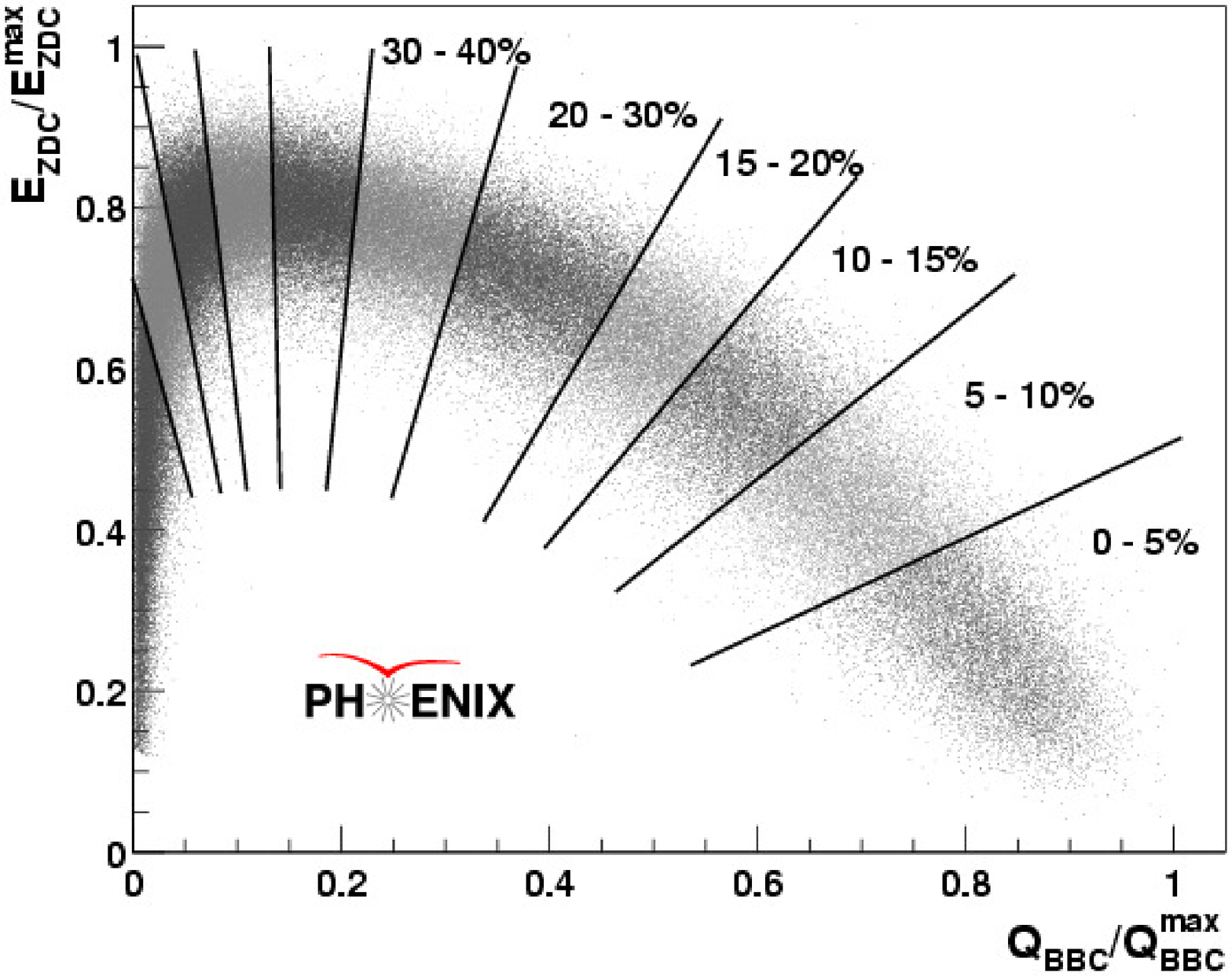} } \\
  \ \ \ \ \ \ \subfloat{\includegraphics[scale=0.5]{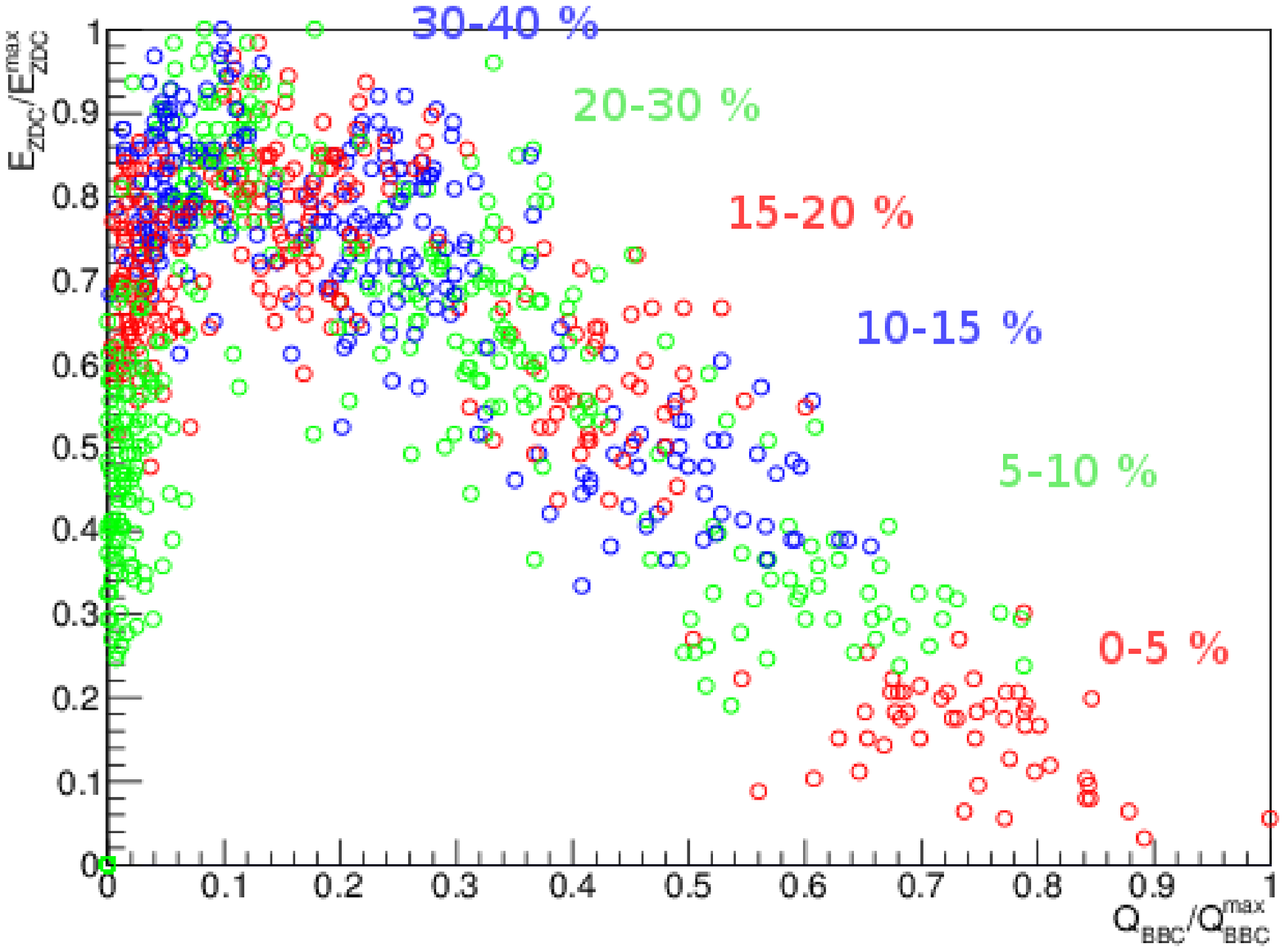} }
  \caption{Centrality determination at PHENIX}
  \label{f: centrality}
\end{figure}

The centrality of a collision is determined from these two detectors; it is defined as the correlation between the BBC and ZDC hits as seen on the top of Fig.\ \ref{f: centrality} (Fig.\ from Ref.~\cite{PHENIXcentrality}). It shows the centrality classes determined of Run 2 PHENIX measurement~\cite{PHENIXcentrality}. The bottom part of Fig.~\ref{f: centrality} indicates the impact parameters obtained from a HIJING~\cite{HIJING} simulation which has been filtered through the PHENIX specific PISA~\cite{PISA} software. Obviously, the centrality classes correlate with the impact parameters, but note that these are two different definitions. Note also that the two dimensional $Q_\textrm{BBC}/Q_\textrm{BBC}^\textrm{max} - E_\textrm{ZDC}/E_\textrm{ZDC}^\textrm{max}$ plots allow for a more precise centrality determination than any of them alone: ZDC measures small number of neutrons both for most central and for most peripheral collisions, while BBC alone can not resolve the centralities below $40-50$ \%.

\pagebreak

The Drift Chambers (DCs) and Pad Chambers (PCs) are in the Central Arm and are used to measure the transverse momenta of the particles from their bending curvature in magnetic field. The DCs are in the radial region from 2.02 m to 2.46 m. PC has 3 layers, the first (PC1) is between the DC and the RICH, PC2 is behind RICH (but only in the west arm) and PC3 is in front of the EMCal. PC1 and PC3 are present in both arms. The magnetic field is produced by the Central Magnet which consists of two pairs of concentric coils able to run separately. It is an axial field magnet, and if both coils are running to the same direction, the momentum resolution for a single particle is better than 1\% in the ($0.2 - 1$) GeV/$c$ range. DC is applied to measure momenta in the plane perpendicular to the axis of collision, while PC1 measures momentum along the collision axis, thus combining the two results in a precise momentum measurement. The additional layers provide pattern recognition and a possibility for systematic checks.

\begin{figure}[H]
  \centering
  \includegraphics[scale=0.6]{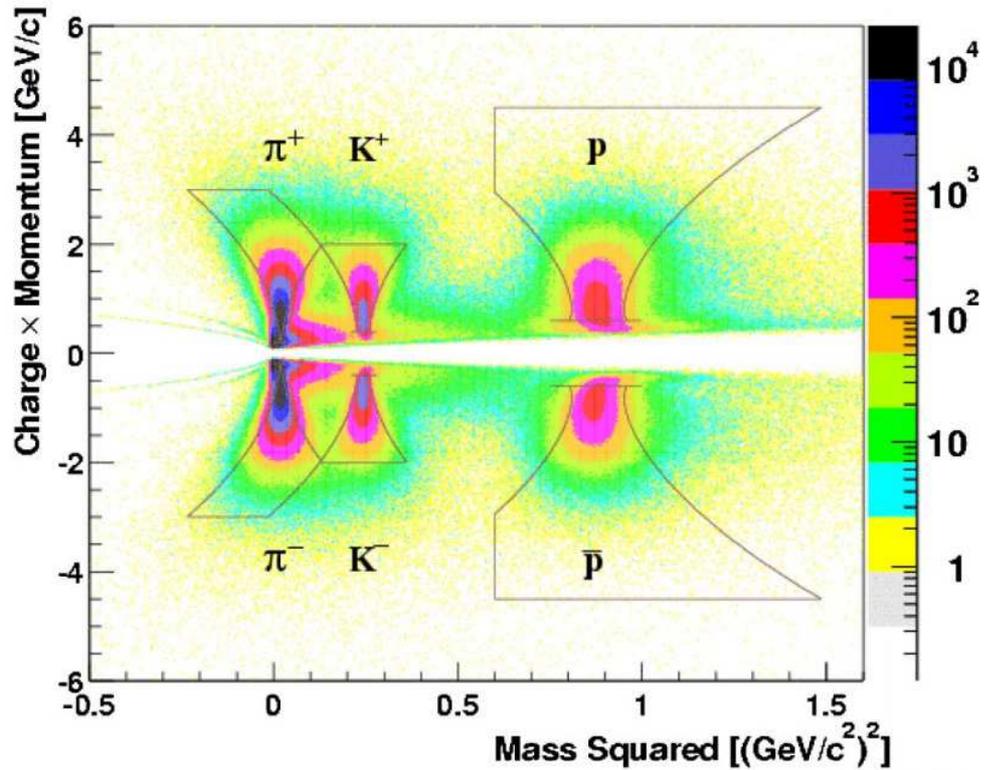} 
  \caption{Particle Identification at PHENIX. The lines are the PHENIX specific cuts inside of which the signal is considered a pion, kaon and proton, respectively. Figure from Ref.~\cite{PPG026}.} \label{f: PID}
  \vspace{40pt} 
\end{figure}

The Time of Flight detector (TOF) measures the particles' speed from their time of flight in the following way. It has multiple tubes with length $l$, and every tube has two Photomultiplier Tubes (PMT's) at its both ends. If it has a hit at position $x$ (measured from PMT1), the time required for the generated photons to reach PMT1 is $T_1 = T_0+\frac{x}{v}$, while the time they reach PMT2 is $T_2 = T_0+\frac{l-x}{v}$, where $v$ is the speed of photons moving in the tube, and $T_0$ is the time it takes for a particle to emerge from a collision and to hit the tube. The starting time, namely the time when the collision happens is measured by the BBC's. Combining the two equations for $T_0$, and to take the arithmetic mean of them one gets the time of flight, $t_\textrm{TOF} \equiv T_0$:
\begin{equation}
t_\textrm{TOF} = \frac{(T_1+T_2)-l/v}{2}.
\end{equation}
From the relationship $t_\textrm{TOF} = \frac{L}{c\beta} = \frac{L}{c}\frac{\sqrt{p^2+m^2}}{p}$ ($L$ being the path length of a particle from the collision vertex to the TOF hit) one gets

\begin{equation}
m^2 = \frac{p^2}{c^2}\left[\left(\frac{t_\textrm{TOF}}{L/c}\right)^2-1\right],
\end{equation}
where the momentum $p$ is measured by the DC. Plotting the momentum as a function of $m^2$, and multiplying it with the charge to separate matter from antimatter, one gets Fig.\ \ref{f: PID}.

This was just one method from the many used at PHENIX to identify particles. Another way is to calculate their mass from the energy measured by the EMCal and the momentum measured by the DC's and the PC's. For single electron identification a separate detector, the Ring Imaging Cherenkov (RICH) has been built, which is located in the radial region 2.5 m $<$ r $<$ 4.1 m of the Central Arm. 

\pagebreak

{\color{white}disappear please from this page}
\thispagestyle{empty}
\pagebreak

\color{black}
\section{PHENIX dilepton spectra}

Dilepton pairs are one of the direct probes of the quark-gluon plasma and the hot, dense medium formed in $\sqrt{s_\textrm{NN}} = 200$ GeV Au+Au collisions, since they can escape from the medium without final state interactions. By analysing the dilepton spectrum, one can probe the time evolution and dynamics of the collision, at least in principle. Dileptons can also be used to study the properties of low-mass vector mesons ($\rho$, $\omega$ and $\phi$) in the medium due to their relatively short lifetime compared to that of the medium (hadronic gas), and also to the pseudoscalar mesons, such as $\eta$ and $\eta'$. Production of photons can be measured through their conversion to dileptons as well. 

The PHENIX Collaboration has measured the invariant mass spectrum (shown on Fig.~\ref{f: PPG088ppcocktail},~\ref{f: PPG088cocktail}), centrality (Fig.~\ref{f: PPG088cocktail_CentrSlices}) and transverse momentum dependence (Fig.~\ref{f: PPG088cocktail_pTSlices}) of dilepton production in both Au+Au and p+p collisions~\cite{PPG088}. This is a detailed investigation and we only highlight few parts which are the most important, and most relevant inputs of our study. PHENIX determined the background dilepton spectrum from a p+p measurement. Different identified particles were measured, their spectra were fitted and parametrized. As for the particles' $p_\textrm{T}$ spectra, in the PHENIX analysis a modified Tsallis formula was used:
\begin{equation} \label{e: PPG088}
 E\frac{d^3\sigma}{dp^3} = A_\textrm{Tsallis} (e^{-(a p_\textrm{T} + b p_\textrm{T}^2)} + p_\textrm{T}/p_0 )^{-n}.
\end{equation}

With the formula above a simultaneous fit was elaborated to nearly all measured particles that decay to dileptons, as shown on Fig.~\ref{f: PPG088ppspectra} for p+p and on Fig.~\ref{f: PPG088spectra} for Au+Au collisions. 

Note, that proton and antiproton spectra has been also measured by TOF, but they are not included in these fits, since Fig.~\ref{f: PPG088spectra} contain only those that decay to dileptons. The proton+antiproton spectrum is important, however, since they have been measured in low transverse momentum range as well. As we shall demonstrate below, low $p_\textrm{T}$ particle decays dominate the min.\ bias 200 GeV Au+Au dilepton spectrum, while the contributions of resonances with $p_\textrm{T} >$ 2 GeV on Fig.~\ref{f: PPG088cocktail} are nearly negligible. Thus it is very important to understand the low $p_\textrm{T}$ part of the meson spectra, and as seen on Fig.~\ref{f: PPG088spectra}, $K^{\pm}$ already indicates some difficulties and incompatibilities with the PHENIX modified Tsallis function of Eq.~\ref{e: PPG088} for Au+Au collisions, while in p+p collisions the $K^{\pm}$ spectra are well described by Eq.~\ref{e: PPG088}, as seen on Fig.~\ref{f: PPG088ppspectra}. 


\pagebreak

\subsection{Dilepton in p+p}

\begin{figure}[H]
  \centering
  \includegraphics[scale=0.6]{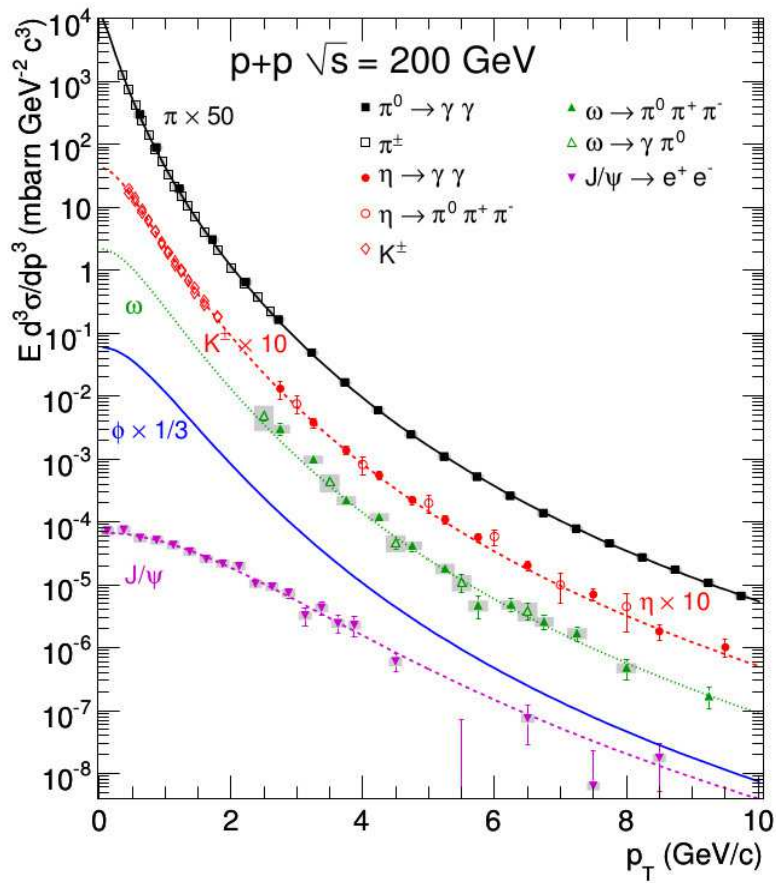} 
  \caption{Different meson's $p_\textrm{T}$ spectra of Eq.~\ref{e: PPG088} for p+p collisions. Figure from Ref.~\cite{PPG088}. } \label{f: PPG088ppspectra}
\end{figure}

\begin{figure}[H]
  \centering
   \vspace{70pt}  
  \includegraphics[scale=0.8]{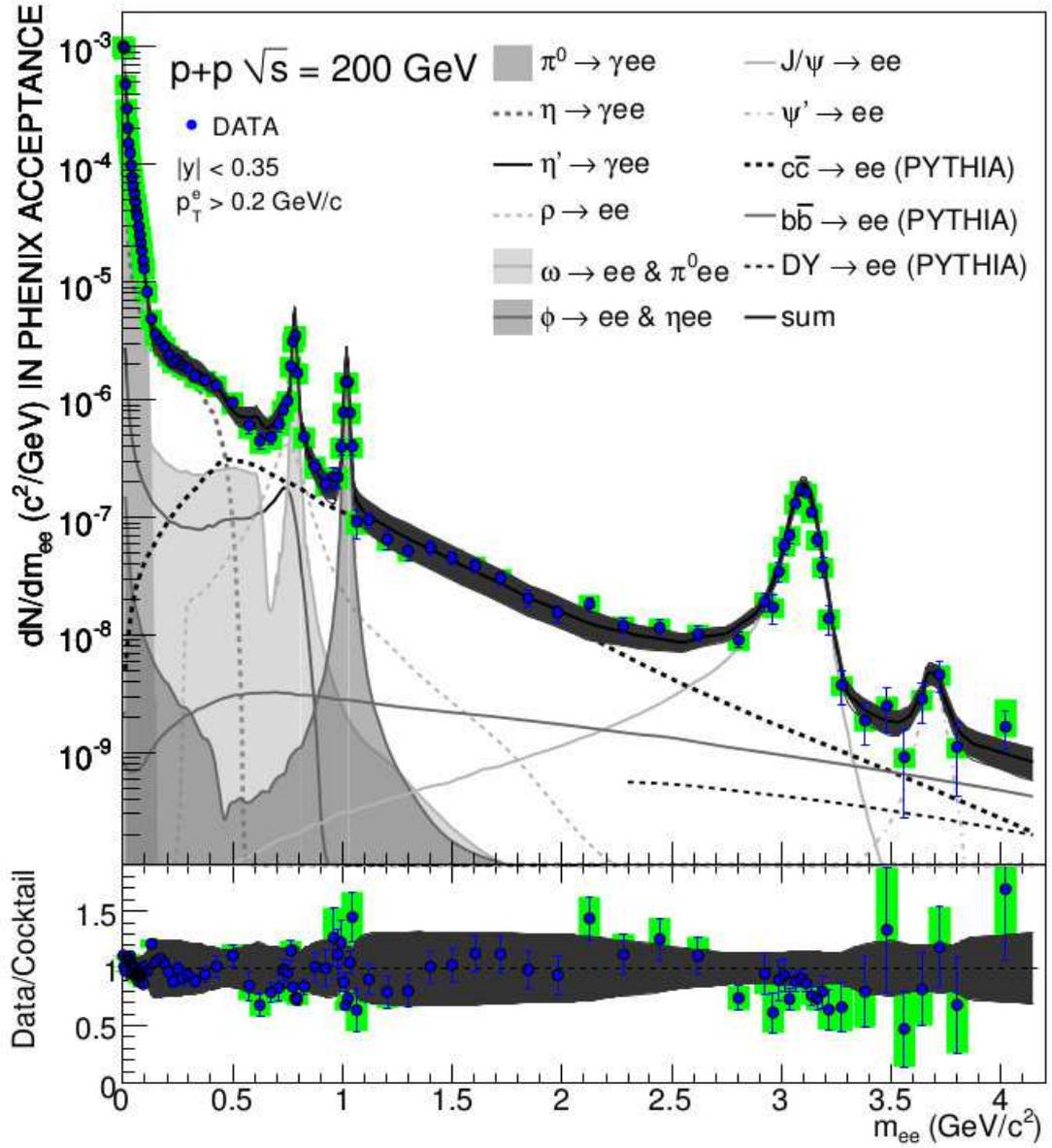}
  \caption{Dilepton cocktail of 200 GeV p+p collision,  Figure from Ref.~\cite{PPG088}.} \label{f: PPG088ppcocktail}
\end{figure}

As for p+p collisions, both the $p_\textrm{T}$ spectra and dilepton cocktail are well described as seen on Figs.~\ref{f: PPG088ppspectra},~\ref{f: PPG088ppcocktail}. Note, that $K^{\pm}$ spectra also fit, which does not apply for Au+Au collisions, as detailed subsequently.

Fig.~\ref{f: PPG088ppcocktail} shows data and simulation results of Ref.~\cite{PPG088}, elaborated with meson  $p_\textrm{T}$ spectra of Eq.~\ref{e: PPG088}. This figure illustrates that for p+p collision the background is well understood. Given that in Au+Au collisions, however, both certain meson spectra and dilepton cocktail are not well described, this might confirm the presence of an $\eta'$ enhancement and/or a radial flow effect in Au+Au collisions, since in p+p no chiral symmetry restoration, nor radial flow effects have been observed so far. Hydro spectra with temperature gradient, however, might be at work.

\pagebreak

\subsection{Dilepton in Au+Au}

\begin{figure}[H]
  \centering
  \includegraphics[scale=0.6]{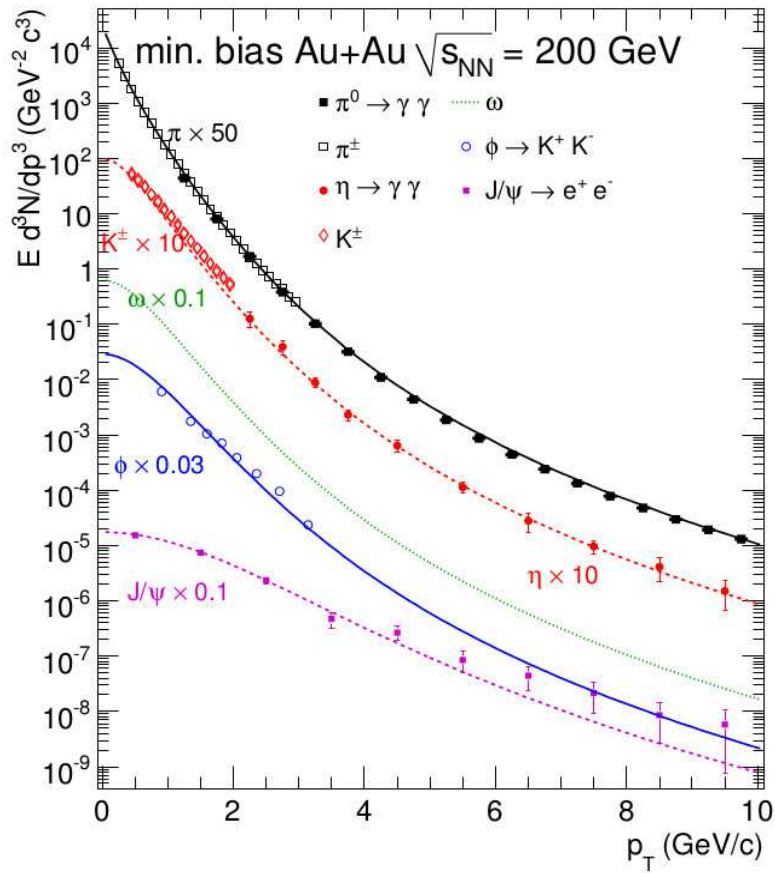} 
  \caption{Different meson's $p_\textrm{T}$ spectra of Eq.~\ref{e: PPG088} for Au+Au collisions. Figure from Ref.~\cite{PPG088}. } \label{f: PPG088spectra}
\end{figure}

\begin{figure}[H]
  \centering
  \vspace{70pt}
  \includegraphics[scale=0.8]{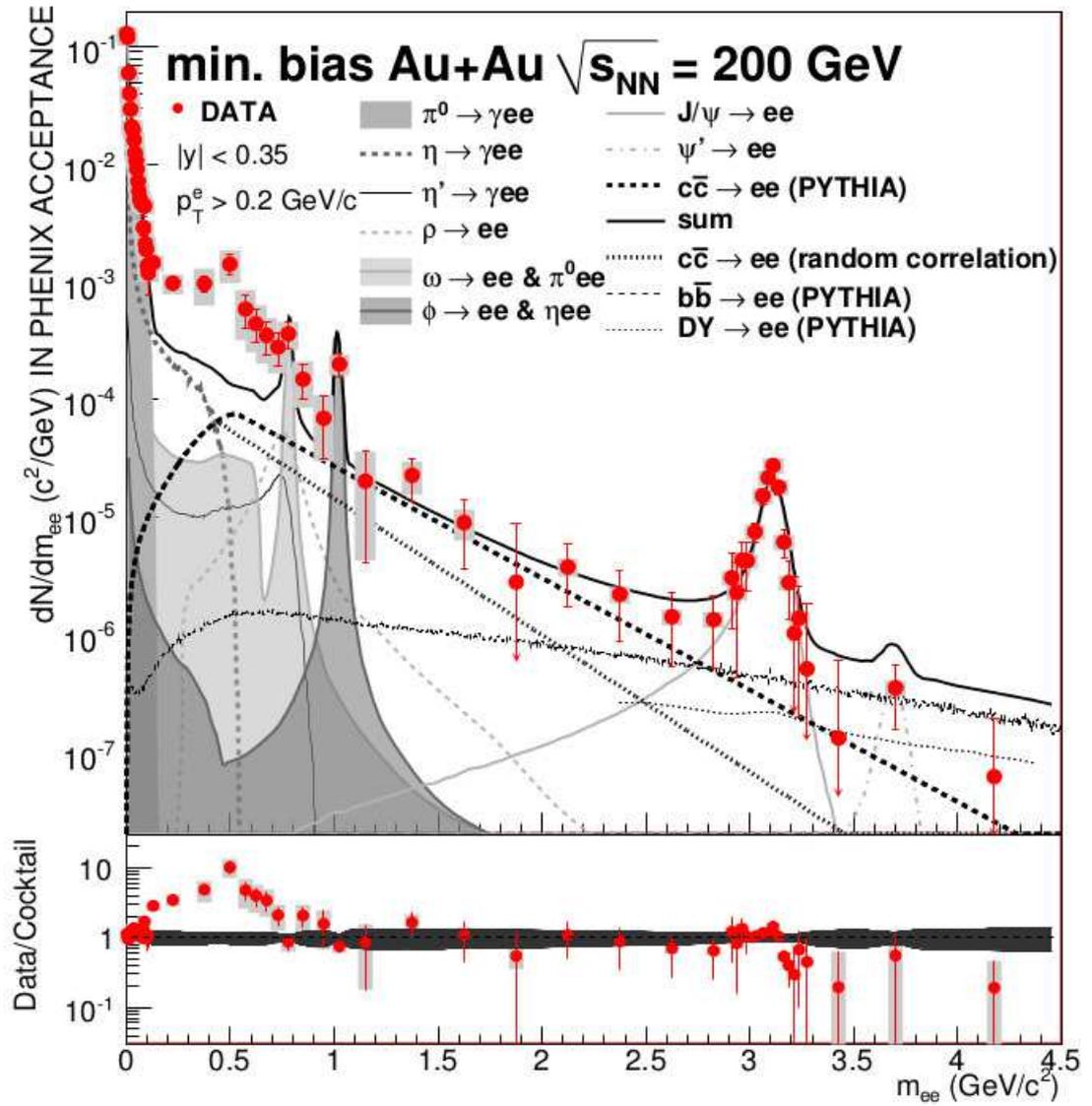}
  \caption{Dilepton cocktail of 200 GeV Au+Au collision, generated with meson  $p_\textrm{T}$ spectra of Eq.~\ref{e: PPG088}.  Figure from Ref.~\cite{PPG088}.} \label{f: PPG088cocktail}
\end{figure}

In the Au+Au, however, where both enhancement of $\eta'$ mesons and radial flow effect can play a significant role, the meson's spectra (as on Fig.~\ref{f: PPG088spectra}) and the dilepton cocktail (Fig.~\ref{f: PPG088cocktail}) are not completely understood yet.

\begin{figure}[H]
  \centering
  \vspace{40pt}
  \includegraphics[scale=0.4]{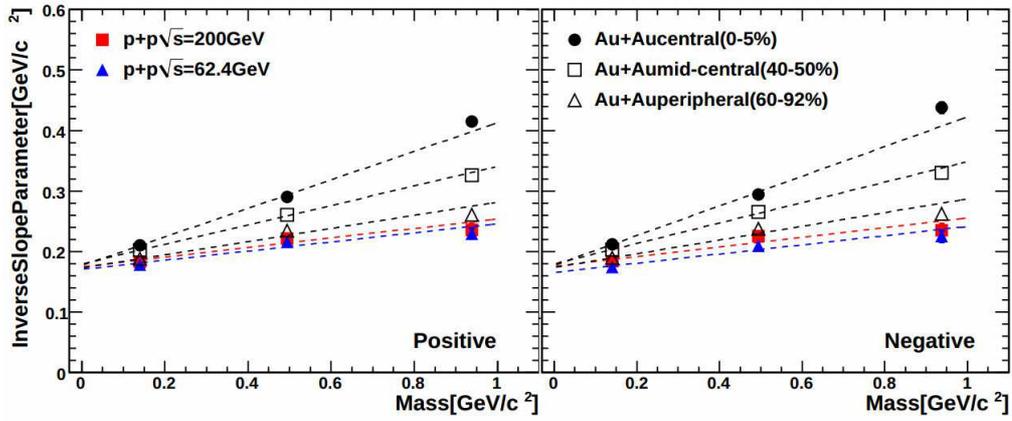}
  \caption{Inverse slope parameter ($T_\textrm{eff} = T_0 + m {\langle u_\textrm{T} \rangle}^2$) obtained both for p+p and Au+Au collision. The fact, that for p+p it is almost constant if plotted as a function of mass indicates that radial flow effects (e.g.\ the average transverse flow, ${\langle u_\textrm{T} \rangle}^2$) is present in Au+Au collisions but it is absent in p+p collisions. Figure from Ref.~\cite{PPG101}.} \label{f: PPG101_slope}
\end{figure}

Note, that $K^{\pm}$ spectra, as on Fig.~\ref{f: PPG088spectra}, are not well described by Eq.~\ref{e: PPG088}, possibly due to the lack of radial flow. An evidence for the radial flow effect in Au+Au collision and its absence in p+p collision is illustrated on Fig.~\ref{f: PPG101_slope} of Ref.~\cite{PPG101}. Note that the linear rise of the effective temperature with particle mass for non-central collisions was first derived in Ref.~\cite{Csorgo:2001xm}.

As for the dilepton cocktail, a significant excess of measured dileptons as compared to the sum of known processes is seen in the ($0.1-1$) GeV region of Fig.~\ref{f: PPG088cocktail}.

\pagebreak

\subsection{Centrality dependence}

\begin{figure}[H]
  \centering
  \includegraphics[scale=0.7]{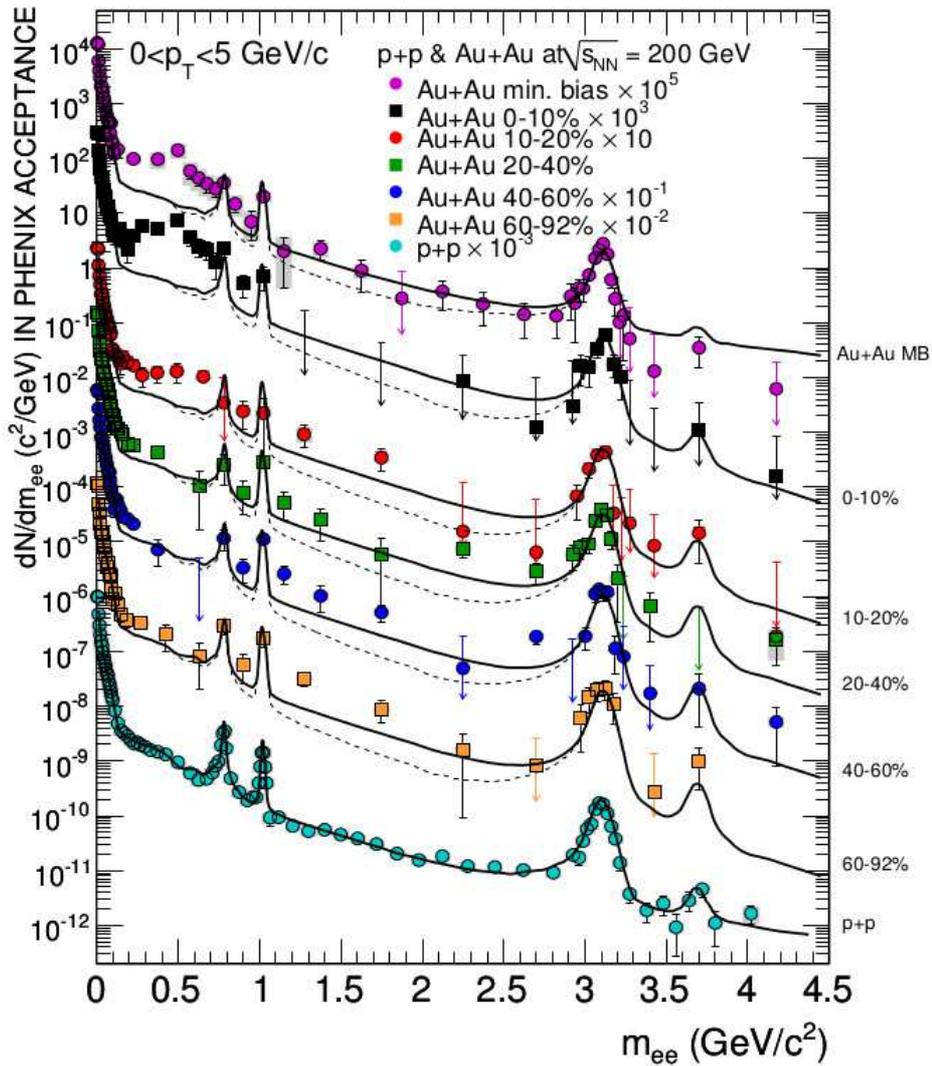}
  \caption{Dilepton cocktail of 200 GeV Au+Au collision, generated with meson  $p_\textrm{T}$ spectra of Eq.~\ref{e: PPG088}, in different centrality classes.  Figure from Ref.~\cite{PPG088}.} \label{f: PPG088cocktail_CentrSlices}
\end{figure}

The dilepton cocktail in different centrality classes has also been published in Ref.~\cite{PPG088}, as seen on Fig.~\ref{f: PPG088cocktail_CentrSlices}.  This plot indicates the possible presence of hydro behaviour, as the excess tends to disappear for peripheral collisions, in accordance with the hydro-picture, where radial flow decreases with the centrality. Thus the excess can be a sum of (at least) two effects, one coming from a spectra which lacks the radial flow term, and the enhanced production of the $\eta'$ mesons, due to the  in-medium mass modification effect.

\subsection{Transverse momentum dependence}

\begin{figure}[H]
  \centering
  \includegraphics[scale=0.6]{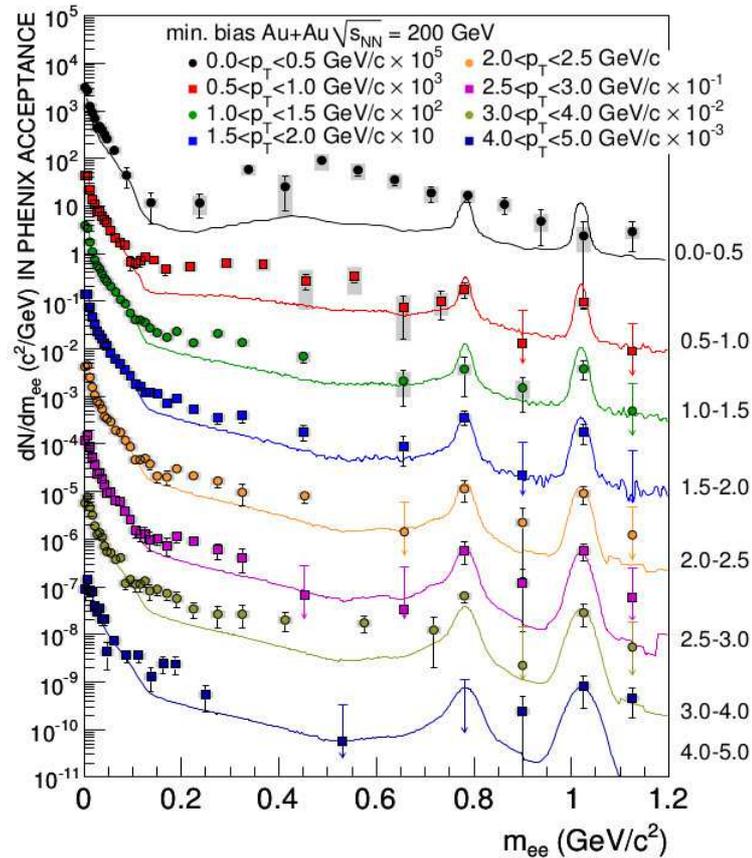} 
  \caption{Dilepton cocktail of 200 GeV Au+Au collision, generated with meson  $p_\textrm{T}$ spectra of Eq.~\ref{e: PPG088} in different $p_\textrm{T}$ slices. Figure from Ref.~\cite{PPG088}.} \label{f: PPG088cocktail_pTSlices}
\end{figure}

Fig.~\ref{f: PPG088cocktail_pTSlices} suggests that hydrodynamics-motivated $p_\textrm{T}$ spectra might result in a better agreement with data, since the input spectrum was fitted to the high $p_\textrm{T}$ region, where the cocktail has negligible contributions from. The soft component of Eq.~\ref{e: PPG088} lacks important physics ingredient, that we identify with the radial flow, thus it lacks to decribe the low-$p_\textrm{T}$ part of meson spectra. If one looks at the $\phi$ meson peak of Fig.~\ref{f: PPG088cocktail_pTSlices} at 1020 MeV, the low $p_\textrm{T}$ inaccuracy of Eq.~\ref{e: PPG088} can be clearly observed, as the $\phi$ meson peak has a different $p_\textrm{T}$ dependence in the model/cocktail calculation as compared to the data. Also note that the low-mass dilepton enhancement is predominant in the soft, 0 GeV $< p_\textrm{T} < 1.5$ GeV region, where the hydrodynamical picture is relevant. 

\pagebreak

\chapter{Theoretical Introduction}
\section{Chiral symmetry of the 3 quark model}

For setting the theoretical background, QCD's approximate chiral symmetry and references to chiral symmetry restoration models are outlined here. 

The term "chiral" comes from the Greek word for "hand", indicating that it is like a mirror symmetry between one's left and right hand. In particle physics, however, the chiral symmetry is defined as the invariance of the theory; the independent transformation of the left-handed and right-handed parts of Dirac fields in the QCD Lagrangian. Left-handed fermions correspond to a state when the particle's spin vector points to the opposite direction than its momentum. Right-handed state is when the momentum and spin vectors point to the same direction. 

To understand this phenomena in details, consider the QCD Lagrangian of the fermion Dirac-fields as an example:
\begin{equation} \label{e: QCD Lagrangian}
L_q^\textrm{QCD} = \sum\limits_{i=1}^{flavour}  \left[\bar{q}_L^i(i\gamma^\mu D_\mu)q_L^i + \bar{q}_R^i(i\gamma^\mu D_\mu)q_R^i\right] - \sum\limits_{i=1}^{flavour} m_q^i (\bar{q}_L^iq_R^i + \bar{q}_R^iq_L^i),
\end{equation}
where $\gamma^\mu$'s are the Dirac matrices, $q_L$ and $q_R$ correspond to the left- and right-handed fermions projected from the fermion fields using $q_L^i = 0.5 (1-\gamma_5) q^i$, and $q_L^i = 0.5 (1+\gamma_5) q^i$
transformation. The $\bar{q}_L$ notation is the Dirac adjoint, e.g.\ $\bar{q}_L = q_L^\dagger \gamma^0$. 

From Eq.\ \ref{e: QCD Lagrangian} it is easy to see, that the quark's masses explicitly break the chiral symmetry, e.g\ the mass terms are responsible for entangling the right- and left-handed parts of the Dirac fields, in accordance with the expectation from special relativity. According to that, massive particles can never reach the speed of light, thus if one considers a coordinate system moving any faster than the velocity of a (e.g.)\ left-handed quark, viewing from that system, the momentum of that quark would point to the opposite direction, while its spin would be unchanged under this coordinate transformation. Thus, massive left-handed fermions can become right-handed after such a Lorentz boost. The quarks have to be massless to avoid the existence of such coordinate transformations. So in the limit of massless quarks, chiral symmetry becomes an exact symmetry of QCD.
So let's restrict ourselves to the massless quark approximation ($m_q = 0$), and consider the $SU(3)$ flavour symmetry, which contains only the $u$, $d$ and $s$ quarks and their antiquarks. 
In this case the transformation of the quark flavours will become unitary transformation, e.g.\ $q_L^i~\rightarrow~U_L^{ij}~q_L^j$ and $q_R^i~\rightarrow~U_R^{ij}~q_L^j$, where $U_L, U_R$ $\epsilon$ $SU(n)$,
thus the $SU(3)$ flavour symmetry of the quark model falls apart to a direct product of the left-handed and right-handed parts;  $SU(3) \rightarrow SU(3)_L \bigotimes SU(3)_R$. Thus, in the $m_q\rightarrow 0$ limit QCD becomes chirally symmetric. \\


The chiral symmetry transformation has a different formulation, written in terms of the axial-vector symmetry. To do that, a new approach has to be introduced. Let $U_L$ and $U_R$ be the left-handed and right-handed unitary transformations defined as:
\begin{equation}
U_L = e^{i\Theta_L^a t^a}, \ U_R = e^{i\Theta_R^a t^a}.
\end{equation}
Thus the quark Dirac fields transform as follows:
\begin{equation}
q=q_L+q_R \rightarrow \left(\frac{1-\gamma_5}{2} e^{i\Theta_L^a t^a} +  \frac{1+\gamma_5}{2} e^{i\Theta_R^a t^a} \right) q.
\end{equation}
The leading order, infinitesimal transformation reads as
\begin{equation}
q=q_L+q_R \rightarrow \left(\mathbb{I} + \frac{i}{2}(\Theta_L^a + \Theta_R^a) t^a + i\frac{\gamma_5}{2}(\Theta_L^a - \Theta_R^a)\right) q.
\end{equation}
Introducing vector $\Theta_V^a$ and axial $\Theta_A^a$ transformations, such as
\begin{equation}
\Theta_V^a = \frac{\Theta_L^a + \Theta_R^a}{2}, \ \Theta_A^a = \frac{\Theta_L^a - \Theta_R^a}{2},
\end{equation}
namely the {\bf vector part} of this transformation which treats the left-handed and right-handed fermions equally, and the {\bf axial part} which treats them differently. Thus, the flavour transformation of a quark's Dirac field becomes
\begin{equation}
q\rightarrow e^{i\Theta_V^a t^a} e^{i\gamma_5 \Theta_A^a t^a} q,
\end{equation}
which means that QCD develops an approximate \\
\begin{equation}
U(3)_V~\otimes~U(3)_A \simeq SU(3)_V~\otimes~SU(3)_A~\otimes~U_V(1)~\otimes~U_A(1)
\end{equation}
symmetry in the chiral limit (e.g.\ in limit of $m_q\rightarrow 0$), where $U_V(1)$ is identified by the barion charge and is conserved exactly, while the so-called $U_A(1)$ symmetry is broken by the topological charge of QCD vacuum. 

This means that parity-partners have to have the same mass, which condition is generally not fulfilled, e.g.\ in the pseudoscalar octet $m_\pi \approx 140$ MeV, while $m_{K,\eta} \approx 500$ MeV. The difference between the quark masses would not result such deviation in the meson masses, only a few MeV difference would occur from that. It has been concluded, that the spontaneous breaking of the axial part $U_A(1)$ of the $SU(3)$ symmetry is responsible for that mass deviation. In addition to that, the outstandingly large mass of the $\eta'$ meson ($m_{\eta'} = 938$ MeV) complicates the picture even more.

Further reading on the chiral symmetry restoration can be found in Ref.~\cite{chiral2}, and also among its references.

\vspace{11cm}

\pagebreak

\section{The $\eta'$ meson in a hot and dense medium}

In case of a chiral symmetry restoration in the hot and dense medium, the mesons can suffer a significant mass modification (and their other properties can also be changed). Here we focus on the $\eta'$ meson, and only on its mass modification effect. Its interaction cross-section is very small and its lifetime is very long, so it can escape from the hot hadronic medium without interacting with it, and will decay in the vacuum after regaining its original mass.  Because of this, we do not have a direct observation channel for the chiral symmetry restoration through $\eta'$, but we have an indirect one by searching for enhanced production of $\eta'$. So we distinguish two $\eta'$ mesons, and we note the properties of the in-medium $\eta'$ with an asterisk, while quantities without further notation correspond to the free $\eta'$ from now on. The Hagedorn formula estimates the enhanced production:
\begin{equation}
f_{\eta'} = \left(\frac{m_{\eta'}^*}{m_{\eta'}}\right)^\alpha e^{\frac{m_{\eta'}-m_{\eta'}^*}{T_\textrm{cond}}} \label{e: etap_enh},
\end{equation}
where  $T_\textrm{cond}$ is the temperature of the condensate. This equation implies, that the lower the mass, the more $\eta'$ we have. This is the basic method applied in this Thesis, looking for signs of enhanced $\eta'$ production.

Ref.~\cite{Vertesi:2009io} suggests a double exponential spectrum of $\eta'$ in the transverse mass\footnote{The transverse mass ($m_\textrm{T}$) is defined from the transverse momentum, $m_\textrm{T} = \sqrt{m^2 + p_\textrm{T}^2}$.}, which predicts more particles to appear in the low $p_\textrm{T}$ region. This double exponential $\eta'$ spetrum will be used as an input in the current simulations, so an overview on the physical picture behind it will be given. 

So there are two components in the $\eta'$ spectrum. Some of the $\eta'$ have large enough transverse momenta so that they can escape from the hot and dense medium to the asymptotic vacuum by decreasing their momenta and increasing their mass, according to the conservation of energy:
\begin{equation} \label{e: cons}
{m_{\eta'}^*}^2 + {p_\textrm{T}^*}^2 = m_{\eta'}^2 + p_\textrm{T}^2.
\end{equation}
Those $\eta'$ mesons, however, that have low transverse momentum\footnote{Low transverse momentum means $\eta'$ mesons with $p_\textrm{T}^* < \sqrt{{m_{\eta'}}^2 - {m_{\eta'}^*}^2}$ as a result of conservation of energy (Eq.~\ref{e: cons})} in the chirally modified vacuum, cannot come on mass shell so they are trapped in that modified vacuum. They are released and come on-shell only when the modified vacuum state decays to the asymptotic vacuum. Both components realize an effective $|0\rangle_*$ + $|\eta'\rangle_* \rightarrow |0\rangle$ + $|\eta'\rangle$ transition, but in a different way.

The first component of the $\eta'$ spertum is the one coming from the condensate of which they were trapped into, to these a random transverse momentum was given: 
\begin{equation} \label{e: etap_low_pT}
f(p_\textrm{T}) = \frac{1}{2\pi m_{\eta'}B^{-1}} e^{- \frac{p_\textrm{T}^2}{2m_{\eta'}B^{-1}} }.
\end{equation}
Note, that most of the $\eta'$ mesons are emitted from the condensate at low $p_\textrm{T}$, and the transverse momentum generated in the $\eta'\rightarrow\eta\pi^+\pi^-\rightarrow(\pi^+\pi^0\pi^-)+\pi^++\pi^-$ chain is small, given the kinematics and the small difference of $m_{\eta'}-5 m_\pi \sim 250$ MeV.

\noindent The second, high $p_\textrm{T}$ part is a hydrodynamical spectrum:
\begin{equation} \label{e: etap_high_pT}
N(m_\textrm{T}) = C m_\textrm{T}^\alpha e^{-m_\textrm{T}/T_\textrm{eff}},
\end{equation}
$\alpha = 1-d/2$, where $d$ is the spatial dimension of the expansion (thus falls between -1/2 and 1). The effective temperature is $T_\textrm{eff} = T_0 + m {\langle u_\textrm{T}\rangle}^2$, where $T_0$ is the freeze-out temperature and ${\langle u_\textrm{T}\rangle}^2$ is the average transverse flow, respectively. In this spectrum, $\eta'$ momenta are shifted according to Eq.~\ref{e: cons}.

The final $\eta'$ spectrum is the sum of Eq.~\ref{e: etap_low_pT} and Eq.~\ref{e: etap_high_pT}, which corresponds to a double exponential spectrum. In the current work the following parametrization was used:
\begin{equation} \label{e: double_exp}
\frac{1}{2\pi m_\textrm{T}} \frac{d N}{d m_\textrm{T}} = A \cdot \exp[-B(m_\textrm{T}-m_{\eta'})]+C \cdot \exp[-D(m_\textrm{T}-m_{\eta'})].
\end{equation}

The $A$, $B$, $C$ and $D$ parameters of the spectra and the enhancement (namely the $\eta$ and $\eta'$ weight factors) coming from this double exponential spectrum were obtained in Ref.~\cite{Vertesi:2009io} for different resonance models (ALCOR~\cite{ALCOR}, Kaneta {\it et al.}~\cite{Kaneta}, Letessier {\it et al.}~\cite{Letessier}, UrQMD~\cite{UrQMD} and Stachel {\it et al.}~\cite{Stachel}) and are listed here in Table \ref{t: etap_models} for $\eta'$, and Table~\ref{t: eta_models} for $\eta$ mesons. 

\begin{figure}[h!]
 \begin{center}
  \vspace{50pt}
  \subfloat{ \includegraphics[width=0.96\textwidth]{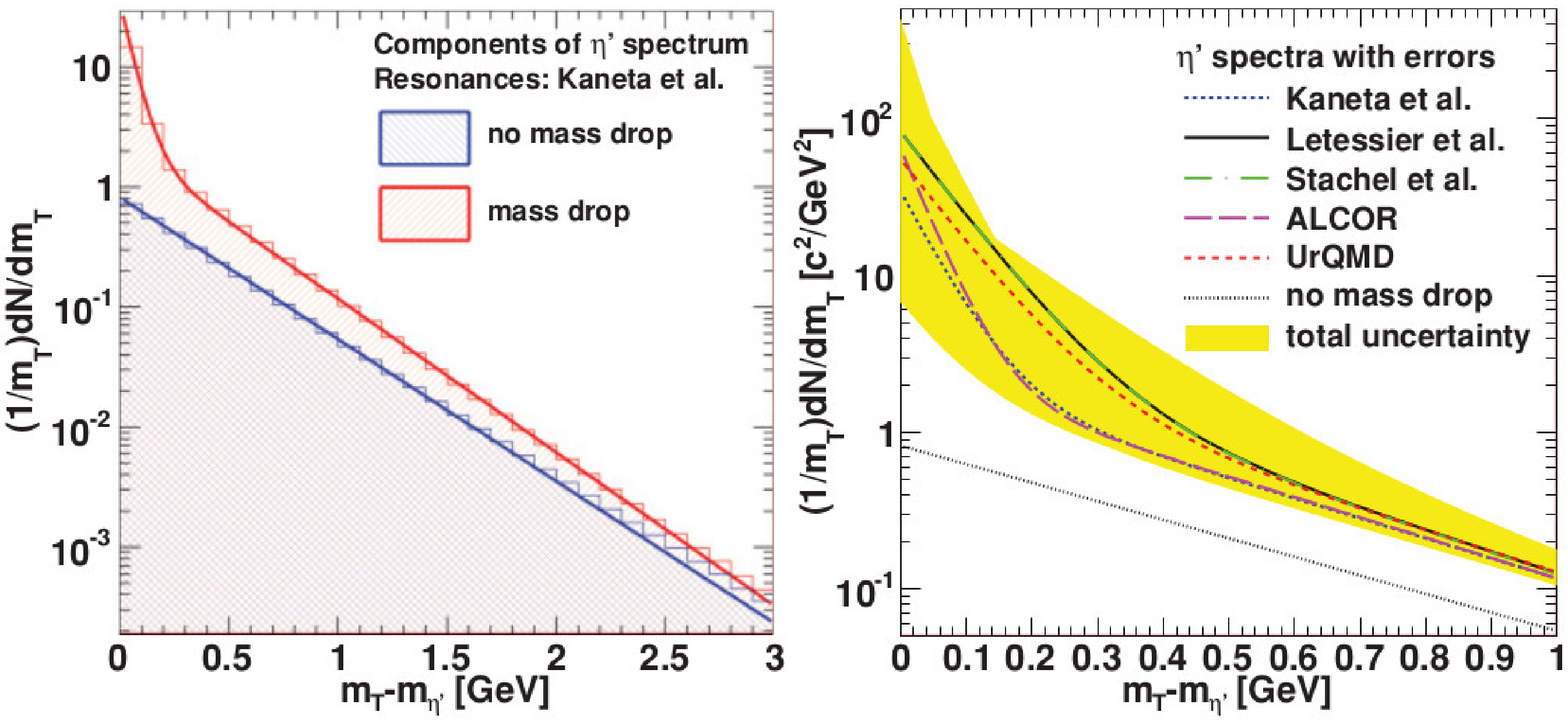} } \\[2cm]
  \subfloat{ \includegraphics[width=0.96\textwidth]{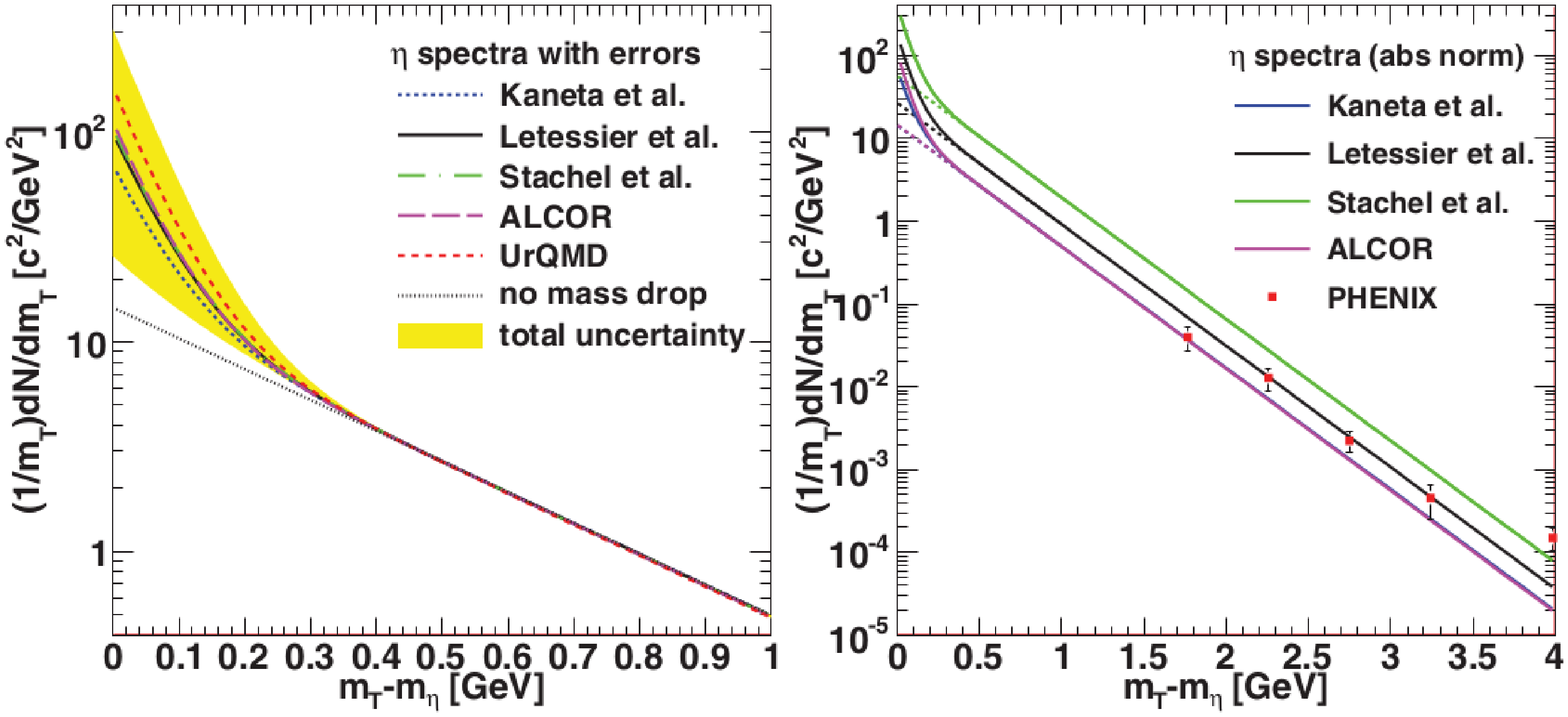}}
 \caption{$\eta'$ and $\eta$ enhancement as observed indirectly from $\pi^{\pm} \pi^{\pm}$ Bose-Einstein correlation measurements at RHIC, in 200 GeV Au+Au collisions, as in Refs.~\cite{Vertesi:2009io, Vertesi:2010si}}   \label{f: eta+etap_PRC} 
 \vspace{40pt}
 \end{center}     
\end{figure}

The input spectra for both $\eta'$ and $\eta$ mesons are shown on Fig.~\ref{f: eta+etap_PRC}. The top left part shows the reconstructed $(m_\textrm{T}-m)$ spectrum of the $\eta'$ mesons using the  resonance ratios of the Kaneta model~\cite{Kaneta}. The blue line corresponds to an $\eta'$ spectrum without in-medium mass reduction, the red line shows the enhanced scenario.
   Top right part presents a comparison of reconstructed $(m_\textrm{T}-m)$ spectra of the $\eta'$ mesons for different models. The shaded (yellow) band represents the total error. Above $m_\textrm{T}-m_{\eta}'$ = 1 GeV, all models result in very similar values, corresponding to an approximate $m_\textrm{T}$ scaling. This figure indicates that the violation of this $m_\textrm{T}$ scaling is model dependent. The bottom left part of Fig.~\ref{f: eta+etap_PRC} shows a comparison of reconstructed $(m_\textrm{T}-m)$ spectra of the $\eta$ mesons from different models. The  bottom right panel shows the  absolute normalized spectra and compares it  to PHENIX 200 GeV central Au+Au collision measurements~\cite{PRC75}. 
\begin{table}
\begin{center}
\begin{tabular}[H]{ l | c c c c | c }
Model&A&B&C&D&$f_{\eta'}$\\ \hline \hline
No enhancement&0.82&2.72&0&0&-- \\
ALCOR &2.30&2.98&62.4&23.5&43.4\\
Kaneta {\it et al.}&2.21&2.94&32.4&18.7&25.6\\
Letessier {\it et al.}&2.91&3.14&80.1&12.8&67.6\\
Stachel {\it et al.}&2.85&3.13&80.0&12.8&67.6\\
UrQMD&2.75&3.07&52.5&12.7&45.0 
\end{tabular}
\end{center}
\caption{$\eta'$ spectrum parameters of Eq.~\ref{e: double_exp} from Ref.~\cite{Vertesi:2009io} for different models. The spectrum without enhancement was obtained from the Kaneta model using the $m^*_{\eta'} = m_{\eta'}$ substitution. } \label{t: etap_models}
\end{table}
\begin{table}
\begin{center}
\begin{tabular}[H]{ l | c c c c | c }
Model&A&B&C&D&$f_\eta$\\ \hline \hline
No enhancement&14.6&3.38&0&0&-- \\
ALCOR &14.6&3.40&97.0&17.8&5.25\\
Kaneta {\it et al.}&14.6&3.38&54.9&16.2&3.47\\
Letessier {\it et al.}&14.6&3.38&84.1&16.9&4.75\\
Stachel {\it et al.}&14.5&3.38&89.2&17.0&4.97\\
UrQMD&14.6&3.41&148&17.9&7.49 
\end{tabular}
\end{center}
\caption{$\eta$ spectrum parameters of Eq.~\ref{e: double_exp} from Ref.~\cite{Vertesi:2009io} for different models} \label{t: eta_models}
\end{table}
\pagebreak

\chapter{Simulations of the $\eta'$ excess}

\section{Introducing the simulation environment}

Our simulation tool, EXODUS is an event generator that has already been tuned to the PHENIX experiment~\cite{PPG088}. It uses a fast Monte-Carlo method to simulate individual particles as well as their decays. It is capable of generating single particles, whole events, and most importantly, it can directly extract the dilepton spectrum in the specific PHENIX acceptance. 
Basically, this means a rapidity cut at $|y|<0.35$, and a $p_\textrm{T} < 0.2$ GeV/$c$ cut in the transverse momenta of electrons and positrons, but further corrections have also been included. \\

\noindent The program reads the particle information, e.g.\ the mass, width, charge and spin, from the "defined\_particles.txt" file. For particle numbering it uses the standard Particle Data Group codes~\cite{PDGnum}. 
The decay definitions are in the "defined\_decays.txt" file. One needs to specify the ID's of the parent and child particles, the number of child particles (NBody) and the branching ratio of the decay. Decays can be turned off by setting the option "Enabled" from 1 to 0. Decay products ("children particles") can suffer further decays if "ChildrenStable" is set to 0 (and if that certain decay is defined). EXODUS can be run either interactively, or using an input file. \\ 

\noindent 
In the input file, one has to choose between some predefined setups (e.g.\ "PHENIX single particles", "PHENIX: complete events" or the desired \\ "PHENIX electron cocktail"). After some obvious steps, which include defining the name of the output file (supports root and text format too) and the number of events to generate, the program requires the weight factor of $\pi^{0}$'s (which is basically the number of pions obtained from the integral of the $\pi^{0}$ $p_\textrm{T}$-distribution). After this, the weight factors have to be given for the other particles as well. In this Thesis minimum bias data was investigated, so the simulation has been done accordingly, by setting the number of binary N+N collisions to 257.8. At the end, the parameters of the  $p_\textrm{T}$-distribution (Eq.~\ref{e: PPG088}) have to be specified for each particle species, that decay to dileptons. PHENIX defaults to use a Tsallis distribution of Eq.~\ref{e: PPG088} with the same parameters for all particles, except the normalization factor, $A_\textrm{Tsallis}$. The values of the parameters are published in Ref.~\cite{PPG088}.

\pagebreak
 
\section{Model predictions for the $\eta'$ and $\eta$ enhancement}
\vspace{1cm}
The first estimation to understand the $\eta'$ meson's role in the dilepton spectrum was elaborated by using different  input spectra from the PHENIX analysis~\cite{PPG088} in EXODUS, for both the $\eta$ and $\eta'$ mesons, of which parameters have been obtained by an indirect measurement~\cite{Vertesi:2009io} and listed here in Table~\ref{t: etap_models} and \ref{t: eta_models}. 

In addition to the rescaling of $\eta$ and $\eta'$ with the obtained weight factors (according to the model predictions), additional $\eta$ mesons appear from the enhanced number of $\eta'$ mesons through the $\eta'\rightarrow\pi\pi\eta$ decay channel. Since none of the $\eta'\rightarrow\eta$ decays are calculated by EXODUS at this point, the $\eta$ weight factor has to be rescaled. The formula below na\"{\i}vely estimates the $\eta$ mesons coming from the enhanced $\eta'$ mesons:
\begin{equation} \label{e: naive_eta}
f_\eta = \frac{N_\eta (\textrm{enhanced})}{N_\eta (\textrm{original})} = 1 + (f_\eta'-1)(N_{\eta'}/N_\eta) \times \textrm{BR}(\eta' \rightarrow \pi\pi\eta),
\end{equation}
where BR stands for branching ratio.

\begin{figure}[H]
 \begin{center}
  \subfloat{ \includegraphics[width=0.9\textwidth]{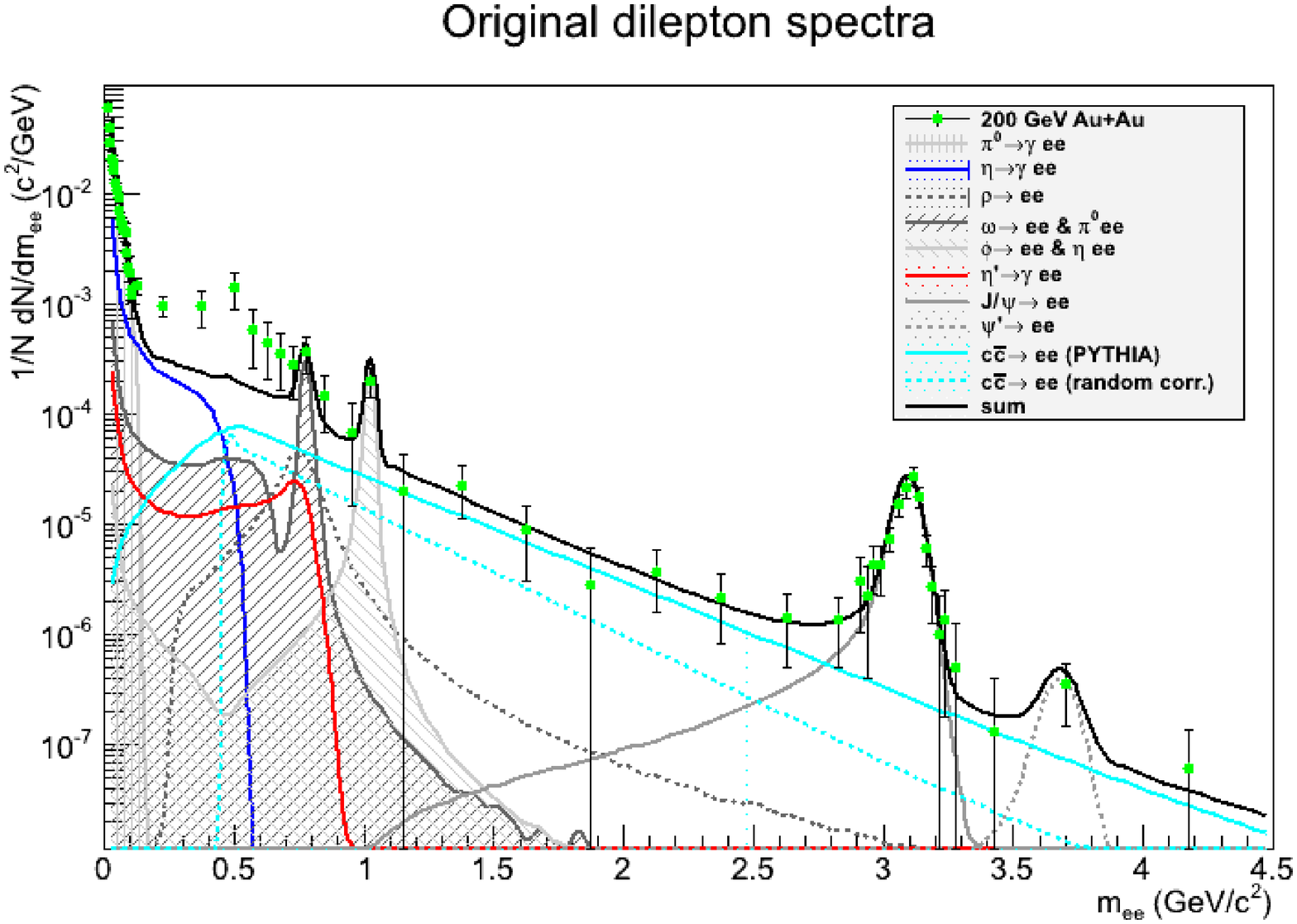} } \\
  \subfloat{ \includegraphics[width=0.9\textwidth]{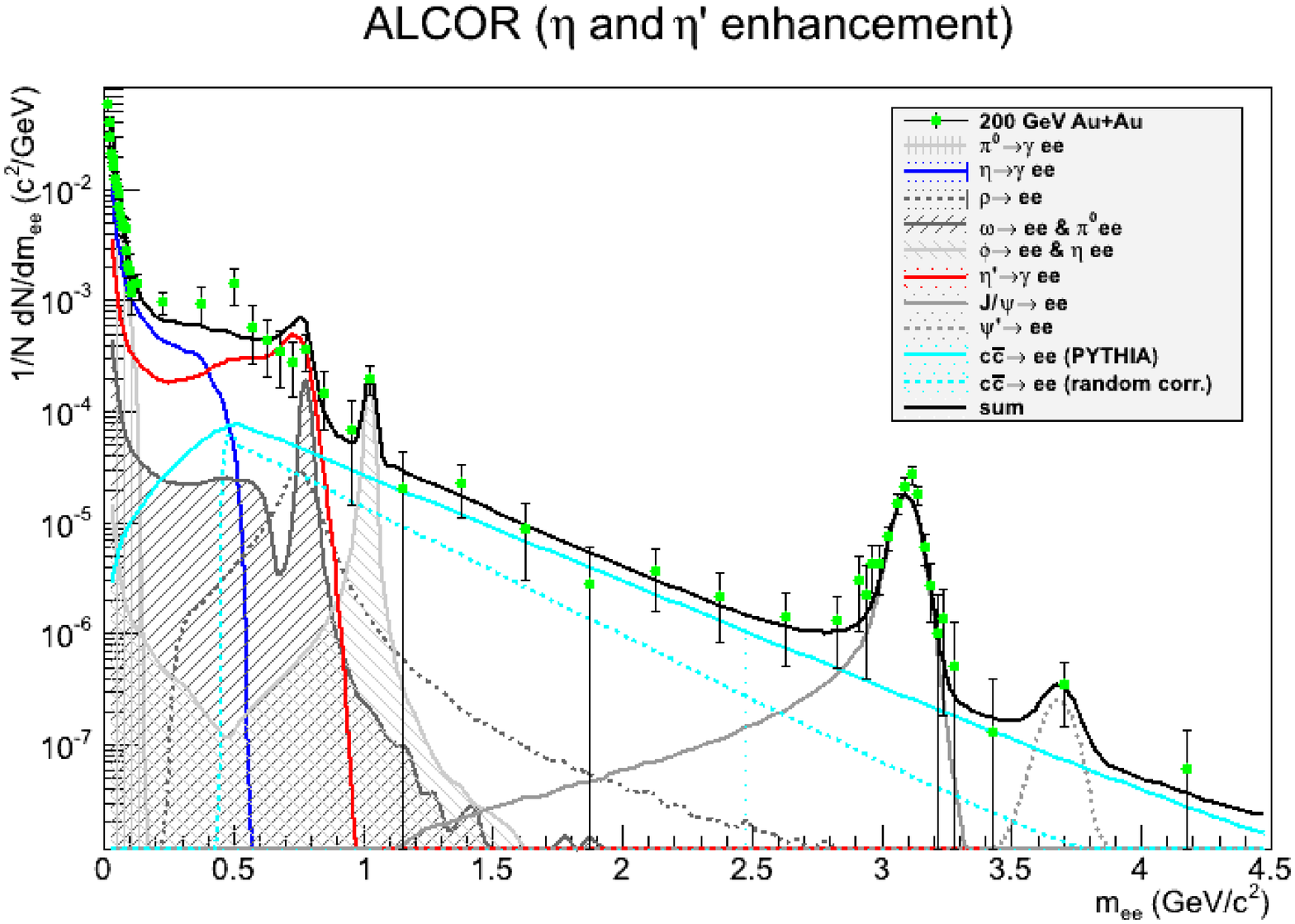}}
 \caption{\\{\bf Top:} EXODUS simulation result reproducing the dilepton spectra of  the PHENIX analysis \cite{PPG088} without $\eta'$ enhancement, \\
 $\chi^2$/NDF (for 0.15 GeV < $m_{ee}$ < 0.7 GeV) = 4.6, CL = 0.011\% \\
 {\bf Bottom:} A promising first simulation including $\eta'$ and $\eta$ enhancement predicted by the ALCOR~\cite{ALCOR} model, 
 $\chi^2$/NDF (for 0.15~GeV~<~$m_{ee}$~<~0.7~GeV) = 2.25, CL = 1.8\% }   \label{f: eta+etaprime_1} 
 \end{center}     
\end{figure}

\begin{figure}[H]
 \begin{center}
 \subfloat{\includegraphics[width=0.9\textwidth]{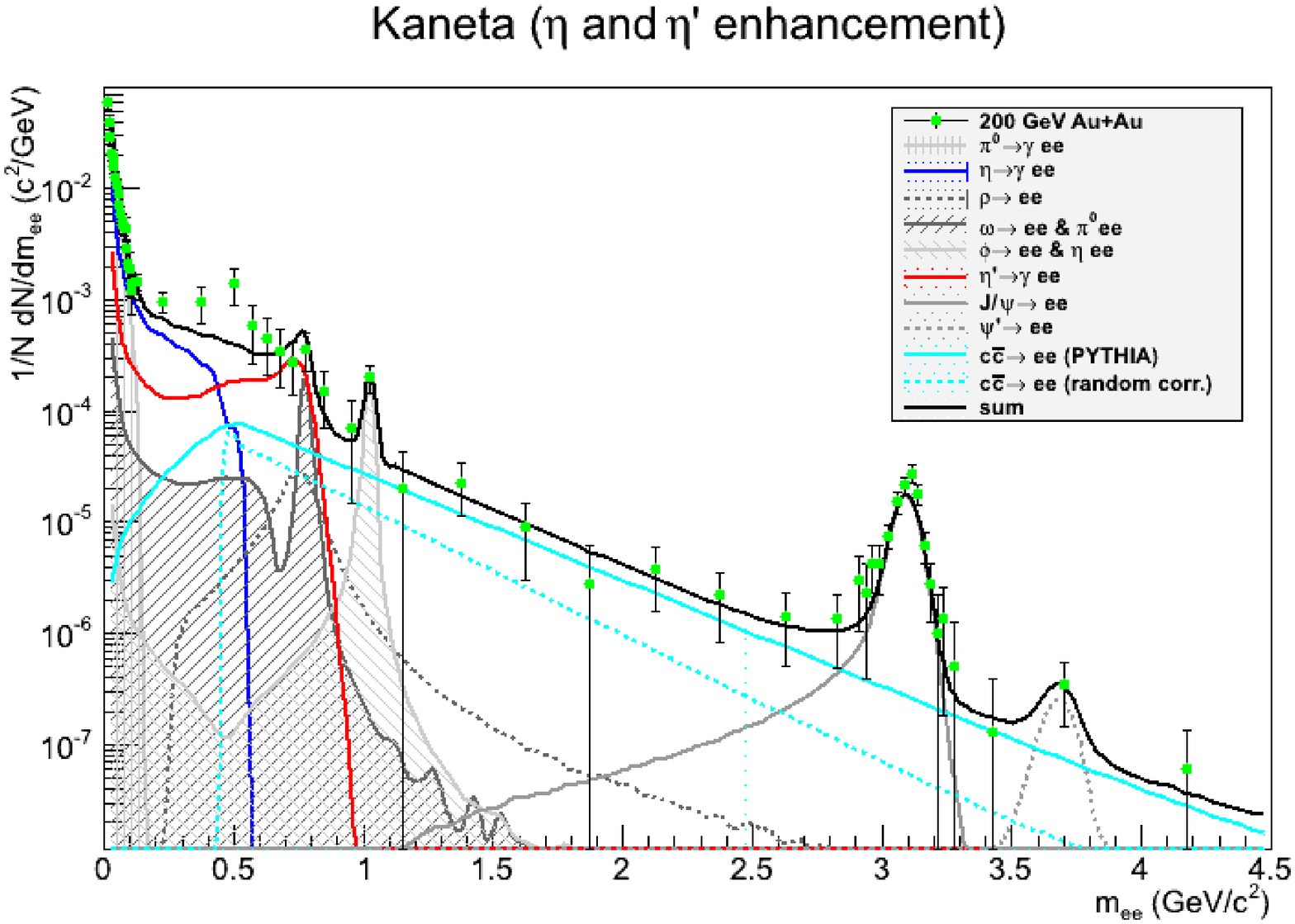} } \\
 \subfloat{ \includegraphics[width=0.9\textwidth]{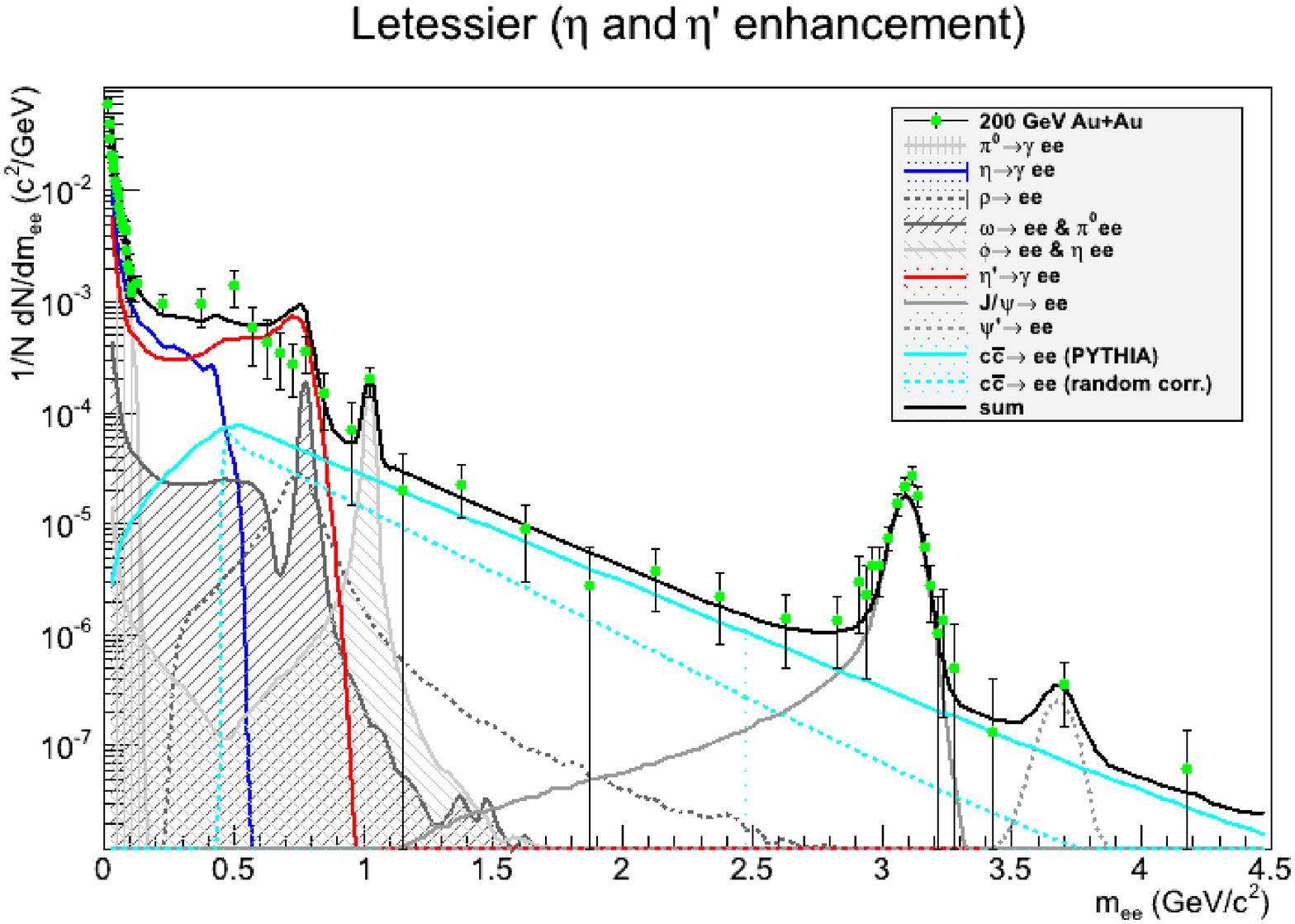} }
 \caption{EXODUS simulation results including $\eta'$ and $\eta$ enhancement. The simulations are based on the results of~\cite{Vertesi:2009io, Rafelski}, using resonance ratios from the Kaneta~\cite{Kaneta} (top) and the Letessier~\cite{Letessier} (bottom) models for multiplicities. \\
 Kaneta: $\chi^2$/NDF (for 0.15 GeV < $m_{ee}$ < 0.7 GeV) = 1.35, CL = 23.16\%. This result is statistically acceptable. \\
 Letessier: $\chi^2$/NDF is larger, than in the case without enhancement as it overestimates the effect.}   \label{f: eta+etaprime_2}
 \end{center}     
\end{figure}

\begin{figure}[H]
 \begin{center} 
 \subfloat{\includegraphics[width=0.9\textwidth]{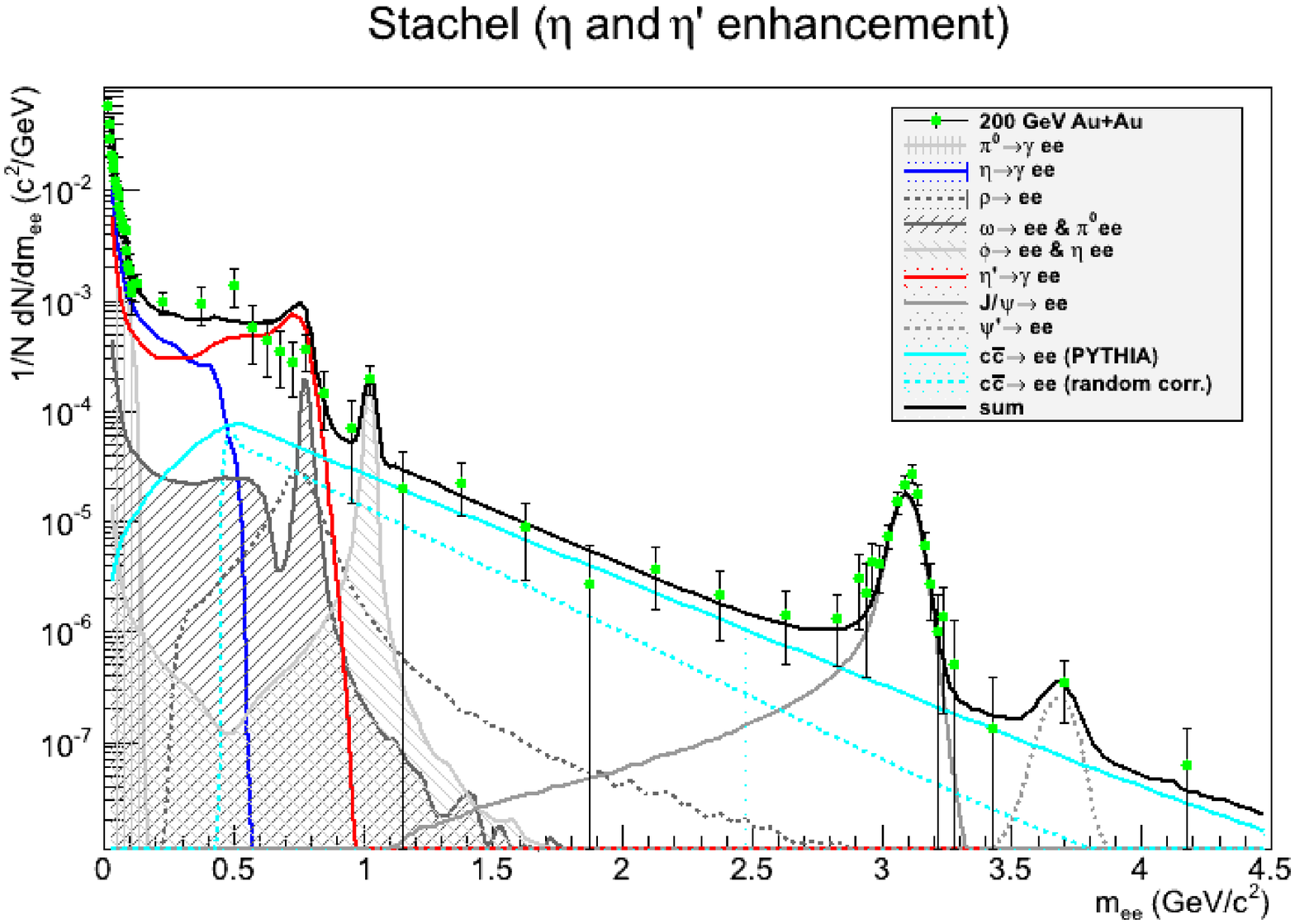} } \\
 \subfloat{\includegraphics[width=0.9\textwidth]{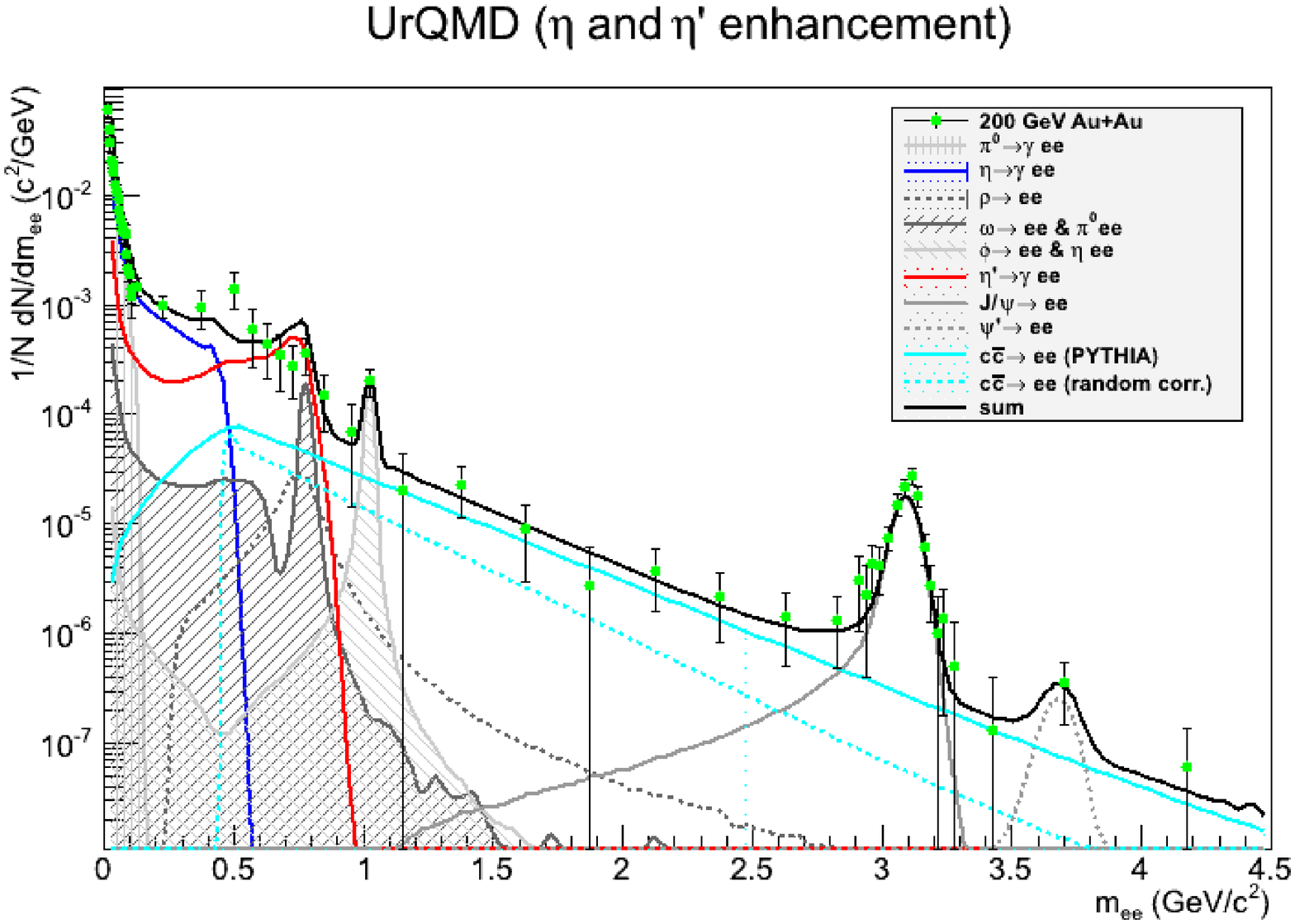} }
 \caption{Simulation result including $\eta'$ and $\eta$ enhancement predicted by the Stachel~\cite{Stachel} (top) and UrQMD~\cite{UrQMD} (bottom) model. Obviously, these calculations are also promising, but they need to be fine-tuned to obtain a quantitative description of the low-mass dilepton enhancement.} \label{f: eta+etaprime_3}
 \end{center}      
\end{figure}

So using the input spectra of Eq.~\ref{e: double_exp} for $\eta'$ and $\eta$, Eq.~\ref{e: PPG088} for all the others an EXODUS simulation was elaborated for the different models. The simulation results for $\sqrt{s_{NN}}=200$ GeV Au+Au collisions with the $\chi^2$/NDF and confidence level values in the region of the excess (0.15 GeV < $p_\textrm{T}$ < 0.7 GeV) are shown on Figs.~\ref{f: eta+etaprime_1}-\ref{f: eta+etaprime_3}. Note, that this simulation has not been optimized yet to the best possible description of the dilepton spectrum, however, its fit quality for some of the models is already acceptable.

This so-called first estimation obviously fails to explain completely the enhancement of the dilepton spectrum, but in some cases (e.g.\ for the Kaneta  model) the results are quite promising. Given that, to our best knowledge, no other, similarly promising explanations up to the time of writing this thesis are publicly available, we follow up to this direction with more detailed simulations which might give a better qualitative-, perhaps even a quantitative description of the low-mass enhancement.
\vspace{2cm}

\pagebreak

\section{Extending EXODUS: Simulation of decay chains}

So far only the direct dileptonic decays were included and calculated by EXODUS, other decays were taken into account via the fact, that inclusive $p_\textrm{T}$ spectra were fitted which already contain the decays chain resulting the current meson. The normalization factors of the mesons are obtained from their $p_\textrm{T}$ spectra and other independent sources (e.g.\ jet fragmentation) if data was not available. For further details, see Ref.~\cite{PPG088}. To explore the dilepton contribution of $\eta'$ in detail, however, specially when using different $p_\textrm{T}$ spectrum for that, the $\eta'$ chain decays, e.g.\ $\eta'\rightarrow \eta \pi^0\pi^0\rightarrow (\pi^+\pi^0\pi^-)\pi^0\pi^0$ decay chains can play a significant role. But adding new decays slows down EXODUS significantly, and more importantly, one has to define the meson's weight factors at the beginning of the simulation (in the input file), which is not an effective way when elaborating a scan through different $\eta'$ weight factors, since its decay product's weights have to be corrected manually in each case.

In this section a different simulation method is presented, e.g.\ the simulation is elaborated in each individual decay chain instead of generating the whole dilepton cocktail every time. This has the advantage that one is able to follow the contribution of individual particles, or simulate only the ones which parameters have been changed, and last but not at least this new approach speeds up the simulation by a factor of 10 (as compared to its original speed). So in addition to the direct dileptonic decays already defined in EXODUS, other meson's decays were defined which result in a significant dilepton contribution, as listed in Table~\ref{t: decays}.

\begin{figure}[h!]
 \begin{center}
 \includegraphics[width=0.8\textwidth]{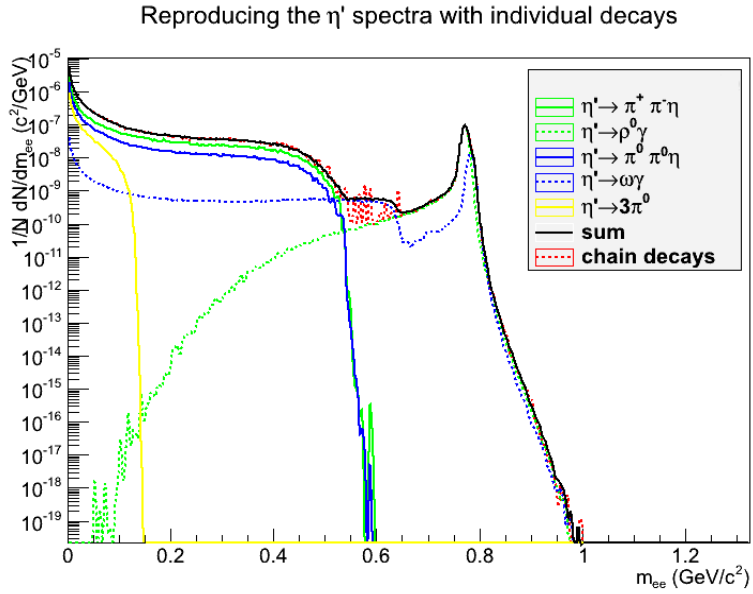} 
 \caption{Reproducing the $\eta'$ spectra in two different ways: by simulating its relevant decays (ID 100-104 \& ID 006) individually and adding them up (black line), and by defining the same decays at once (red line). The result shows that from individual decays the $\eta'$ spectrum can be reproduced faithfully. Thus this method will be applied to other mesons too, as its numerical implementation is much faster than that of the Monte Carlo simulation of each decay chains simultaneously.} \label{f: etap_cons}
 \end{center}      
\end{figure}

The consistency of this approach is checked via the $\eta'$ dilepton spectrum; it was first generated with all their decays defined at once (previous approach), then its individual decays were simulated and added up. The two results were cross-checked and are in accordance with each other, as seen on Fig.~\ref{f: etap_cons}. This indicates that using this method, generating individual decays instead of generating every decay of a given particle at once are two consistent methods and both are well suited to study $\eta'$ mass modification effects.

\vspace{2cm}
\pagebreak

\vspace{2cm}

\begin{table}[h!]
\begin{center}
\begin{tabular}{c|l|l|c}
particle&      decay&                                      BR&    ID \\ \hline \hline
$\eta'_{331}$  &$\pi^+_{211}$ $\pi^-_{-211}$ $\eta_{221}$&  0.434& 100 \\
            &  $\rho^0_{113}$ $\gamma_{22}$&               0.29&  101 \\
	    &  $\pi^0_{111}$ $\pi^0_{111}$ $\eta_{221}$&   0.216& 102 \\
	    &  $\omega_{223}$ $\gamma_{22}$&               0.027& 103 \\
	    &  $3\pi^0_{111}$&                             $1.68 \cdot 10^{-3}$&104 \\
	    &  $\mathbf {e}^+_{-11}$ $\mathbf{e}^-_{11}$ $\mathbf\gamma_{22}$&$9\cdot 10^{-4}$& 6\\ \hline
$\omega_{223}$& $\pi^+_{211}$ $\pi^-_{-211}$ $\pi^0_{111}$& 0.892& 106 \\
	     & $\pi^0_{111}$ $\gamma_{22}$&		   0.0828&107 \\
	     & $\eta_{221}$ $\gamma_{22}$&		   $4.6\cdot 10^{-4}$&108 \\
	     & $\pi^0_{111}$ $\mathbf {e}^+_{-11}$ $\mathbf{e}^-_{11}$&$7.7\cdot 10^{-4}$&9\\
	     & $\mathbf {e}^+_{-11}$ $\mathbf{e}^-_{11}$&  $7.28\cdot 10^{-5}$&8\\ \hline
$\rho^0_{113}$& $\pi^0_{111}$ $\gamma_{22}$&                 $6\cdot 10^{-4}$&  109\\
	     & $\eta_{221}$ $\gamma_{22}$&                 $3\cdot 10^{-4}$&  110\\
	     & $\pi^0_{111}$ $\pi^0_{111}$ $\gamma_{22}$&    $4.5\cdot 10^{-5}$&111\\
             & $\mathbf {e}^+_{-11}$ $\mathbf{e}^-_{11}$&  $4.72\cdot 10^{-5}$&7 \\ \hline
$\eta_{221}$ &  $3\pi^0_{111}$&              		   0.3257&14 \\
	    &  $\pi^0_{111}$ $\gamma_{22}$ $\gamma_{22}$&  $2.7\cdot 10^{-4}$&112 \\
            &  $\pi^+_{211}$ $\pi^-_{-211}$ $\pi^0_{111}$& 0.2274&113 \\
            &  $\mathbf {e}^+_{-11}$ $\mathbf{e}^-_{11}$ $\gamma_{22}$&$7\cdot 10^{-3}$&4\\
            &  $\mathbf {2e}^+_{-11}$ $\mathbf{2e}^-_{11}$&$6.9\cdot 10^{-5}$&24\\ \hline
$\pi^0_{111}$ & $\mathbf {e}^+_{-11}$ $\mathbf{e}^-_{11}$ $\mathbf\gamma_{22}$&0.0174&2\\ \hline
$\varphi_{333}$ & $\eta_{221} \gamma_{22}$ & $1.309\cdot 10^{-2} $                & 115\\
              & $\pi^0_{111} \gamma_{22}$ & $1.27\cdot 10^{-3}$                  & 116\\ 
              & $\mathbf {e}^+_{-11}$ $\mathbf{e}^-_{11}$ &$2.954\cdot 10^{-4}$  &117 \\ 	
              & $\eta_{221}\mathbf {e}^+_{-11}$ $\mathbf{e}^-_{11}$&$1.15\cdot 10^{-4}$&118 \\ \hline
J$/\Psi_{443}$ & $\mathbf {e}^+_{-11}$ $\mathbf{e}^-_{11}$ & $5.94\cdot 10^{-2}$ & 119 \\ \hline
$\Psi'_{444}$ & $\mathbf {e}^+_{-11}$ $\mathbf{e}^-_{11}$ & $9.7\cdot 10^{-6}$ & 120 
\end{tabular}
\end{center}
\caption{Table of the relevant decays contributing to the dilepton spectrum. The lower indices are the particles' ID's, according to the Particle Data Group convention~\cite{PDGnum}, "BR" stands for branching ratio. "ID" is the decay identification number defined in EXODUS. Originally, only the mainly direct dileptonic decays were defined in EXODUS, while ID $\geq$ 100 corresponds to the newly added decays and decay chains.}\label{t: decays}
\end{table}

\pagebreak

\section{The $m^*-B^{-1}$ scan for $\eta'$ mesons}

So far only model predictions were used for the $\eta'$ spectrum, in this section we attempt to find the best $\eta'$ spectrum while scanning through the parameters of the $\eta'$ spectra. 

The double exponential spectrum of Eq.\ \ref{e: double_exp} can be written in a different form (according to Ref.~\cite{Vertesi:2009io}). The shape of the low-$p_\textrm{T}$ part is governed by the temperature-like variable, $B^{-1}$, characterising the momentum distribution of the $\eta'$ mesons coming from the $\eta'$ condensate, the in-medium mass of $\eta'$ is responsible for the normalization (see Eq.~\ref{e: etap_enh}).

The $B^{-1}$ and $m^*$ parameters define a certain $\eta'$ spectrum, so a scan through their physically reasonable values was elaborated with the two different approaches. In the first one only the direct dileptonic decays are defined, and this result will be contrasted to the case when all the $\eta'$ decays of Table~\ref{t: decays} are included. The second approach estimates the number of $\eta$ mesons coming from the enhanced number of $\eta'$ mesons better than the first one, the na\"{\i}ve estimation of Eq.~\ref{e: naive_eta}, and changes the shape of the $\eta'$ dilepton spectrum as well. 

\begin{figure}[H]
 \begin{center}
 \subfloat{\includegraphics[width=0.8\textwidth]{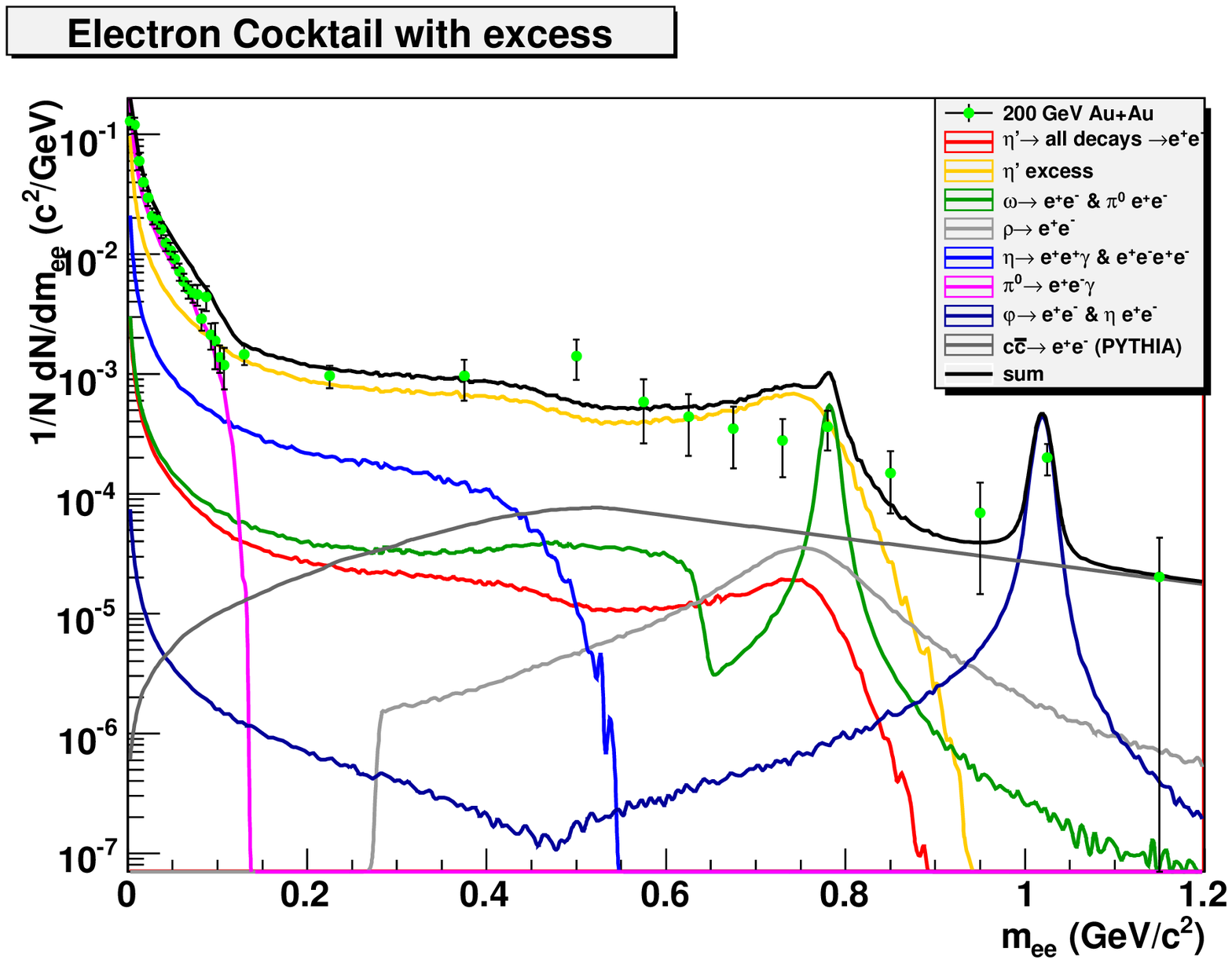}} \\
 \subfloat{\includegraphics[width=0.8\textwidth]{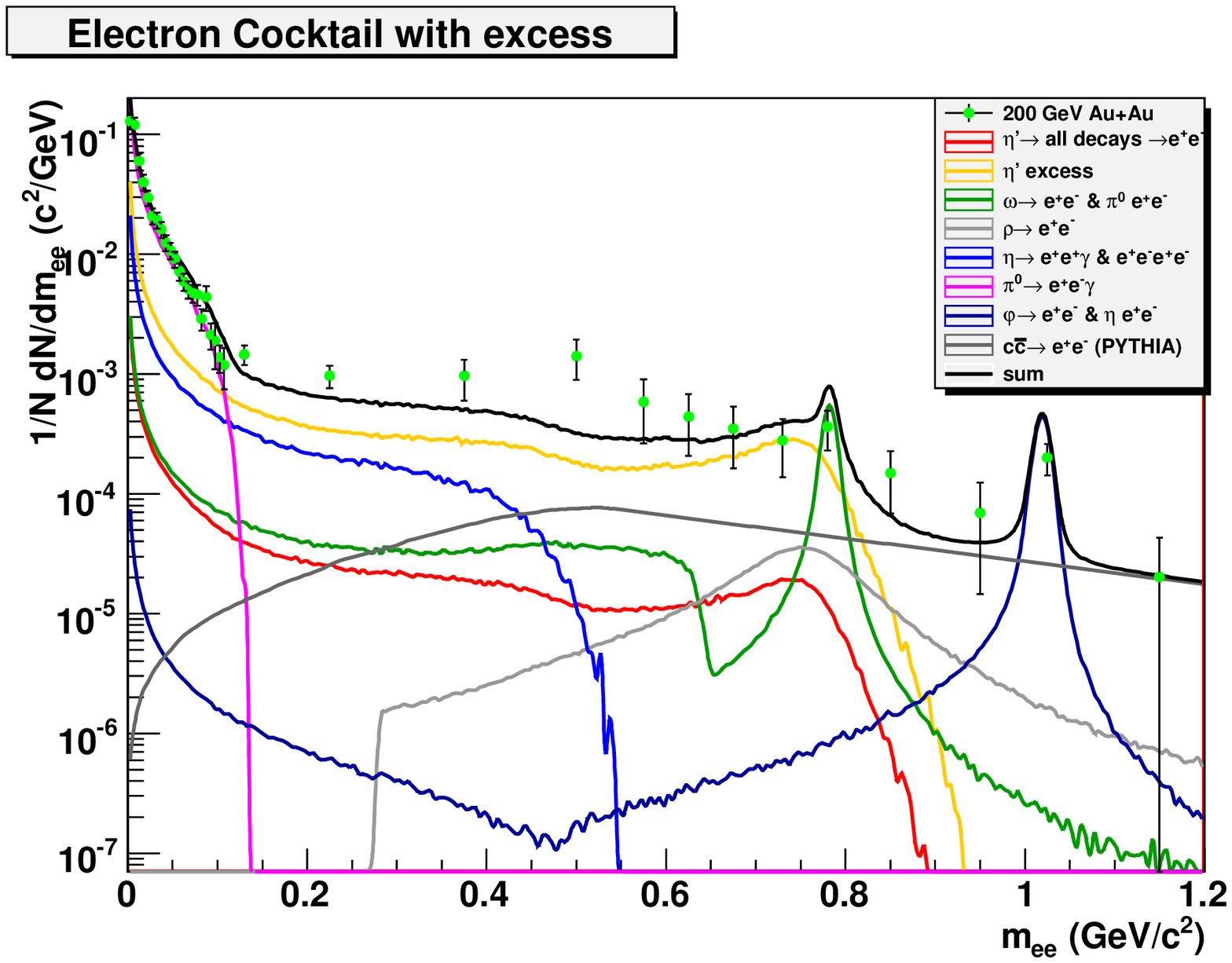}} 
  \caption{Examples from the result of the $\eta'$ input spectra scan for $B^{-1} = 350$ MeV, with mass values (from top to the bottom): 300 MeV, 450 MeV} \label{f: scan_cocktail1}
 \end{center}      
\end{figure}

\begin{figure}[H]
 \begin{center}
 \subfloat{\includegraphics[width=0.8\textwidth]{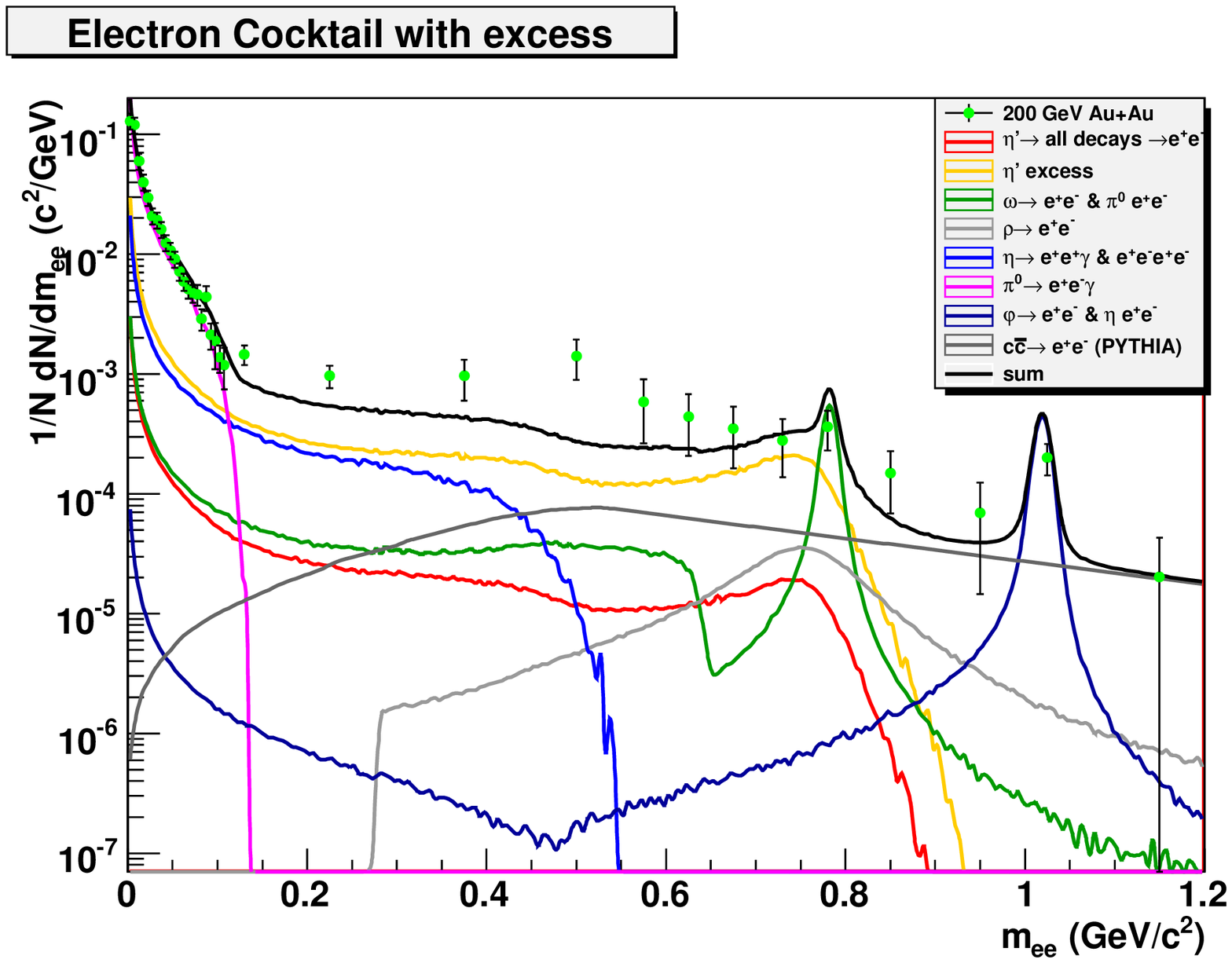}} \\
 \subfloat{\includegraphics[width=0.8\textwidth]{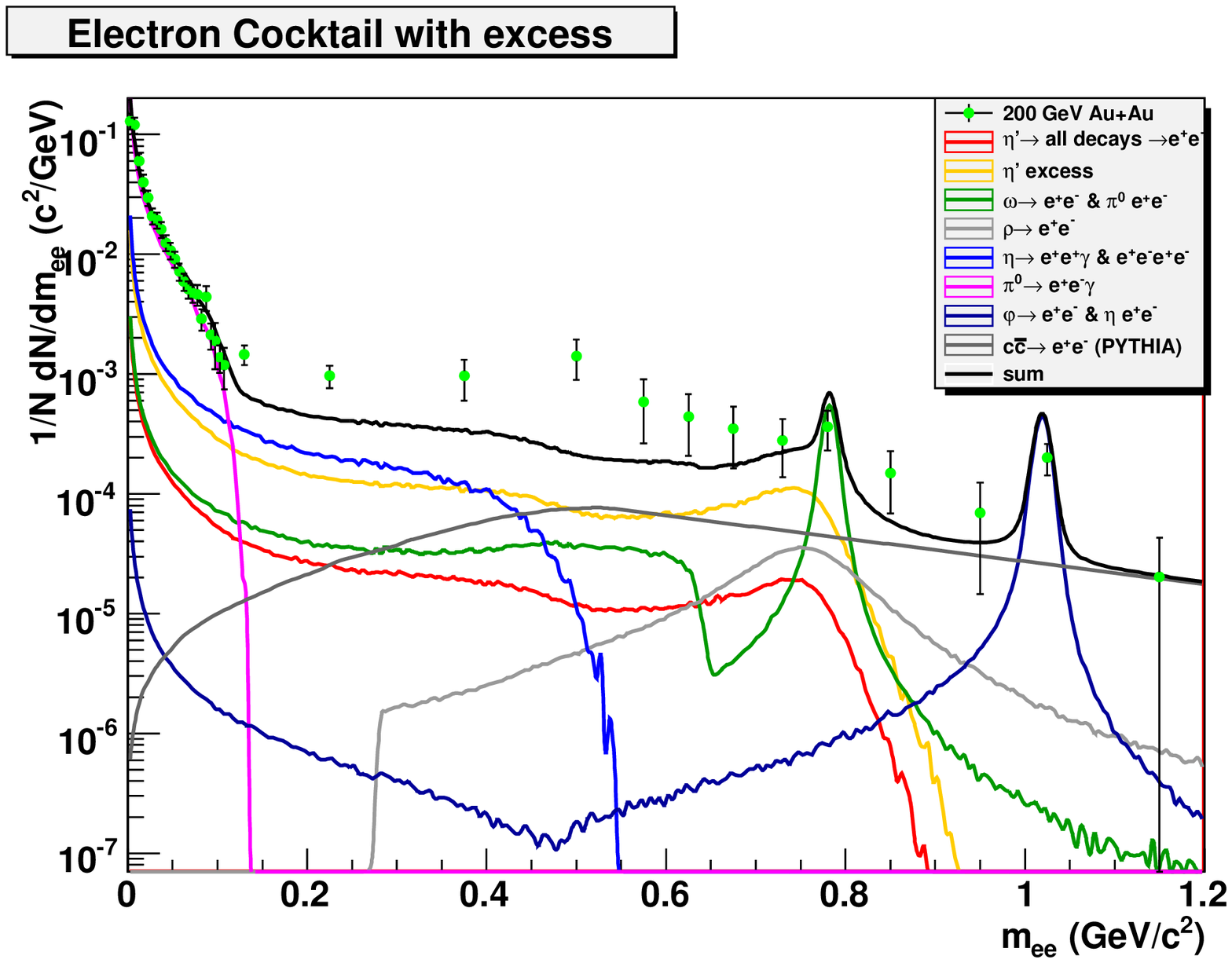}} 
  \caption{Examples from the result of the $\eta'$ input spectra scan for $B^{-1} = 350$ MeV, with mass values (from top to the bottom): 500 MeV, 600 MeV} \label{f: scan_cocktail2}
 \end{center}      
\end{figure}

\begin{figure}[H]
 \begin{center}
 \subfloat{\includegraphics[width=0.8\textwidth]{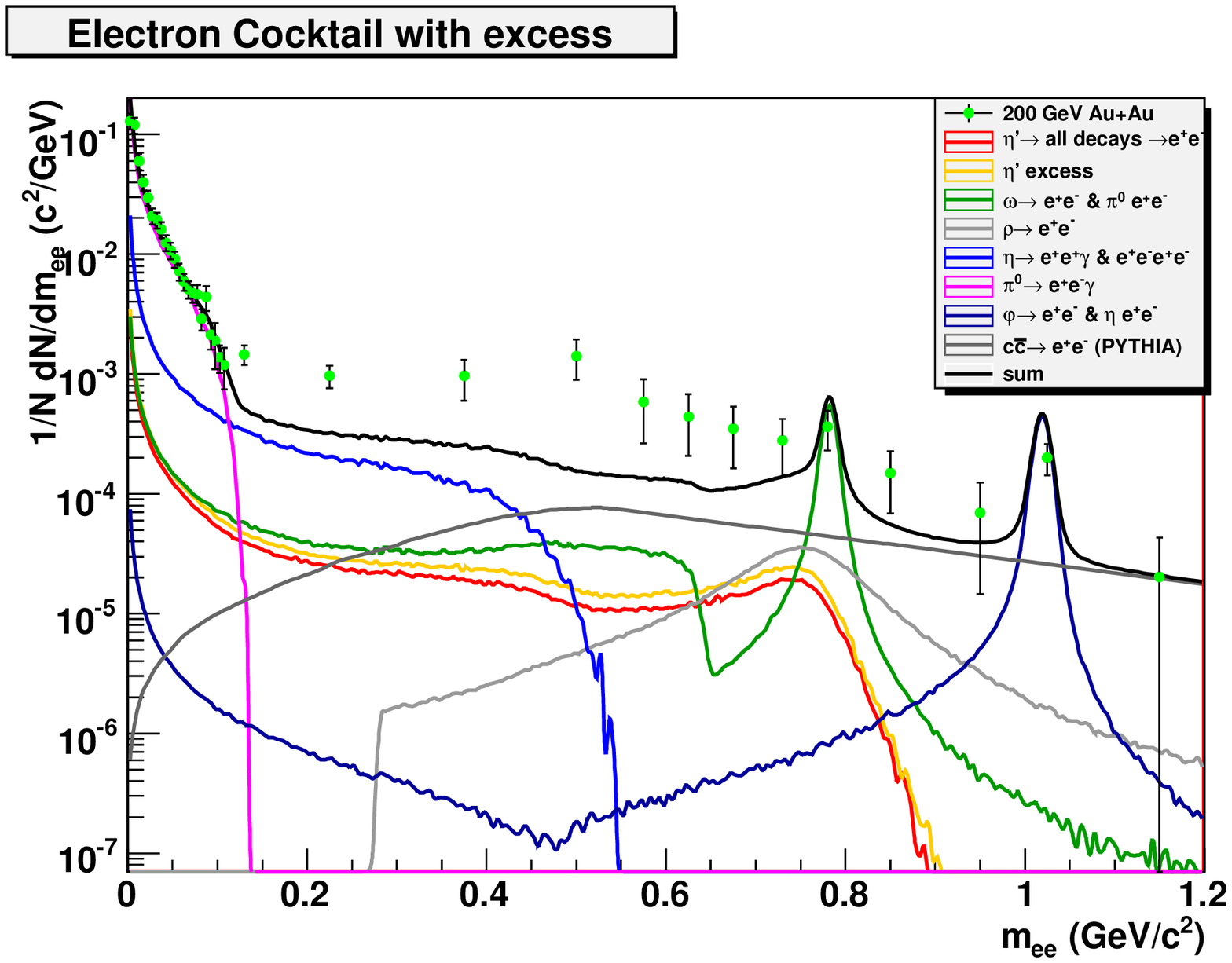}} \\
 \subfloat{\includegraphics[width=0.8\textwidth]{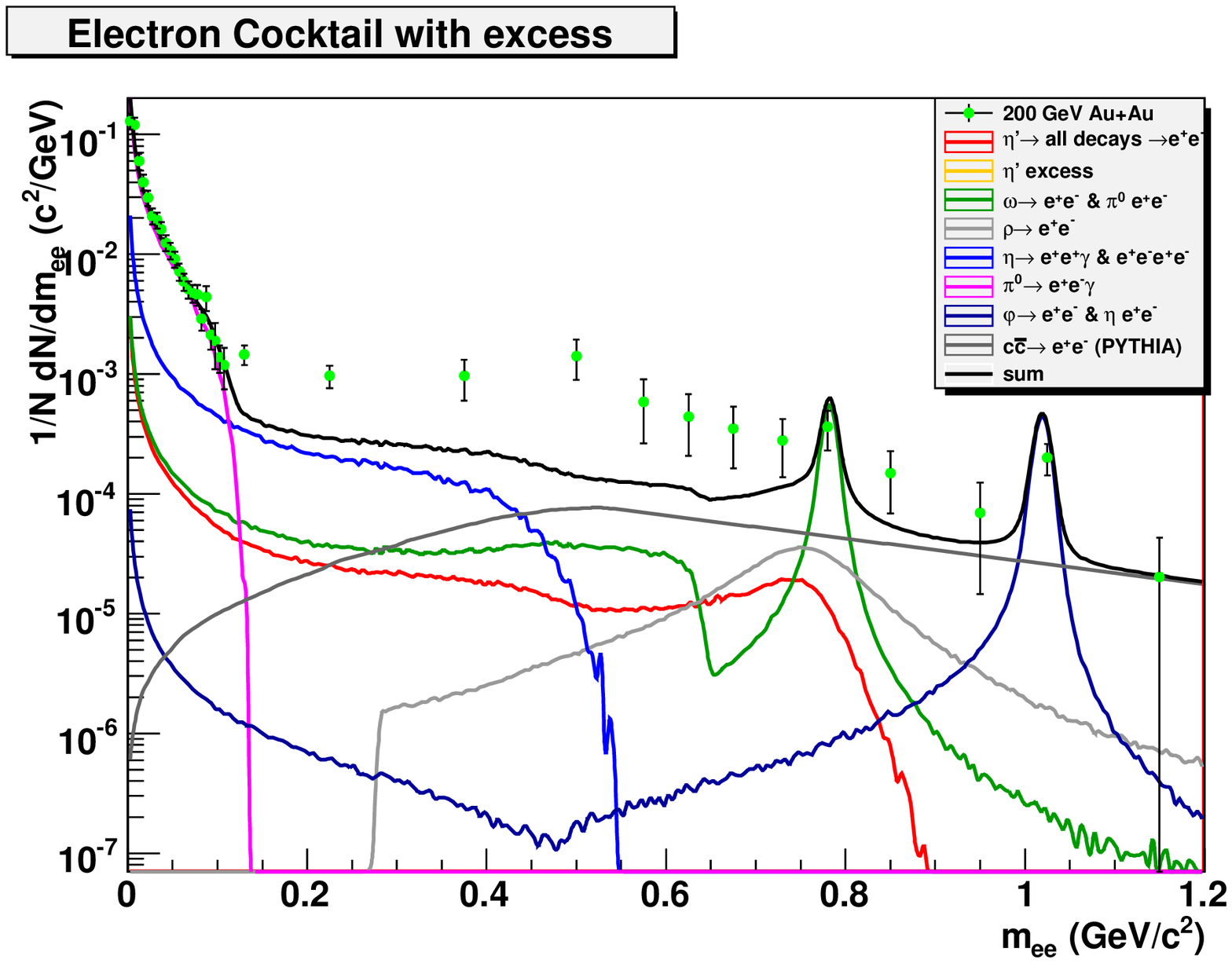}} 
 \caption{Examples from the result of the $\eta'$ input spectra scan for $B^{-1} = 350$ MeV, with mass values (from top to the bottom): 800 MeV and with the original mass, 958 MeV (in this case, no excess is expected)} \label{f: scan_cocktail3}
 \end{center}      
\end{figure}

\begin{figure}[H]
 \begin{center}
 \subfloat{\includegraphics[width=0.8\textwidth]{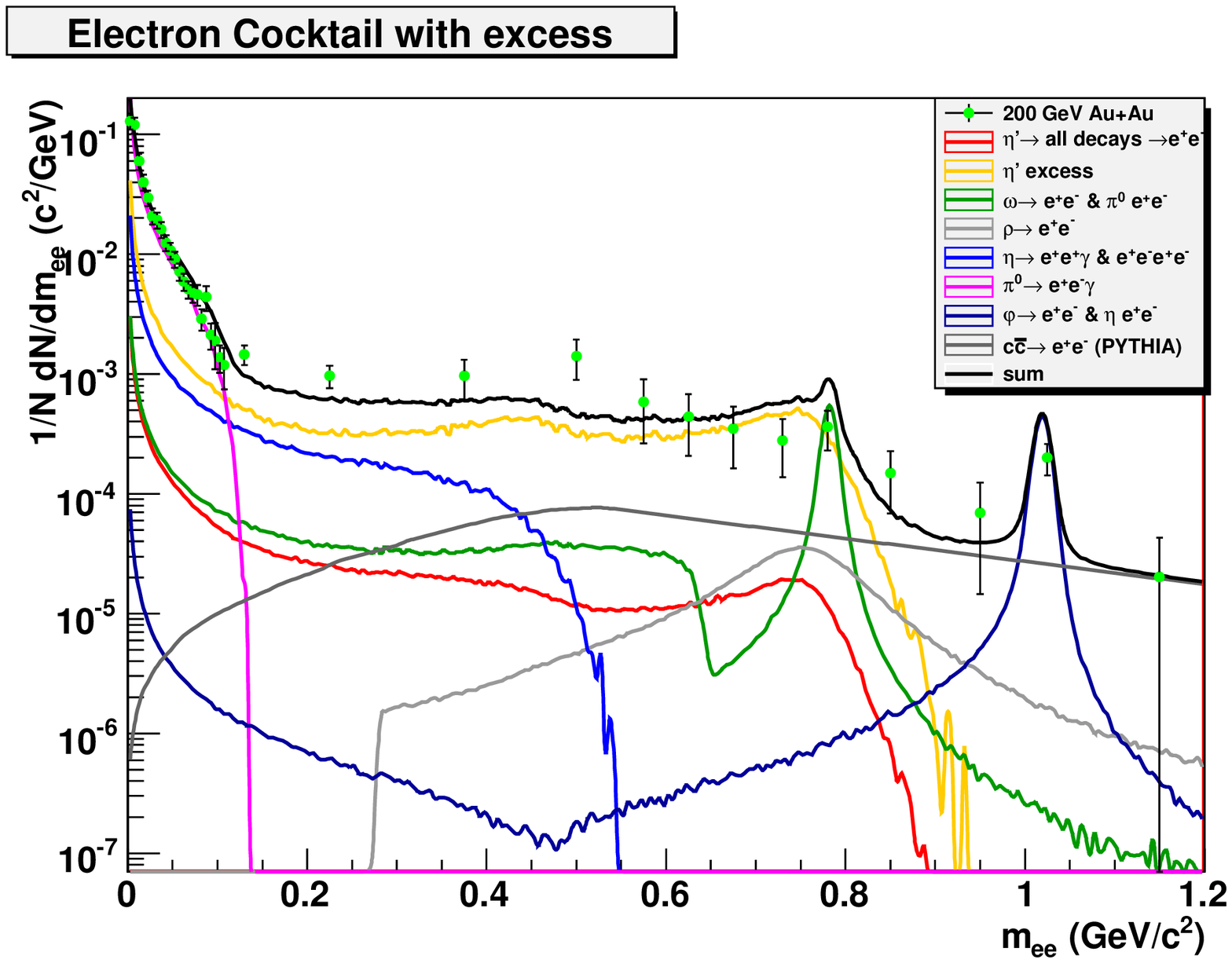}} \\
 \subfloat{\includegraphics[width=0.8\textwidth]{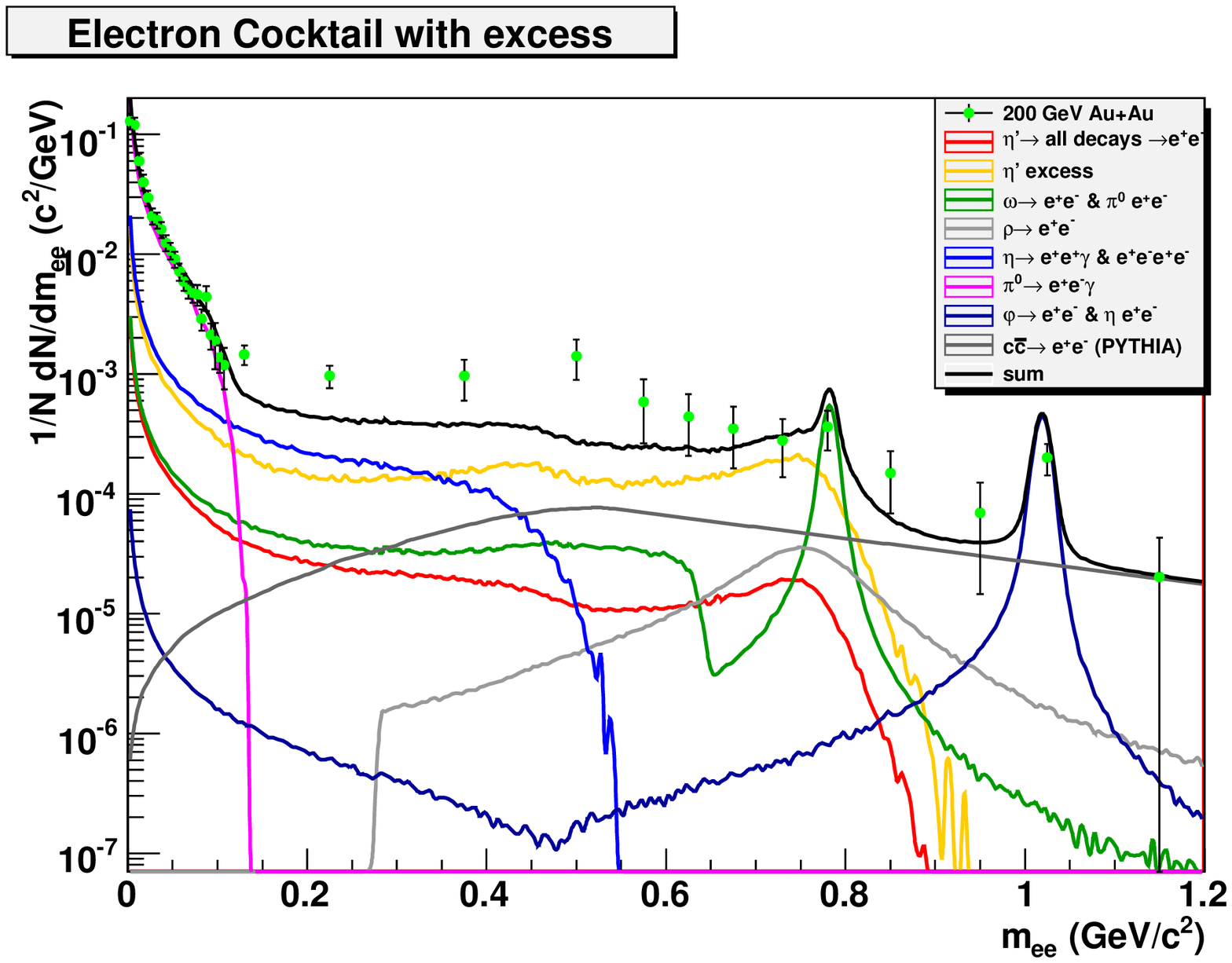}} 
  \caption{Examples from the result of the $\eta'$ input spectra scan for a more physical $B^{-1} = 86.22$ MeV, obtained from~\cite{Vertesi:2009io}. The mass values are (from top to the bottom): 300 MeV and 450 MeV.} \label{f: scan_cocktail4}
 \end{center}      
\end{figure}

\begin{figure}[H]
 \begin{center}
 \subfloat{\includegraphics[width=0.8\textwidth]{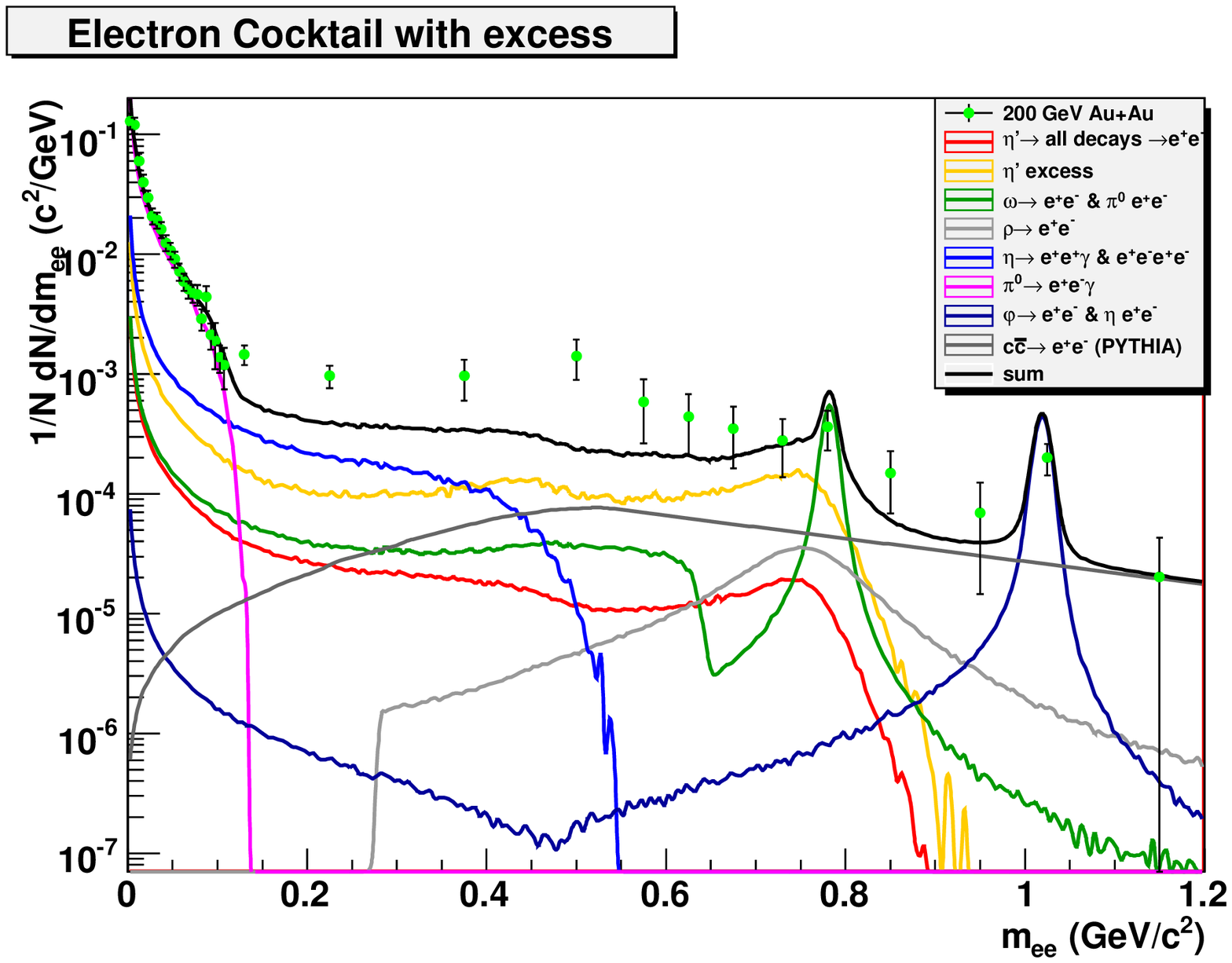}} \\
 \subfloat{\includegraphics[width=0.8\textwidth]{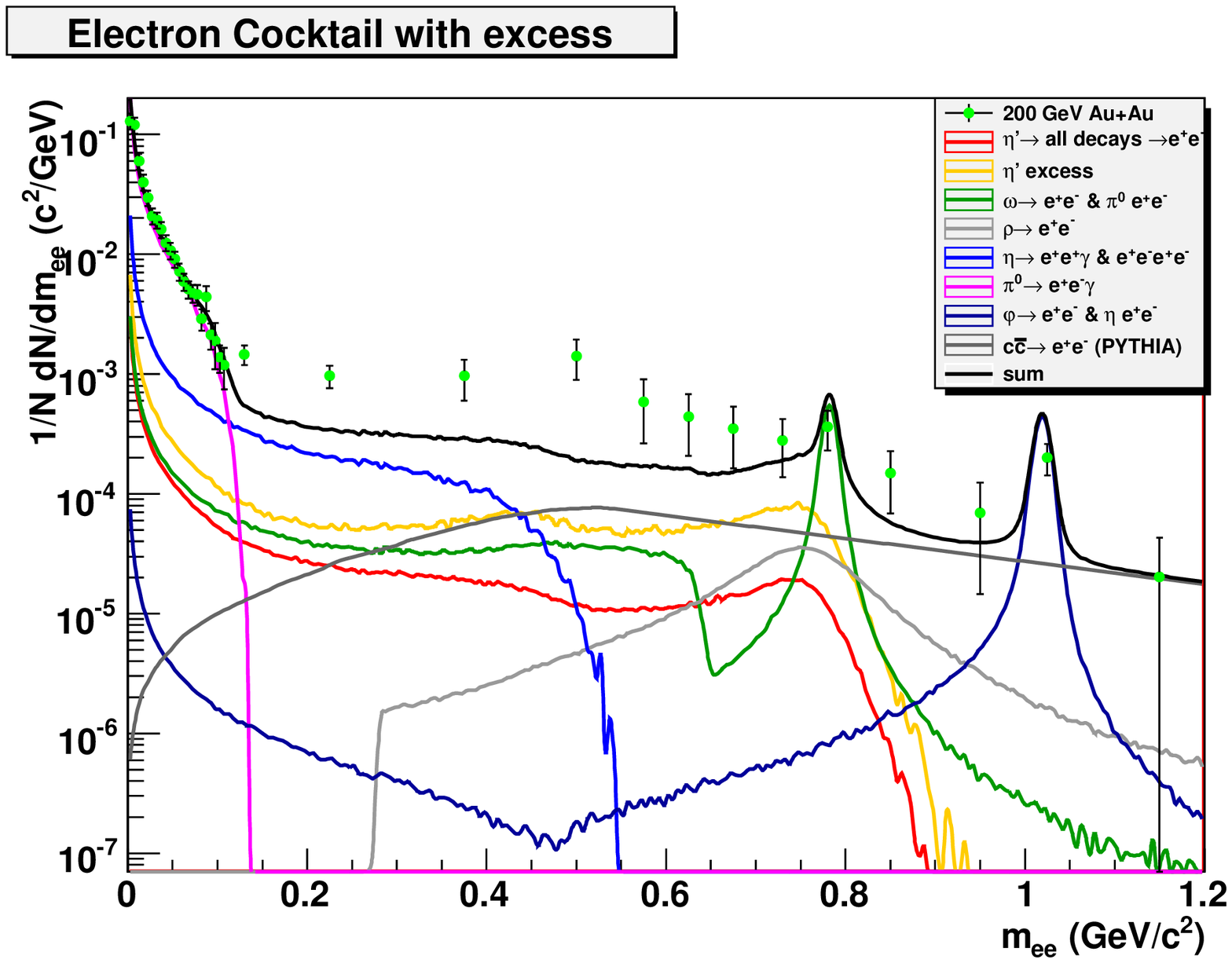}} 
  \caption{Examples from the result of the $\eta'$ input spectra scan for a more physical $B^{-1} = 86.22$ MeV, obtained from~\cite{Vertesi:2009io}. The mass values are (from top to the bottom): 500 MeV and 600 MeV.} \label{f: scan_cocktail5}
 \end{center}      
\end{figure}

\begin{figure}[H]
 \begin{center}
 \subfloat{\includegraphics[width=0.8\textwidth]{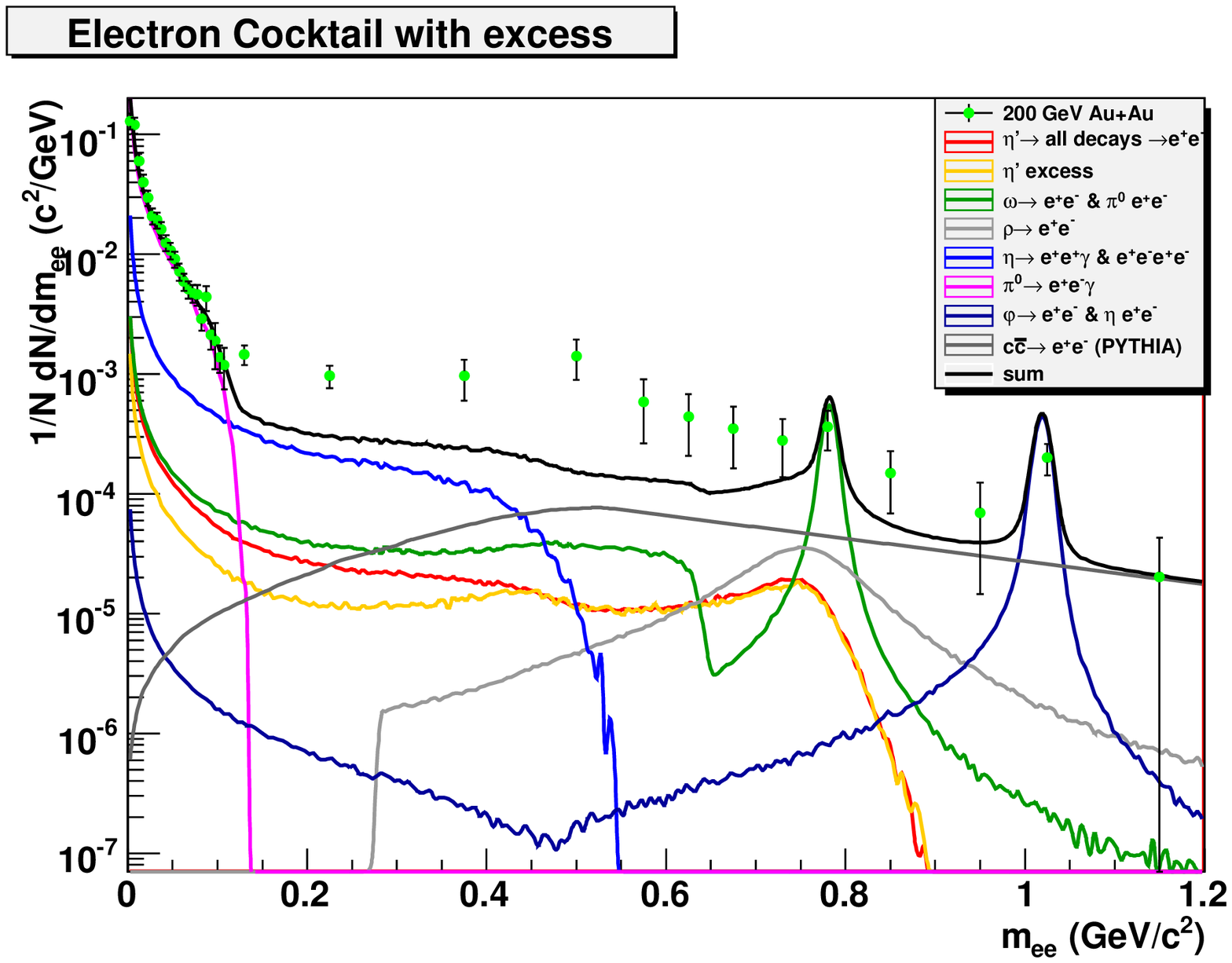}} \\
 \subfloat{\includegraphics[width=0.8\textwidth]{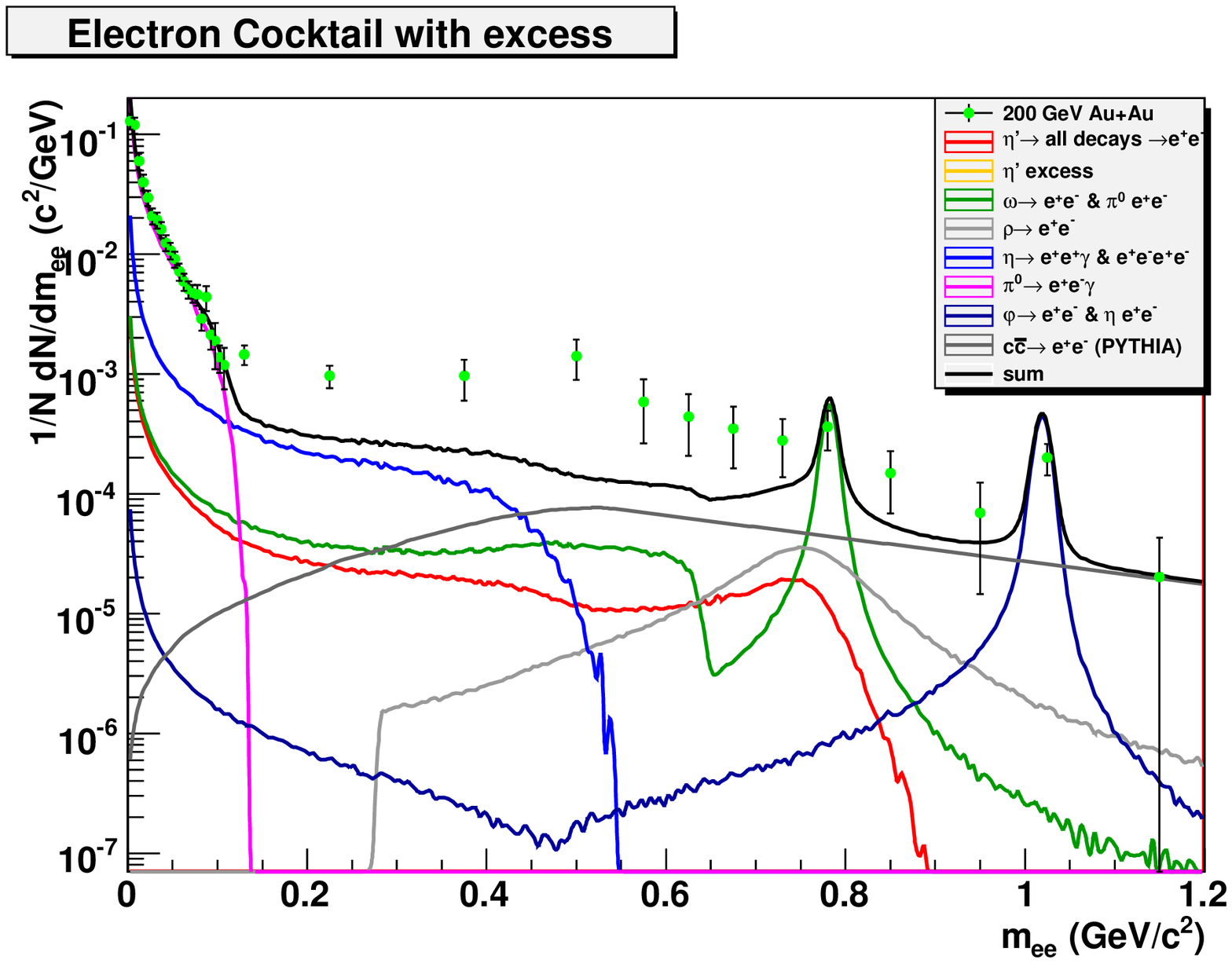}} 
 \caption{Examples from the result of the $\eta'$ input spectra scan for a more physical $B^{-1} = 86.22$ MeV, obtained from~\cite{Vertesi:2009io}. The mass values are (from top to the bottom): 800 MeV and with the original mass, 958 MeV (in this case, no excess is expected).} \label{f: scan_cocktail6}
 \end{center}      
\end{figure}

\begin{figure}[H]
 \begin{center}
 \subfloat{\includegraphics[width=0.6\textwidth]{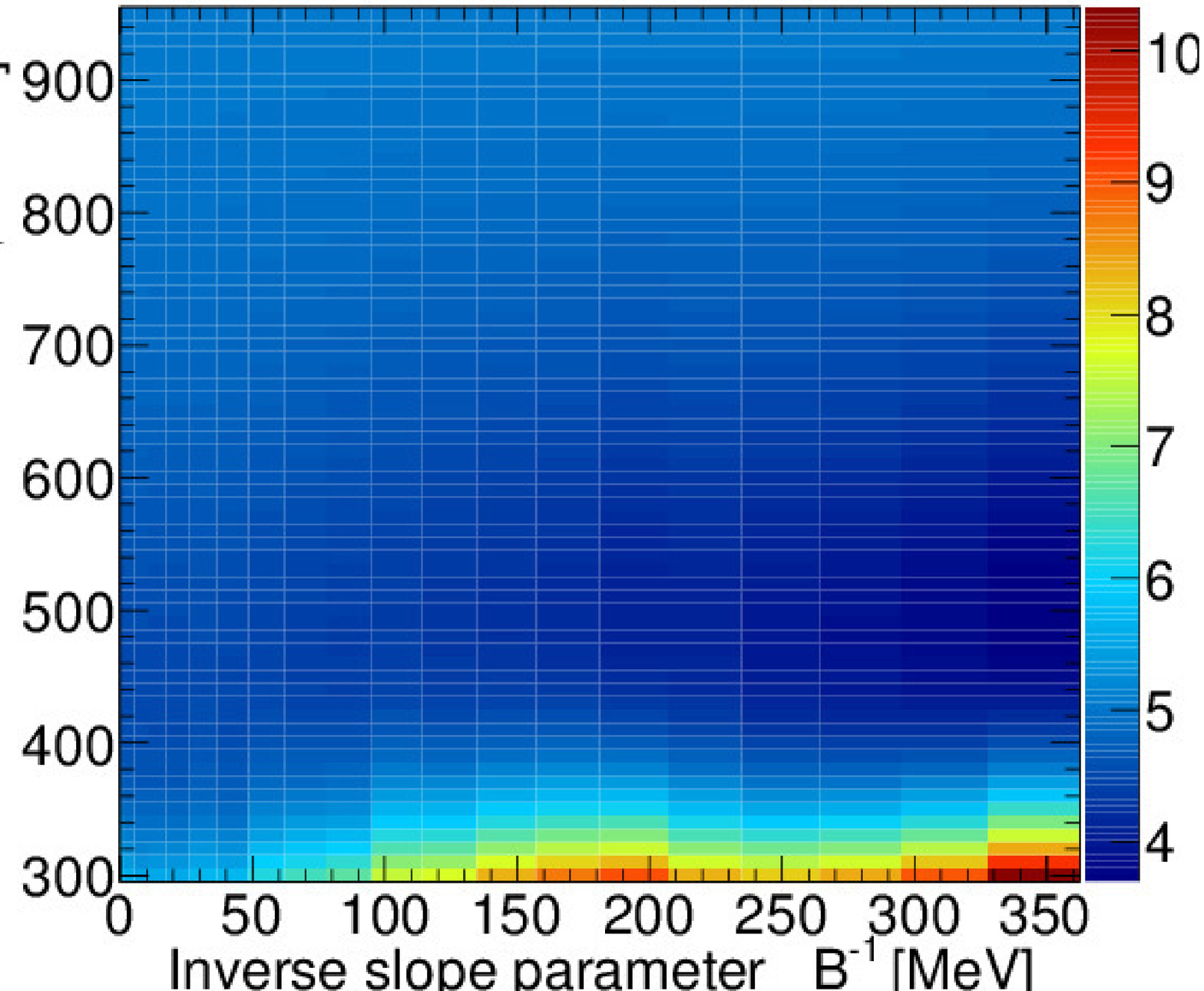}} \\
 \subfloat{\includegraphics[width=0.6\textwidth]{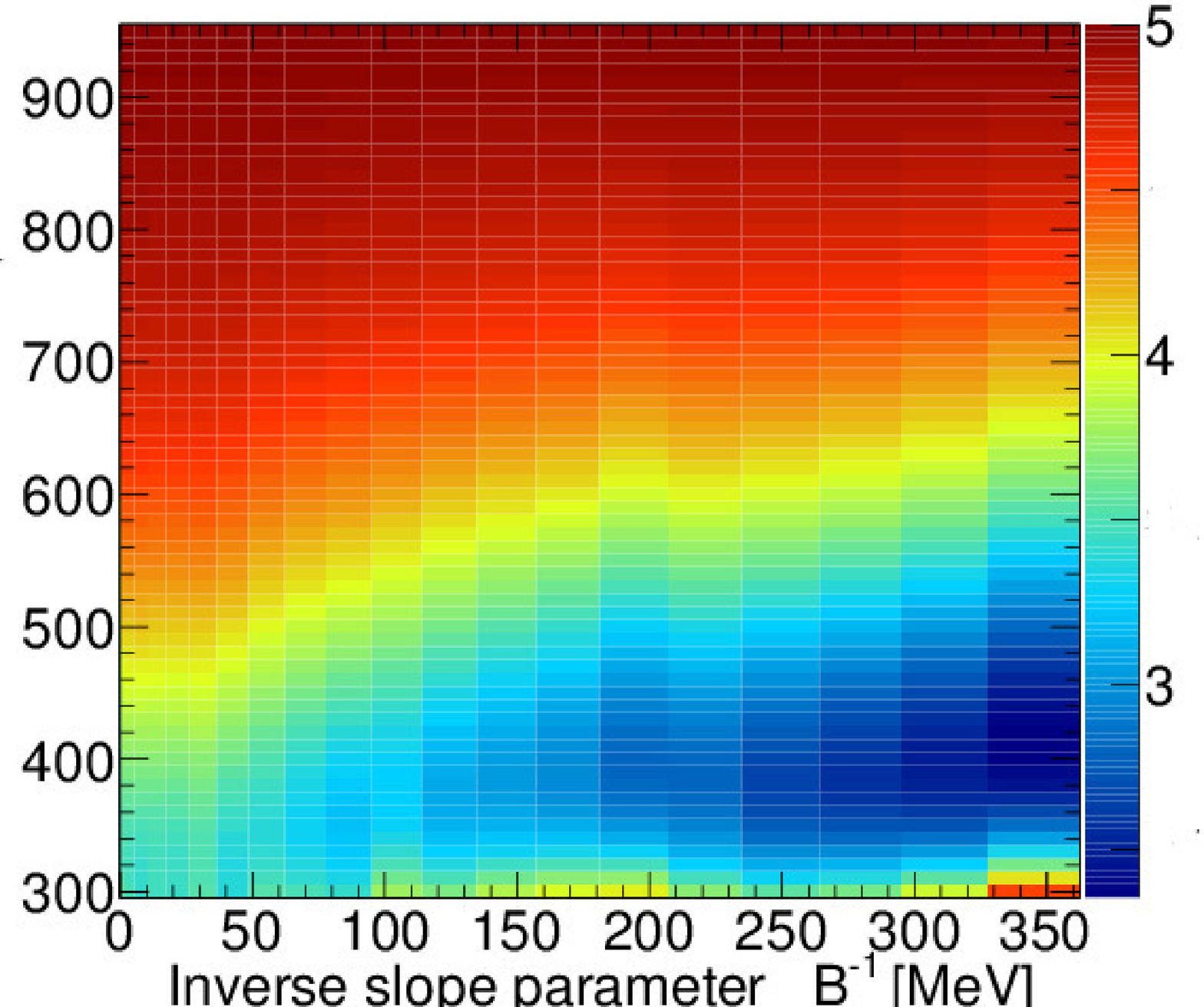}}
 \vspace{20pt}
 \caption{$\chi^2$ map for the two different approach. On the top, from simulations with the $\eta'$ meson's direct dileptonic decay-, on the bottom all the $\eta'$ decays of Table~\ref{t: decays} are defined. On the vertical axis the $B^{-1}$ parameter (which is responsible for the shape of $\eta'$ spectrum), on the horizontal axis the in-medium mass of $\eta'$ ($m^{*}_{\eta'}$) is plotted, both in MeV units. The $\chi^2$ values are calculated for the region of the excess, e.g.\ from 0.12 GeV to 0.8 GeV, NDF = 12.} \label{f: chi_map}
 \end{center}      
\end{figure}

The $\chi^2$ maps of the results are on Fig.~\ref{f: chi_map}. Some dilepton spectra examples for the best $B^{-1}$ parameter obtained by this scan (e.g.\ for $B^{-1} = 350$ MeV) for different mass values ($m_{\eta'}^*$) are on Figs.~\ref{f: scan_cocktail1}-\ref{f: scan_cocktail3}, while Figs.~\ref{f: scan_cocktail4}-\ref{f: scan_cocktail6} show the dilepton cocktail for $B^{-1} = 86.22$ MeV (which was one of the best values of the previous analysis of Ref.~\cite{Vertesi:2009io}), both with the same mass-drop values, respectively. These figures were simulated with the second approach,  which involves every decays of Table~\ref{t: decays} for $\eta'$. The original $\eta'$ contribution and the one coming from the enhanced number from its mass-drop are plotted separately. Note that because of the additional decays of $\eta'$, the shape of its dilepton spectrum has also been changed. \\


\noindent The conclusion from these scans is that the $\eta'$ dileptonic contribution is not very sensitive to the slope of the low-$p_\textrm{T}$ enhancement, the $B^{-1}$ parameter as can be seen on Fig.~\ref{f: chi_map}. It is a conservative estimation, however, that this will play an important role when examining the cocktail in different $p_\textrm{T}$ slices. The important parameter is the in-medium mass of the $\eta'$, from the second approach it is expected to be around the $\eta$ mass. An other important learning from this section is that not only the direct dileptonic decays of $\eta'$ has to be included, as involving the resonance chain decays changes the dilepton spectrum, so that it gives a better agreement with PHENIX data.

The right panel of Fig.~\ref{f: chi_map} indicates that for any reasonable value of the slope of the $\eta'$ spectrum, $B^{-1}$, at least 200 MeV mass drop in the medium is needed to describe the low-mass part of the 200 GeV Au+Au dilepton spectrum, confirming the conclusion of Ref.~\cite{Vertesi:2009io}, but based on a different channel of observation. This plot also indicates that it is important to take into account resonance decay chains of $\eta'$ properly. 

\pagebreak

\chapter{The radial flow effect}
\section{Why to use different spectra}
\begin{figure}[H]
 \begin{center}
 \includegraphics[width=1.0\textwidth]{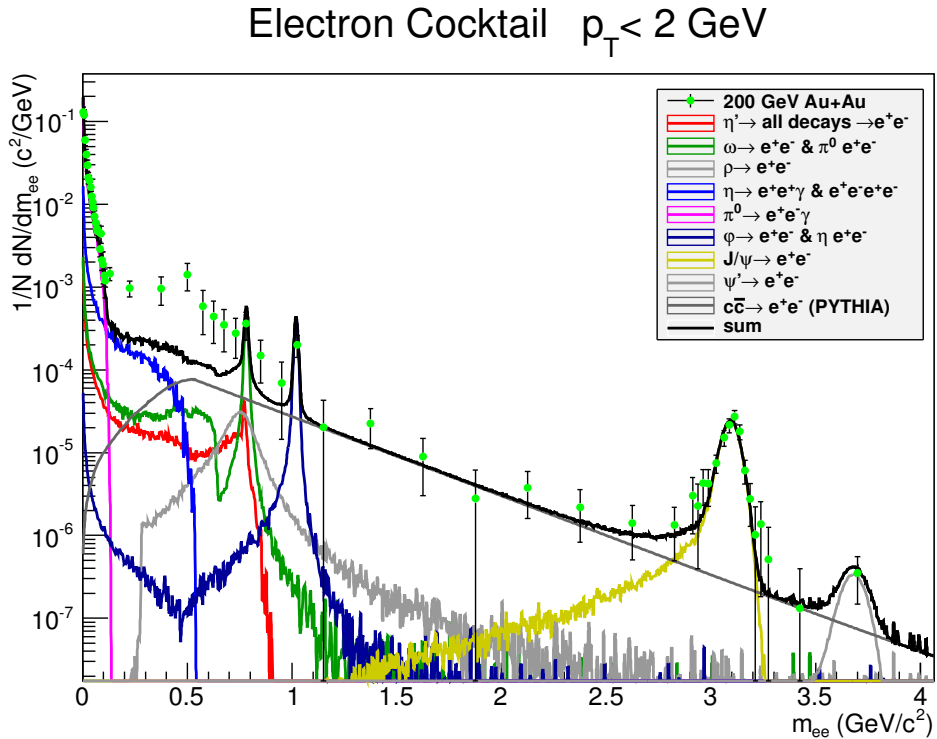} 
 \caption{To demonstrate the importance of the low-$p_\textrm{T}$ part of meson spectra, only dileptons with $(p_{\textrm{T}})_{ee} < 2$ GeV are plotted. They reproduce the dilepton, as this low-$p_{\textrm{T}}$ part dominates the min.\ bias Au+Au dilepton cocktail. The simulation is new, while the PHENIX data points are from Ref.~\cite{PPG088}.} \label{f: pT2GeV}
 \end{center}      
\end{figure}

\begin{figure}[H]
 \begin{center}
  \vspace{10pt} 
  \subfloat{ \includegraphics[width=0.37\textwidth]{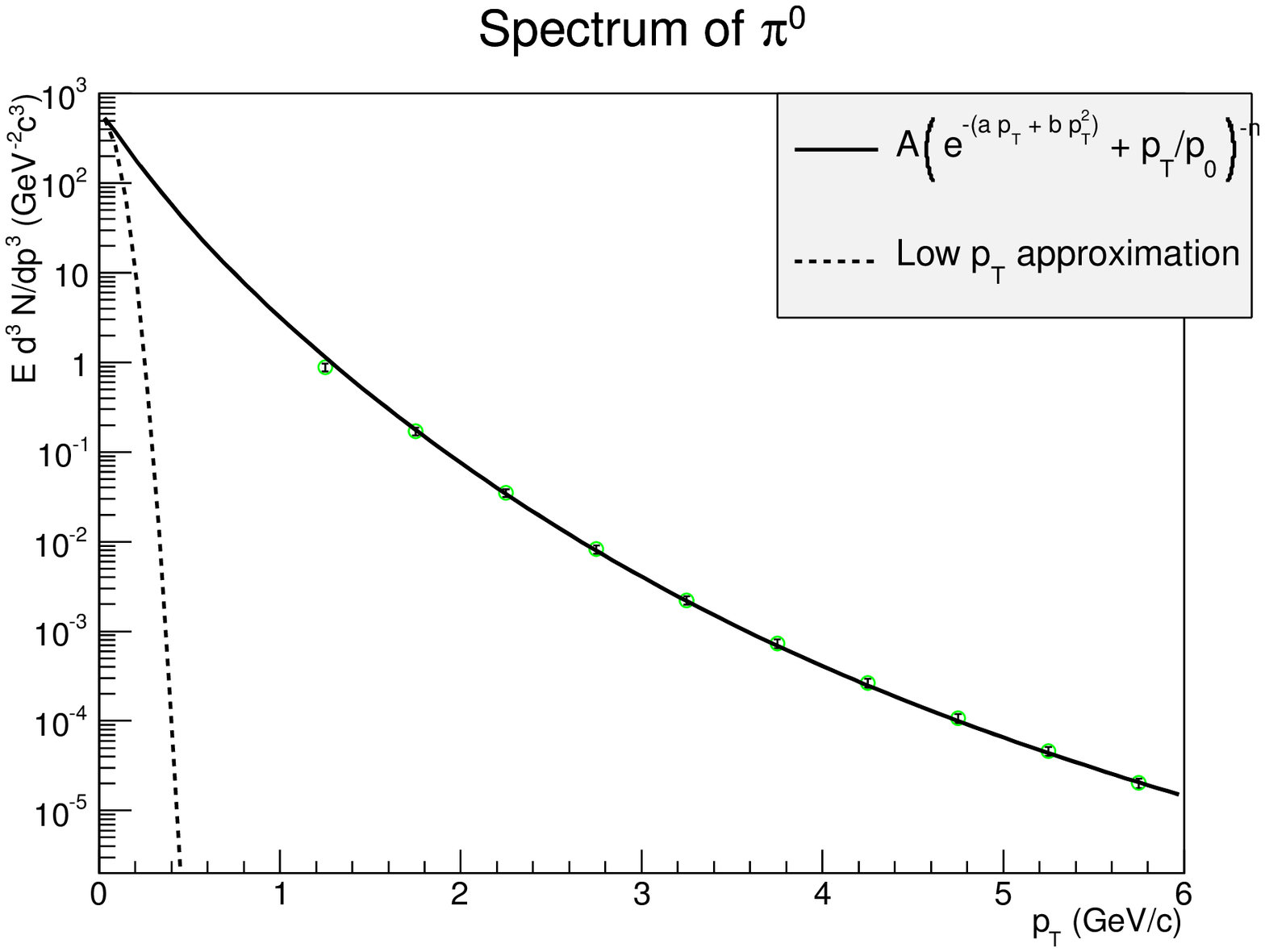} } \
  \subfloat{ \includegraphics[width=0.37\textwidth]{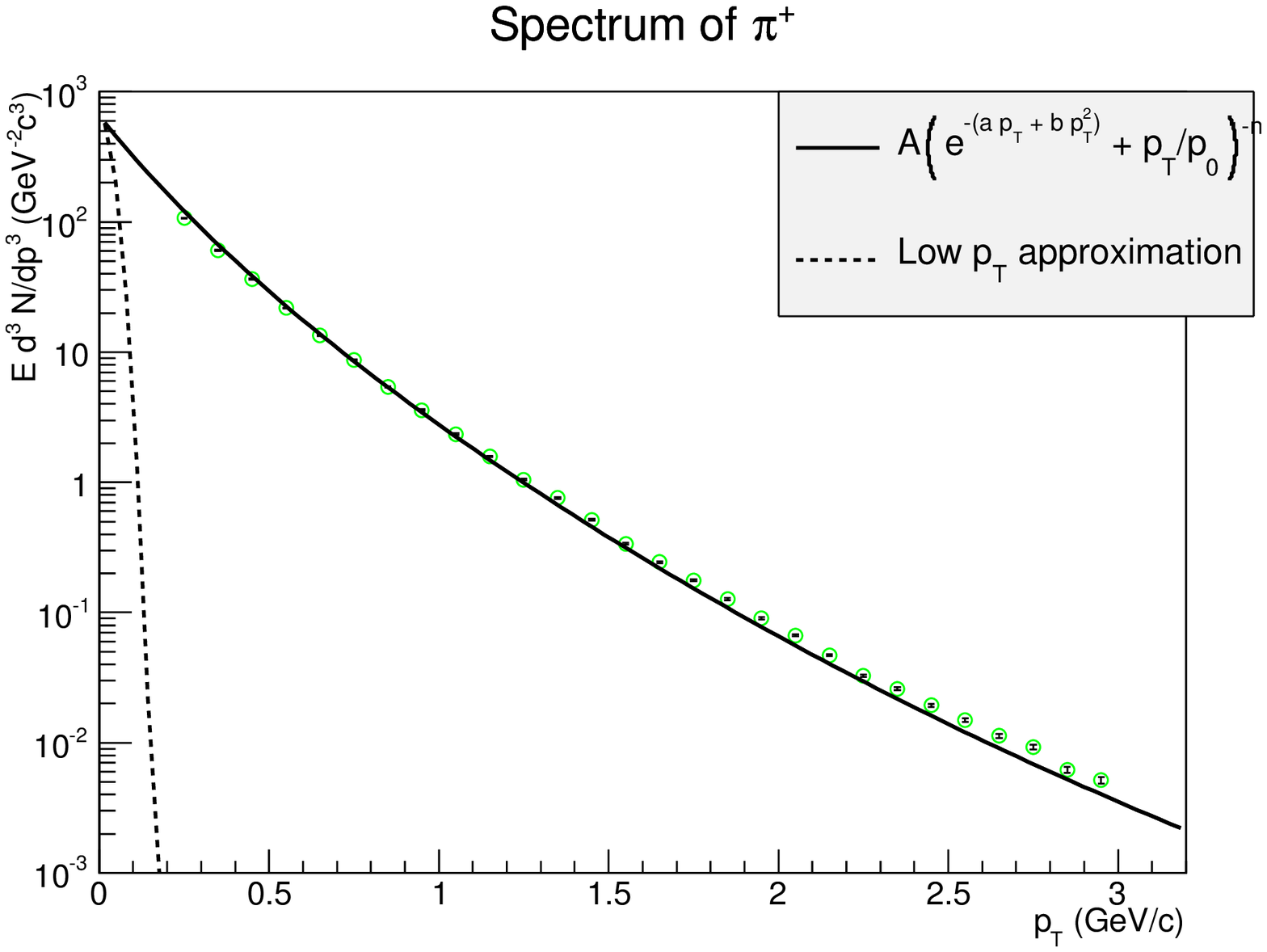} } \\
  \vspace{-20pt}
  \subfloat{ \includegraphics[width=0.37\textwidth]{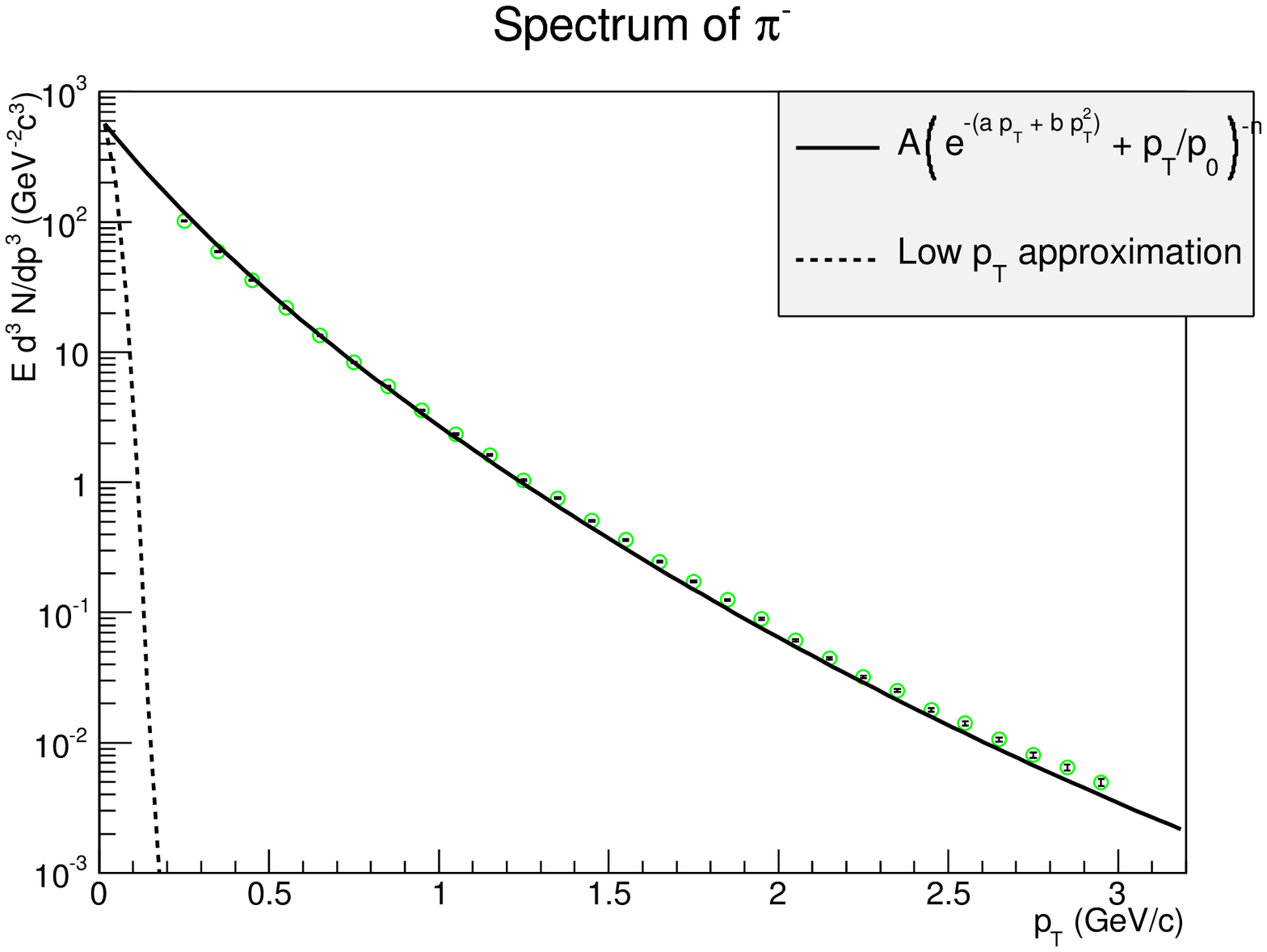} } \
  \subfloat{ \includegraphics[width=0.37\textwidth]{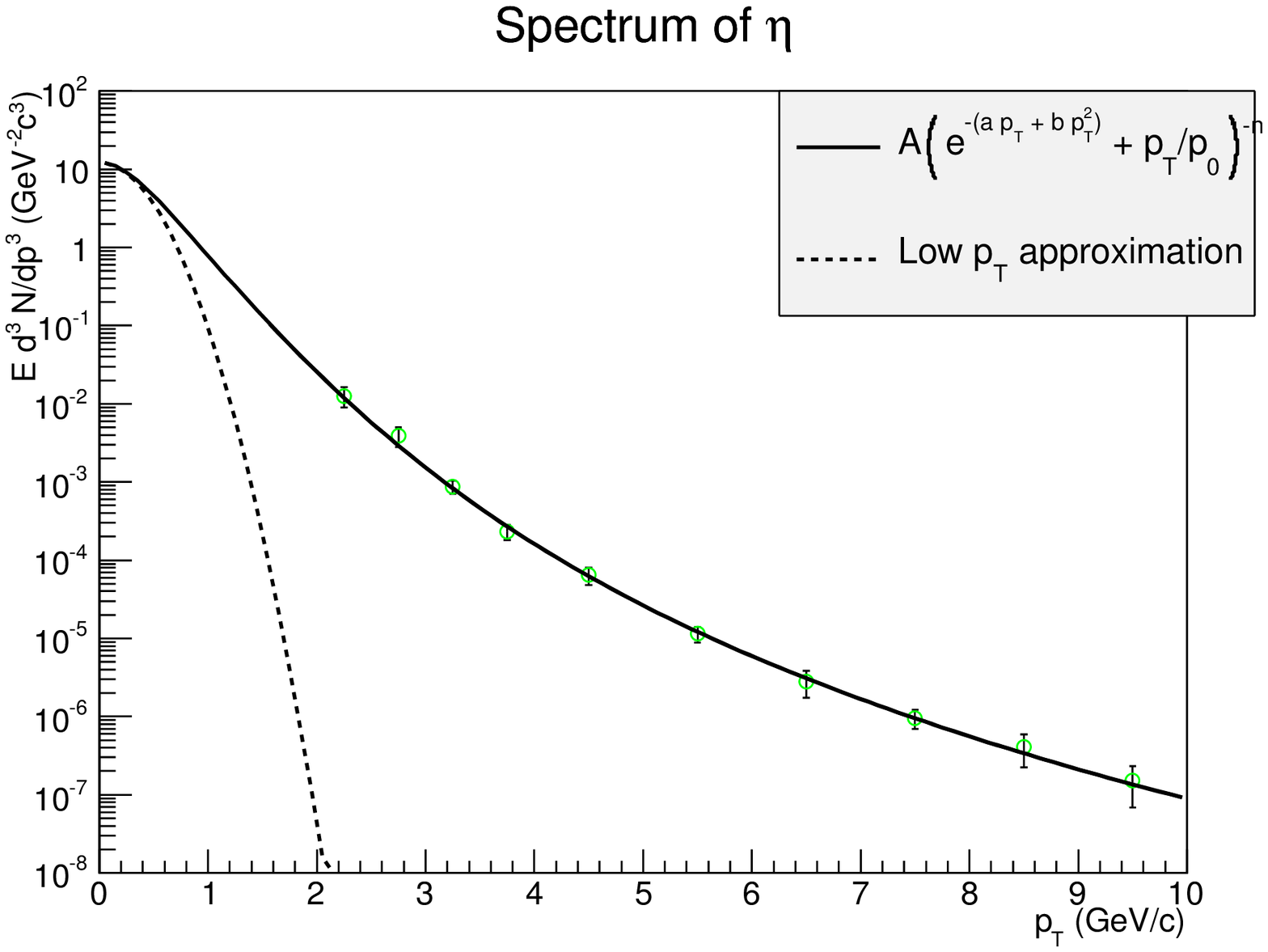} } \\
  \vspace{-20pt}
  \subfloat{ \includegraphics[width=0.37\textwidth]{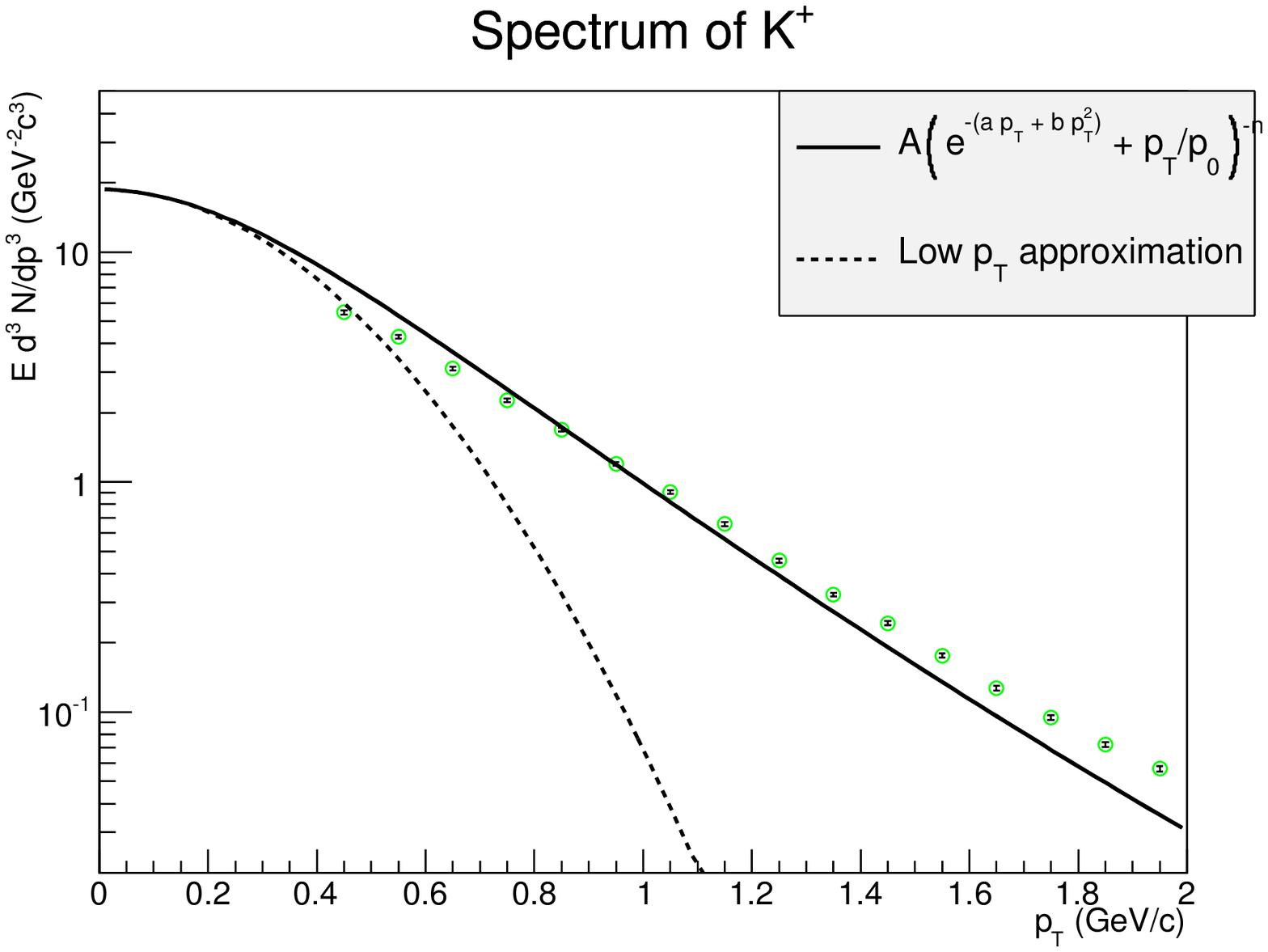} } \
  \subfloat{ \includegraphics[width=0.37\textwidth]{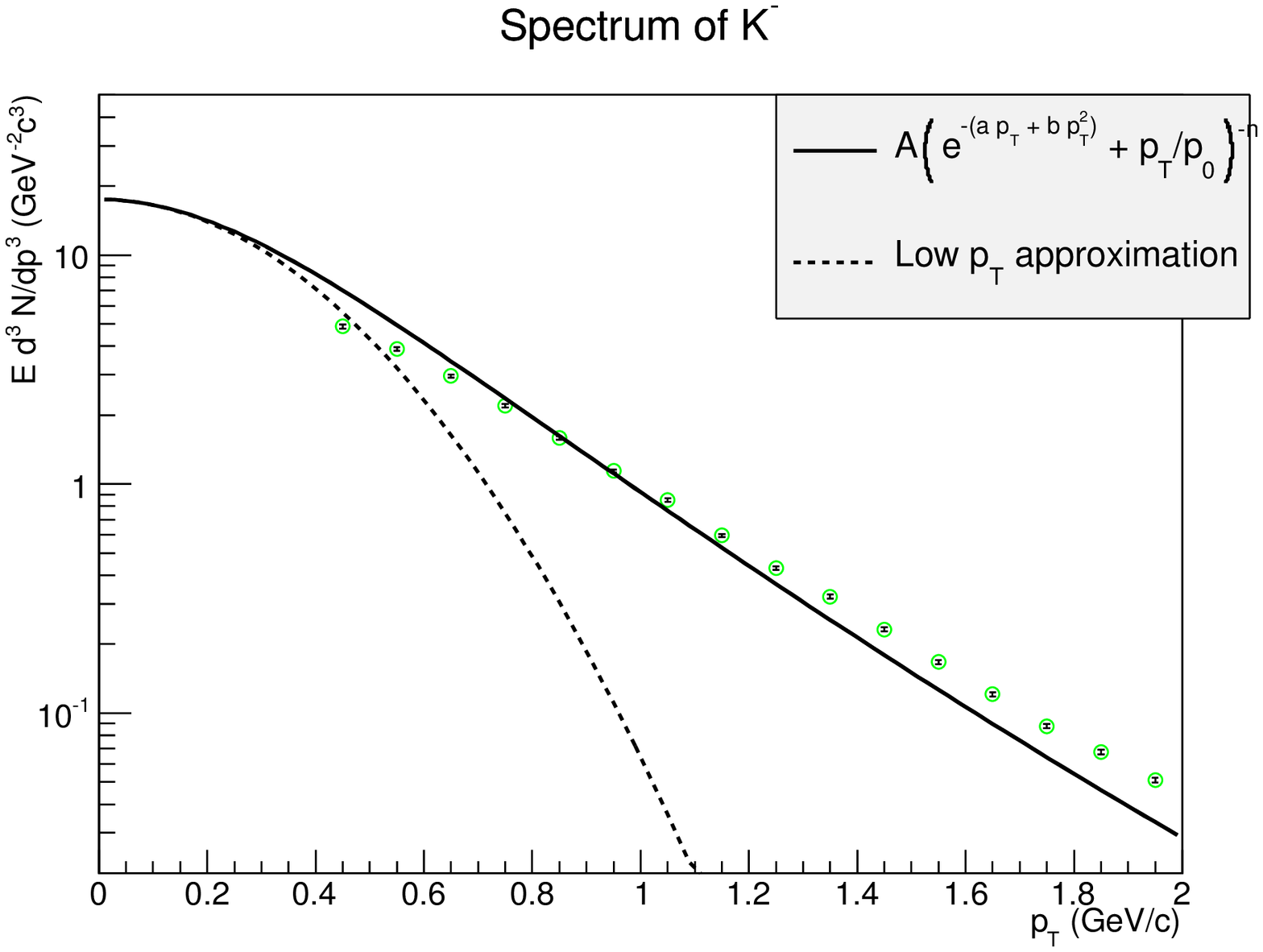} } \\
  \vspace{-20pt}
  \subfloat{ \includegraphics[width=0.37\textwidth]{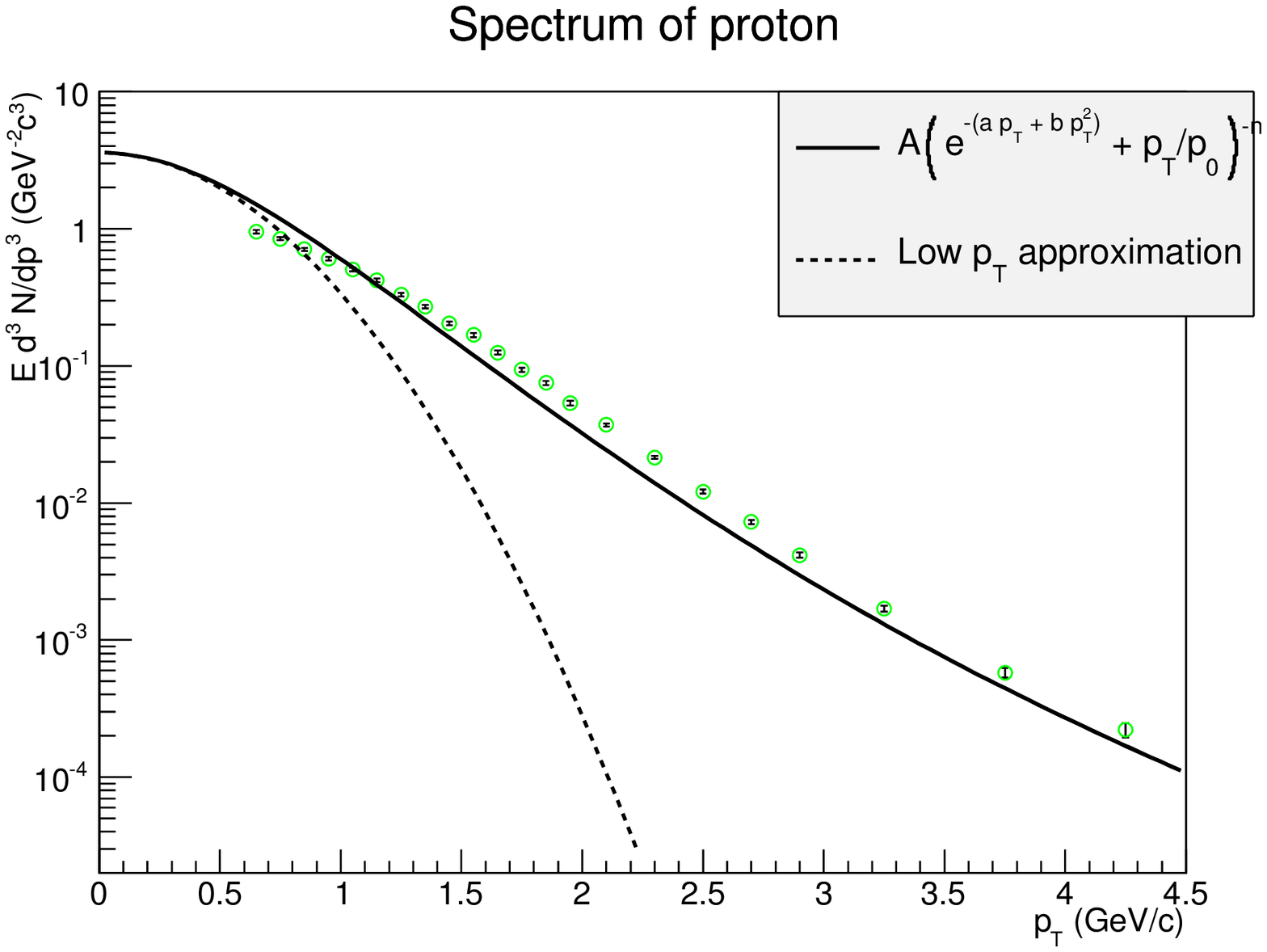} } \
  \subfloat{ \includegraphics[width=0.37\textwidth]{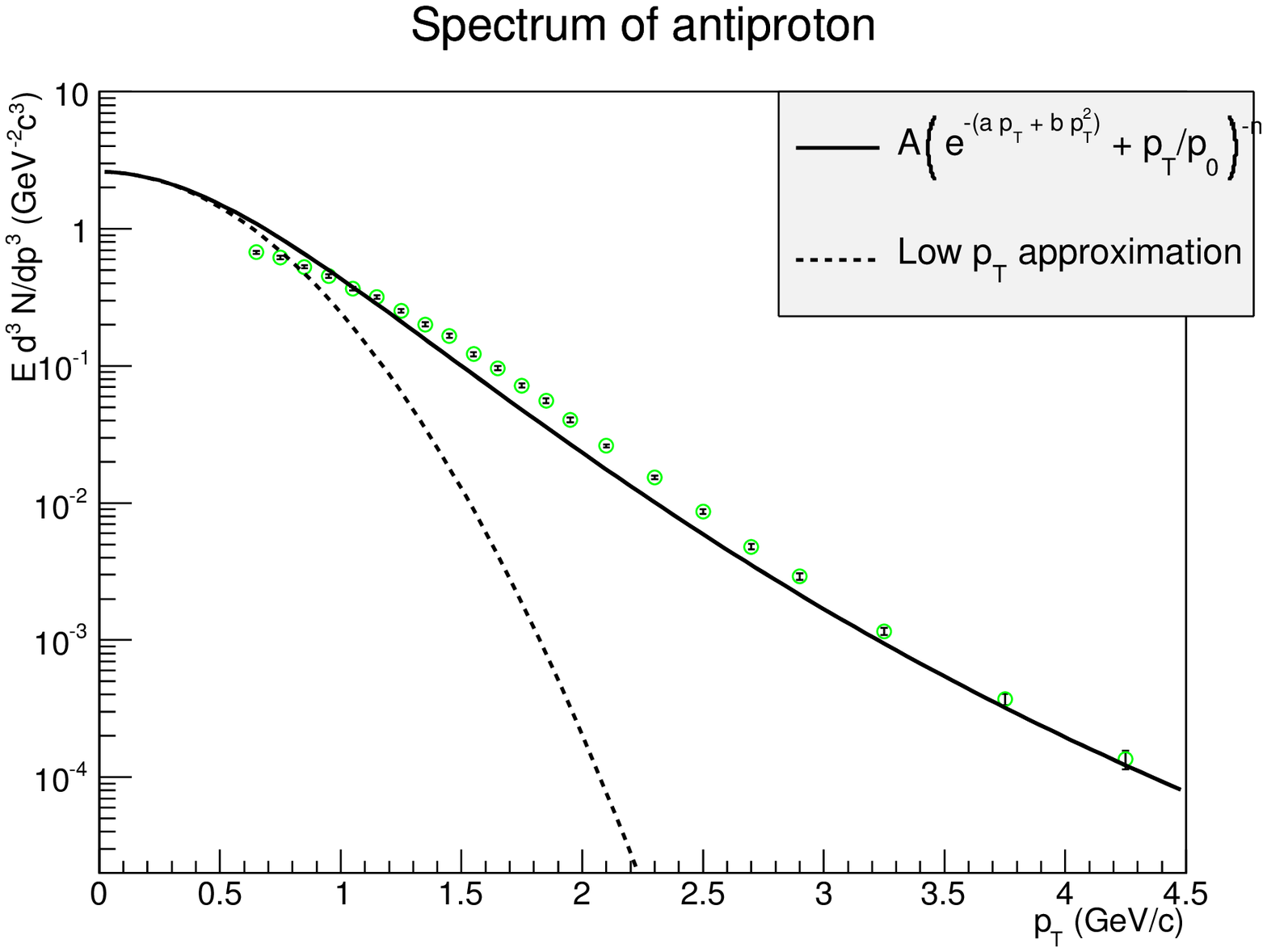} } \\
  \vspace{-20pt}
  \subfloat{ \includegraphics[width=0.37\textwidth]{PPG088_approxFOR_5.eps} } \
  \subfloat{ \includegraphics[width=0.37\textwidth]{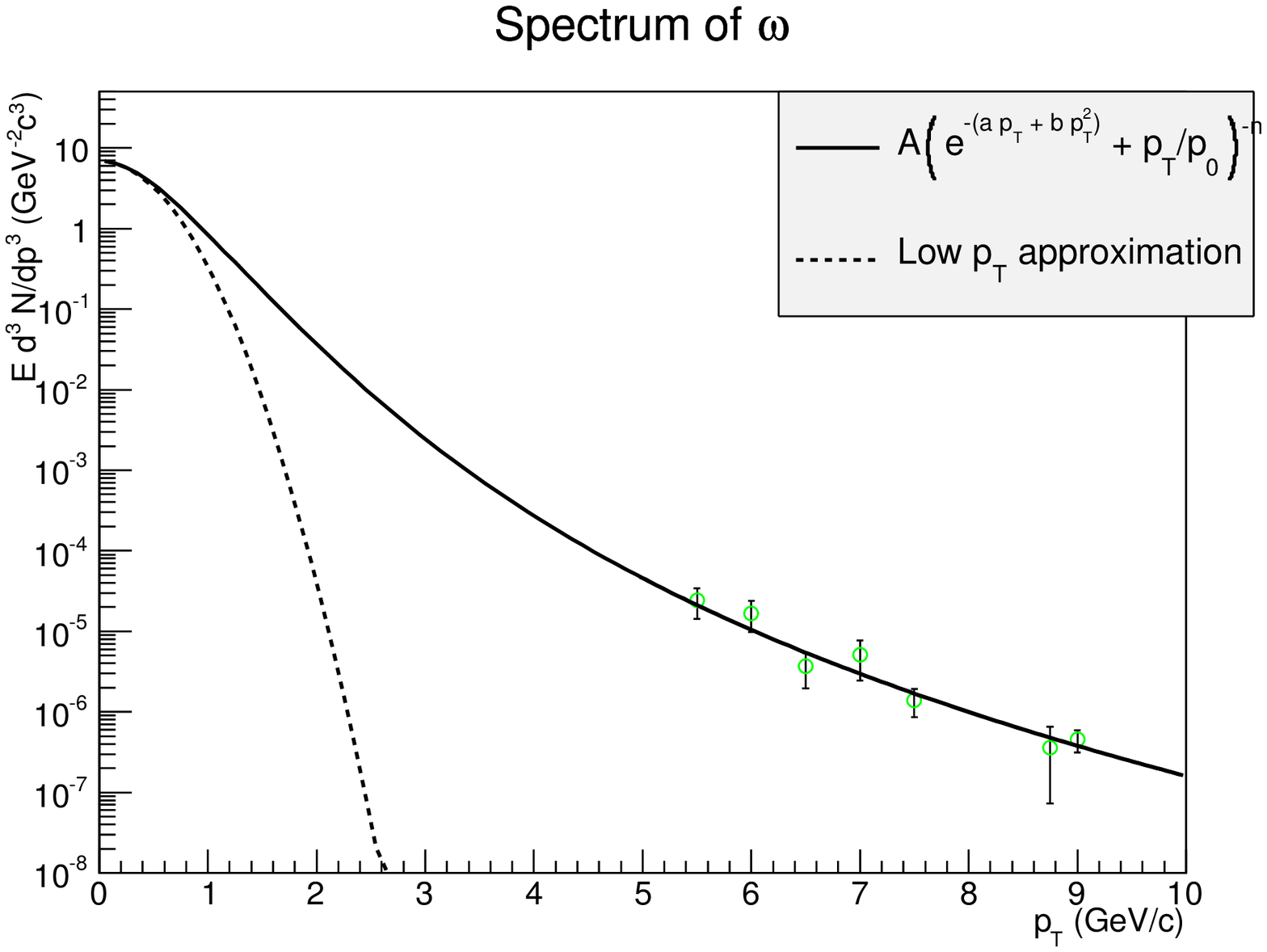} } 
 \caption{Comparison of the PHENIX spectra~\cite{PPG088} (Eq.~\ref{e: PPG088}) for different particles. The low-$p_\textrm{T}$ approximation is the leading order of the exponential part of that spectra. It obviously lacks the radial flow effects, the mass-scaling inverse slope of the spectra, namely $T_\textrm{eff} = T_0+m {\langle u_\textrm{T}\rangle}^2$ would imply that the more massive the particle is, the larger the tail in its exponential spectra (indicated by the dashed line), which is obviously not fulfilled here.}   \label{f: radial_check} 
 \end{center}     
\end{figure}

The $p_\textrm{T}$ spectra of the PHENIX analysis~\cite{PPG088} was fitted to the high-$p_\textrm{T}$ region, as more data were available in that range.  In the dilepton cocktail, however, unfortunately the low-$p_\textrm{T}$ part is dominant, for which a hydrodynamical spectrum including radial flow would be more a accurate estimation.
But their formula (Eq.~\ref{e: PPG088}) lacks the radial flow term as shown on Fig.~\ref{f: radial_check}.
To evaluate the importance of the contribution from low-$p_\textrm{T}$ mesons, look at Fig.~\ref{f: pT2GeV}, where only dilepton pairs with $(p_{\textrm{T}})_{ee} < 2$ GeV has contributed, hence, the original dilepton cocktail was reproduced only with them. Compare it with Fig.~\ref{f: PPG088cocktail}, when the whole $p_\textrm{T}$ range of the resonances up to 10 GeV was utilized. One can see that the addition of the 2 GeV  $ < p_\textrm{T} < $ 10 GeV region does not lead to extra enhancement.

Another signature of the possible radial flow effect is indicated by Figs.\ \ref{f: PPG088cocktail_CentrSlices} and \ref{f: PPG088cocktail_pTSlices}. Fig.\ \ref{f: PPG088cocktail_CentrSlices} shows the PHENIX cocktail in different centrality slices. The excess seen in the low-mass region tends to disappear for peripheral collisions, according to the hydro picture, where the ${\langle u_\textrm{T} \rangle}^2$ parameter which governs the $p_\textrm{T}$ distribution also decreases with increasing centrality. Fig.\ \ref{f: PPG088cocktail_pTSlices} shows the cocktail in different $p_\textrm{T}$ slices, and the introduction of a hydro spectrum is even more motivating here, since it shows that the excess is disappearing for higher transverse momenta. Thus the low-mass dilepton enhancement partially may be due to the lack of radial flow effect in the original background cocktail evaluation of Ref.~\cite{PPG088}.
A different input spectrum, which estimates better the low-$p_\textrm{T}$ part of the resonances that decay to dileptons, may be an important tool in the upcoming systematic investigations.
In the next section a more detailed examination of the PHENIX $p_\textrm{T}$ spectrum will be presented, and a dilepton cocktail generated with a hydro-motivated distribution will be shown, of which parameters have been obtained from a simultaneous fit for $\pi^{\pm}$, $K^{\pm}$, $p$ and $\bar{p}$.
Due to the work-in-progress nature of the current results, in the next section we focus entirely on the effects of radial flow and the effects coming from resonance chain decays will not be considered in this section. For a final analysis, both radial flow and $\eta'$ chain decay effects have to be considered simultaneously, but these studies go well beyond the scope of this M.Sc.\ Thesis.  

\pagebreak

{\color{white} please disappear from this page}
\thispagestyle{empty}
\pagebreak

\color{black}
\section{Simultaneous hydro fit to identified particle spectra}

As previous examinations suggested, a hydro-motivated $p_\textrm{T}$ spectrum might give a better estimation for the low-$p_\textrm{T}$ part of the dilepton cocktail. The simplified hydro spectrum which is detailed in this section is:
\begin{equation} \label{e: hydro}
E\frac{d^3N}{d p^3} = A  \left(\frac{m_\textrm{T}}{m}\right)^\alpha  e^{-\frac{m_\textrm{T}-m}{T_0 + m {\langle ut\rangle}^2}}.
\end{equation}

\begin{figure}[H]
 \begin{center}
 \vspace{-10pt}
 \includegraphics[width=0.8\textwidth]{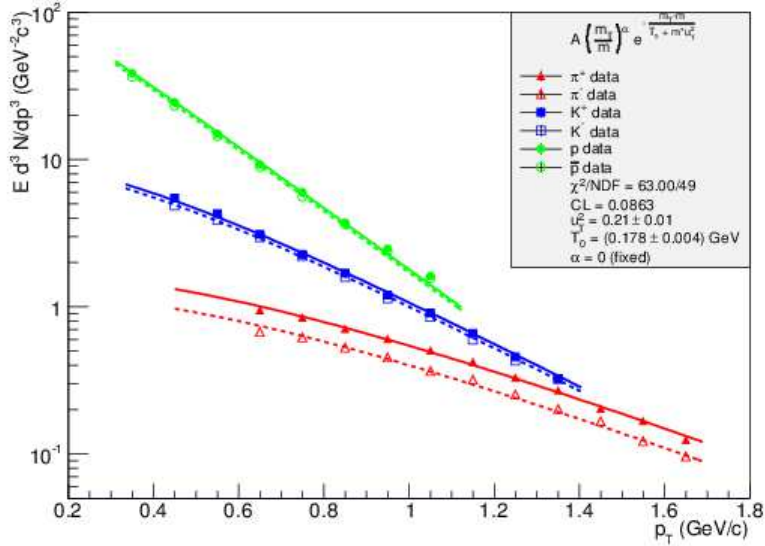} 
 \caption{Simultaneous fit of Eq.\ \ref{e: hydro}.\ for  $\pi^{\pm}$, $K^{\pm}$, $p$ and $\bar{p}$ data with the constraint from Ref.~\cite{PPG026}, e.g.\ the fit ranges are ($0.2-1.0$) GeV/$c^2$ for $\pi^{\pm}$ and ($0.1-1.0$) for $K^{\pm}$, $p$ and $\bar{p}$ in $m_\textrm{T}-m$.} \label{f: simultan fit}
 \end{center}      
\end{figure}

The formula above was simultaneously fitted to available $\pi^{\pm}$, $K^{\pm}$, $p$ and $\bar{p}$ data, similarly to an earlier analysis of Ref.~\cite{PPG026}, e.g.\ the fit ranges are ($0.2-1.0$) GeV/$c^2$ for $\pi^{\pm}$ and ($0.1-1.0$) for $K^{\pm}$, $p$ and $\bar{p}$ in $m_\textrm{T}-m$. The simultaneous fit to $\pi^{\pm}$, $K^{\pm}$, $p$ and $\bar{p}$ data is shown on Fig.~\ref{f: simultan fit}, as a function of $p_\textrm{T}$.

The best fit parameters are: ${\langle u_\textrm{T}\rangle}^2$ = $0.21 \pm 0.01$, and $T_0 = (0.178 \pm 0.004)$ GeV, while the $\alpha$ parameter was consistent with 0 within errors, hence it was fixed to 0. Hopefully, this spectrum, when used as an input for generating of dilepton decays, and when extrapolated for heavy particles like $\eta'$, $\omega$ and $\phi$ will produce better agreement with PHENIX data.


\section{Dilepton cocktail simulated with hydro spectra}

\begin{figure}[H]
 \begin{center} 
 \vspace{45pt}
  { \includegraphics[width=1.1\textwidth]{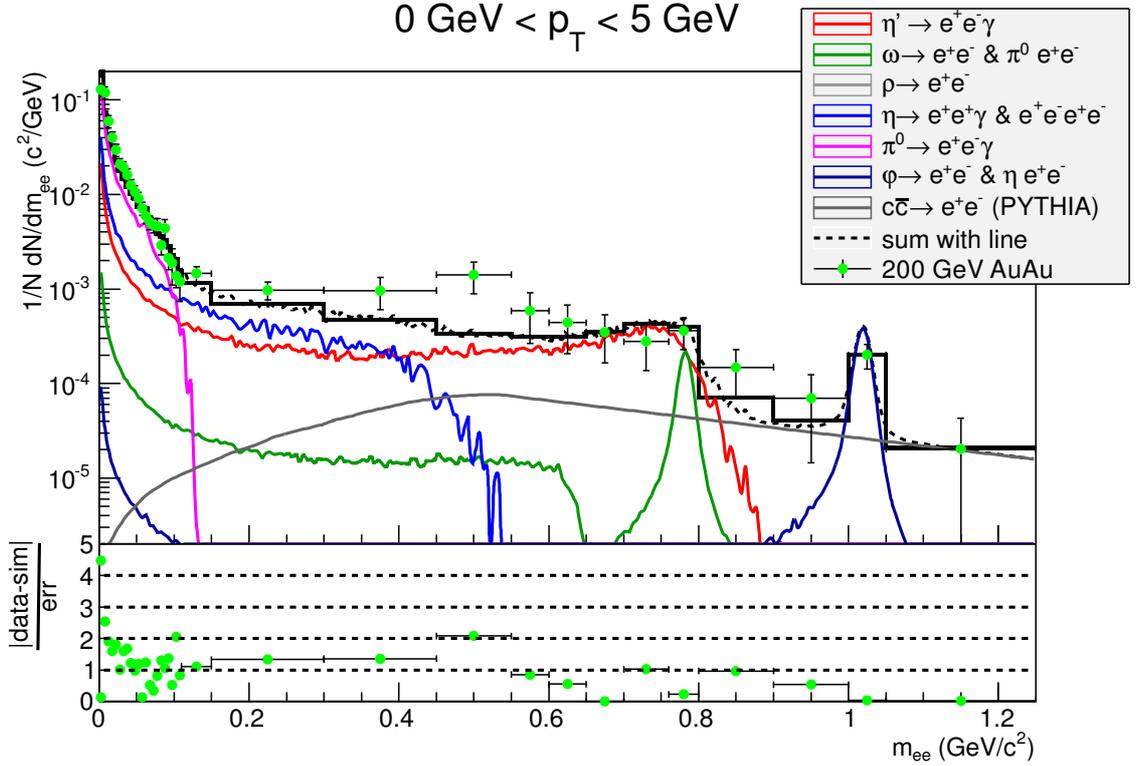} } \
\caption{Dilepton spectrum simulated with the hydro distribution of Eq.\ \ref{e: hydro}. The meson yields were obtained by fitting the meson contributions as a linear combination to PHENIX data. Note the significant increase of $\eta'$ production, as also shown in Table~\ref{t: etap_models},~\ref{t: eta_models} and ~\ref{t: meson_yields}.  As compared to neutral pions, a tenfold enhancement of $\eta'$ mesons appear as compared to the one of the PHENIX analysis~\cite{PPG088}.}   \label{f: radial_result1} 
 \end{center}     
\end{figure}

\begin{figure}[H]
 \begin{center}
  \vspace{40pt} 
  \subfloat{ \includegraphics[width=0.8\textwidth]{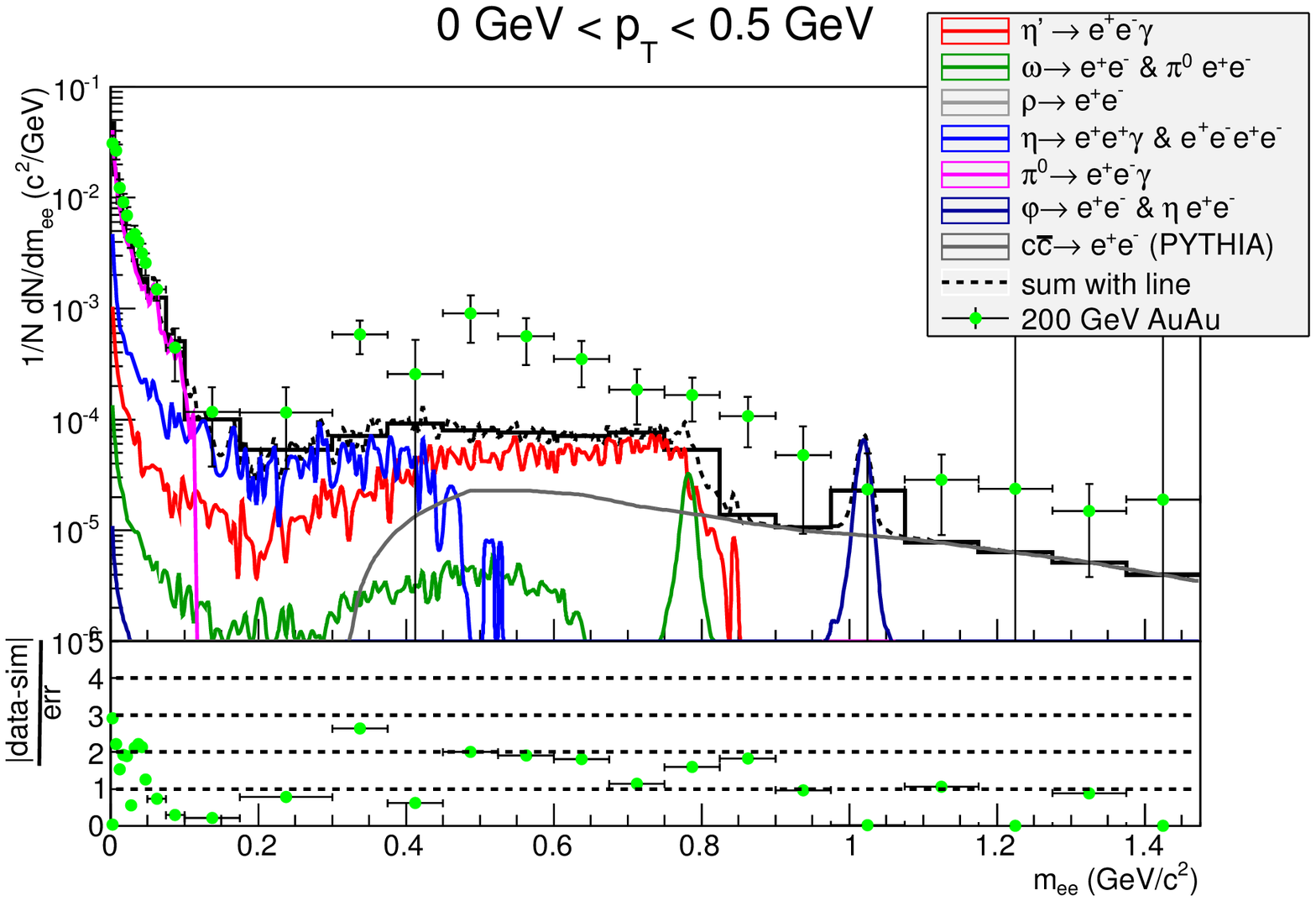} } \\
  \subfloat{ \includegraphics[width=0.8\textwidth]{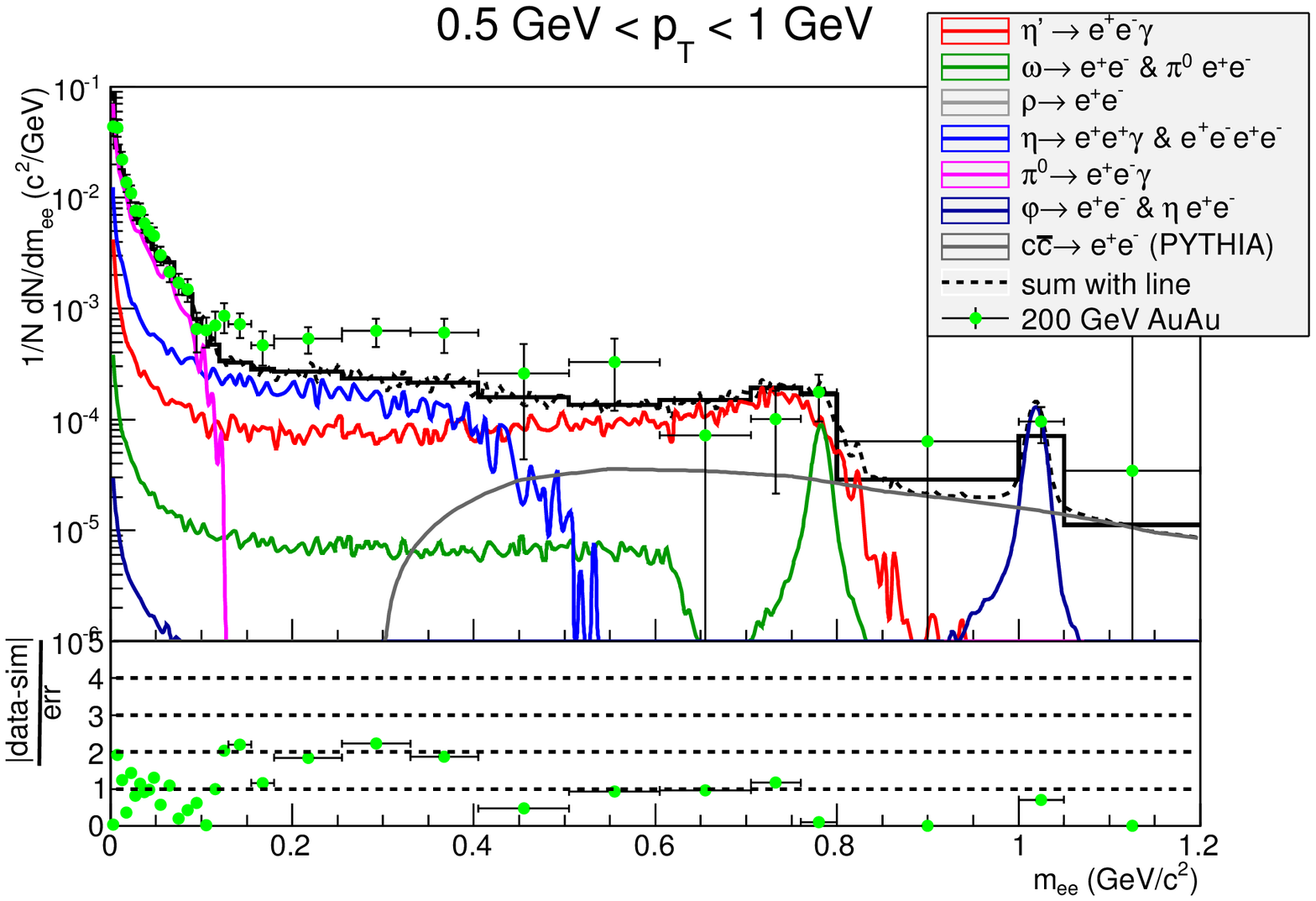} } 
 \caption{Radial flow effects on the low-mass dilepton spectrum in different $p_\textrm{T}$ slices}   \label{f: radial_result2a} 
 \end{center}     
\end{figure}

\begin{figure}[H]
 \begin{center}
 \vspace{40pt}   
   \subfloat{ \includegraphics[width=0.8\textwidth]{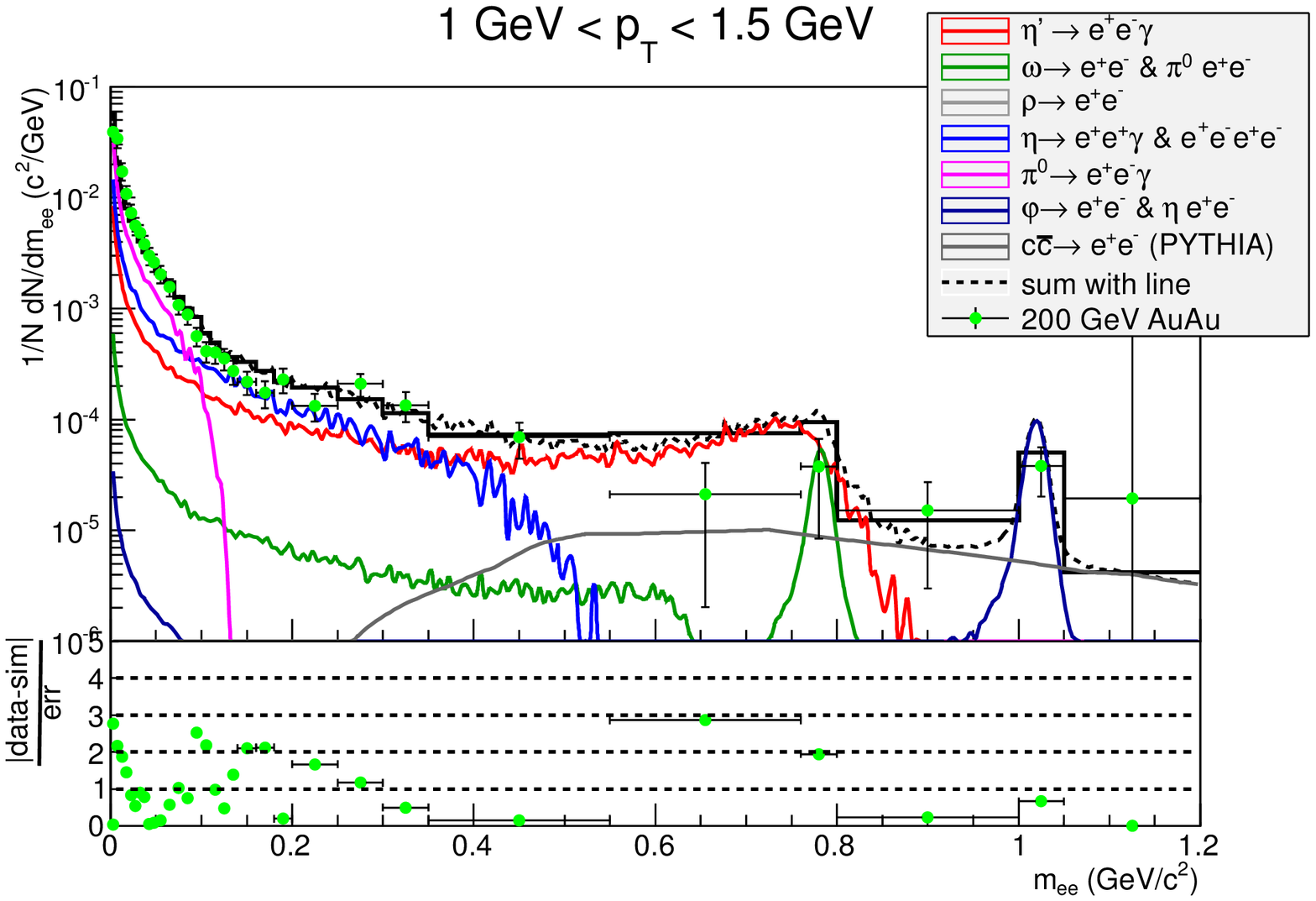} } \\
   \subfloat{ \includegraphics[width=0.8\textwidth]{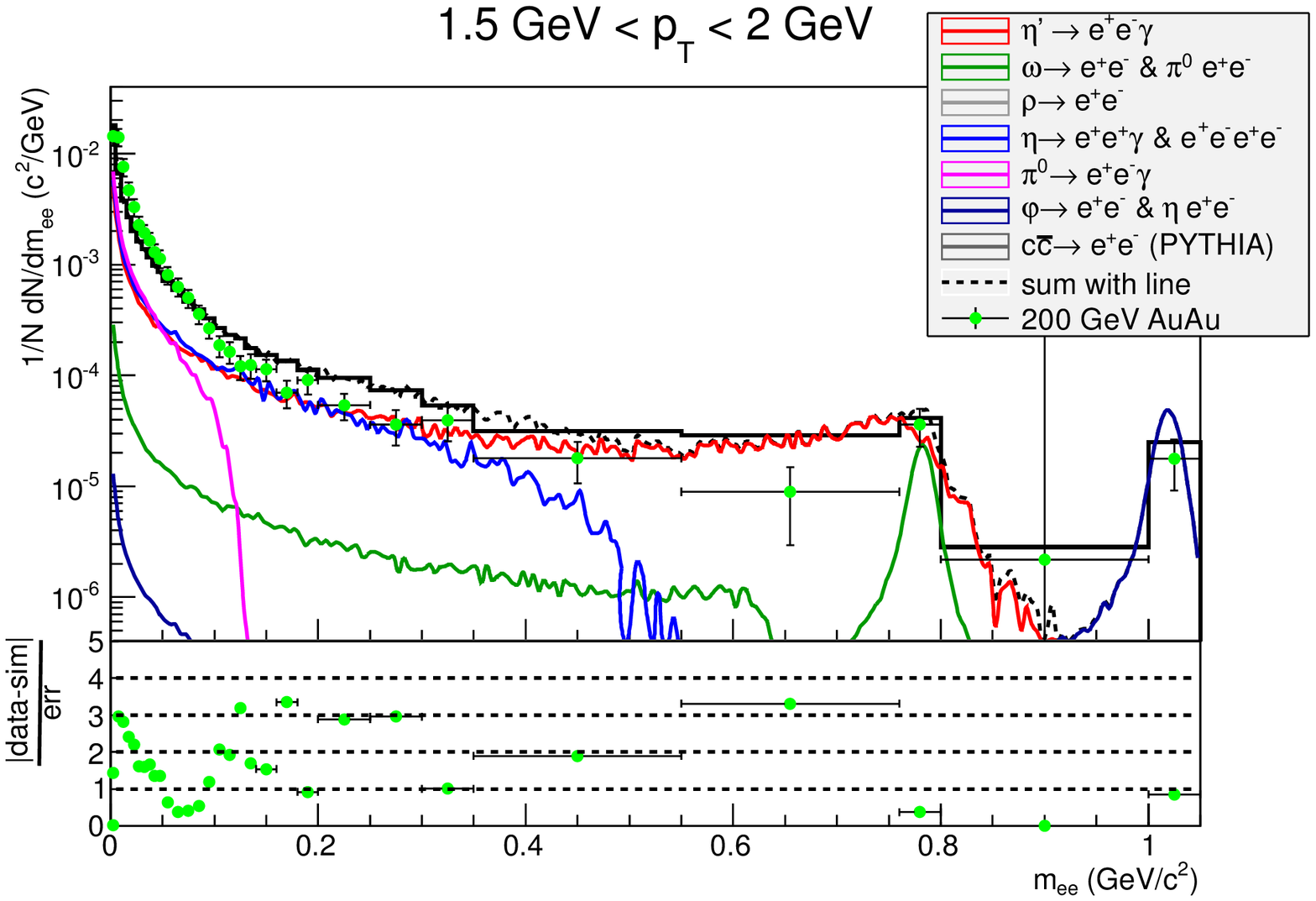} } 
\caption{Radial flow effects on the low-mass dilepton spectrum in different $p_\textrm{T}$ slices}   \label{f: radial_result2b} 
 \end{center}     
\end{figure}

EXODUS simulations using Eq.\ \ref{e: hydro} hydro spectrum as an input  were elaborated, and are shown on Fig.\ \ref{f: radial_result1}, while Figs.~\ref{f: radial_result2a} and \ref{f: radial_result2b} show its $p_\textrm{T}$ slices up to 2 GeV. If $p_\textrm{T}$ is beyond the hydrodynamical domain, and Eq.~\ref{e: hydro} is not appropriate any more, the power-law tail QCD particle spectrum starts to take over then. Hence the $p_\textrm{T} > 2$ GeV region has been left out from this investigations, at present. 

The meson weight factors were obtained by fitting meson contributions as a linear combination to PHENIX data. It would have been better to obtain these weight factors from data, since the weight factors should come from the integral of their $p_\textrm{T}$ spectra, but for all mesons contributing the dilepton spectrum ($\rho$, $\omega$, $\phi$, $\eta$, $\eta'$ and even $\pi^{0}$) no (or not enough) data were available in the $p_\textrm{T} < 2$ GeV region, where the hydro formula of Eq.\ \ref{e: hydro} is valid, so the normalization parameter ($A$ of Eq.~\ref{e: hydro}) could not be determined for them. The meson/pion yields coming from this fit are summarized in Table \ref{t: meson_yields}. Comparing $\eta'$/$\pi^{0}$ ratios, for current simulations  a tenfold enhancement can be obtained if compared with the PHENIX analysis of Ref.~\cite{PPG088}. Using $\phi$ mesons as a basis, e.g.\ calculating $\eta'$/$\phi$ ratios again for both cases results in even bigger enhancement. The physical picture behind this is that in the first case we divided the enhanced number of $\eta'$-s with an enhanced number of pions as well, as the $\eta'$ meson has a significant branching ratio to pions. The  enhancement obtained from the comparison of $\eta'$/$\phi$ ratios is 38.1 for the $\eta'$, which is a same order of magnitude as of that reported in Ref.~\cite{Vertesi:2009io}, based on indirect measurement of the $\eta'$ spectrum.

If one cross-checks the resonance ratios of Table~\ref{t: meson_yields}  with Eq.~\ref{e: naive_eta}, will not get any contradiction, in fact, from that equation (using $f_{\eta'} = 38.13$) $f_\eta = 5.5$ can be obtained, while from the data (comparing $\eta$/$\phi$ ratios of Table~\ref{t: meson_yields} and Ref.~\cite{PPG088}) an $\eta$ enhancement of 9.87 is resulted. The enhancements calculated from different methods are within a range of 1 magnitude. Possible additional effects, e.g.\ $\eta$ production cross-section might also be increased and of course, chain decays need to be more carefully simulated.

Note, that the 1.5 GeV $< p_\textrm{T} <$ 2 GeV slice of Fig.~\ref{f: radial_result2b} is fully explained. More work is needed to be done to combine $\eta'$ decay chains of Chapter 4 and radial flow effects of Chapter 5.

\begin{table}
\begin{center}
\begin{tabular}[H]{ c | c | c}
meson & meson/$\pi^{0}$ & meson/$\pi^{0}$ of Ref.~\cite{PPG088}\\ \hline \hline
$\eta$   &	3.48 $\times 10^{-1}$ & 1.12 $\times 10^{-1}$\\
$\rho$   &	3.70 $\times 10^{-8}$ & 8.98 $\times 10^{-2}$\\
$\omega$ &	1.96 $\times 10^{-2}$ & 1.03 $\times 10^{-1}$\\
$\phi$	 &	6.74 $\times 10^{-3}$ & 2.14 $\times 10^{-2}$\\ 
$\eta'$  &	2.57 $\times 10^{-1}$ & 2.15 $\times 10^{-2}$\\ 
\end{tabular}
\end{center}
\caption{The meson/pion yield ratios obtained by fitting the different meson contributions as linear combination. Note, that this method predicts a significant, tenfold enhancement of $\eta'$ mesons as compared to $\pi^0$-s. } \label{t: meson_yields}
\end{table}

\pagebreak


\chapter{Summary}

The experimentally discovered low-mass dilepton enhancement is one of the most difficult puzzles of relativistic  Au+Au  collisions at 200 GeV. To date, this effect has not yet been described successfully. 

In this Thesis, I investigated if this dilepton excess in the low-mass region can be described by two independent effects; the enhanced production of the $\eta'$ mesons coming from the restoration of the chiral symmetry in the hot, dense medium, and the effect of the radial flow which has not been included in the PHENIX analysis (Ref.~\cite{PPG088}). 

As for including the radial flow term, first it was observed on Fig.~\ref{f: pT2GeV}, that only dileptons with $p_\textrm{T} < 2$ GeV result in a significant contribution to the dilepton cocktail, thus finding a good description of the low-$p_\textrm{T}$ part of the resonances' spectra seems to be inevitable. It was also observed on Fig.~\ref{f: radial_check} that the input spectra used in Ref.~\cite{PPG088} (Eq.~\ref{e: PPG088}) lacks the radial flow term, thus it estimates the (most important) low-$p_\textrm{T}$ part of meson's spectra inaccurately. Hence, a simplified hydrodynamically motivated spectrum (Eq.~\ref{e: hydro}) was introduced and fitted simultaneously to available $\pi^{\pm}$, $K^{\pm}$, $p$ and $\bar{p}$ data. The fit result for the freeze-out temperature ($T_0$) and radial flow term ${\langle u_\textrm{T}\rangle}^2$ are in accordance with previous measurements and the fit resulted an acceptable confidence level of 8.6 \%. The dilepton cocktail simulated with this spectra was then fitted to the dilepton cocktail of 200 GeV Au+Au collisions of Ref.~\cite{PPG088}. The conclusion from this fit is that simulating resonances with a hydro $p_\textrm{T}$ distribution results in a good agreement with data, even for certain $p_\textrm{T}$ slices ($ 1.0$ GeV $-$ 2.0 GeV). Additional effects have to be considered as well to understand lower $p_\textrm{T}$ regions.

As for the $\eta'$ excess, first dilepton cocktails with $\eta'$ spectra of Ref.~\cite{Vertesi:2009io} were simulated, as seen on Figs.~\ref{f: eta+etaprime_1}-\ref{f: eta+etaprime_3}. These parameters were not fine-tuned to the current dilepton simulations, they resulted in a better agreement for the low-mas excess for certain resonance models, however. Later, two scans through all physically reasonable $\eta'$ spectrum parameters ($B^{-1}$ and $m^*$) were elaborated, for simulations including only the direct dileptonic decays of $\eta'$ and with the extended version of the EXODUS event generator, when every significant decays of $\eta'$ were taken into account (the significant decays are listed in Table~\ref{t: decays}). The $\chi^2$ map of the scan, for both methods is on Fig.~\ref{f: chi_map}. These studies gave no restriction for the slope\footnote{The quest for the slope parameter, $B^{-1}$ might succeed, however, if dilepton spectra were fitted to different $p_\textrm{T}$ slices.},
but indicated an at least a 200 MeV mass-drop of the $\eta'$, which is in accordance with the indirect measurement of Ref.~\cite{Vertesi:2009io}. Note that the mass-drop of Ref.~\cite{Vertesi:2009io} was elaborated in an independent observation channel, e.g.\ via $\pi^{\pm}\pi^{\pm}$ Bose-Einstein correlations.

The examination of the radial flow effect resulted in an at least tenfold enhancement of the $\eta'$ meson (when obtained from $\eta'$/$\pi^0$ ratio), such an enhancement might also be a signature of the restoration of the chiral symmetry. When the $\eta'$ enhancement was obtained from comparing $\eta'$/$\phi$ ratios, however, an even higher, 40 times excess has been obtained (comparing to the PHENIX background from Ref.~\cite{PPG088}), which is in the possible region of $\eta'$ enhancements obtained in Ref.~\cite{Vertesi:2009io}. 

We also found that proper simulations of $\eta'$ decay chains play an important role in obtaining the best values of the in-medium $\eta'$ mass, hence radial flow effects and $\eta'$ chain decays have to be combined, which is a subject of an upcoming investigation. 


\pagebreak
\chapter*{Acknowledgement}
\begin{quotation}
The author is grateful to the PHENIX Hungary group, particularly to Róbert Vértesi for the continuous assistance in the simulations (sometimes even on a daily basis), and to Tamás Csörgő for the support both in the current thesis and in the broader field of science. The author is also grateful to his mother and to his friend, András Kovács for their administrational assistance.
\end{quotation}
\thispagestyle{empty}

\pagebreak



\thispagestyle{empty}


\begin{thebibliography}{100}
\thispagestyle{empty}
 \bibitem{Vertesi:2009io}
   T.~Cs\"org\H o, R.~V\'ertesi, J.~Sziklai, Phys.\ Rev.\ Lett.\ {\bf 105} (2010) 182301 
  
 \bibitem{PPG088}
   A.\ Adare {\it et al.} [PHENIX Coll.], Phys.\ Rev.\ C {\bf 81}, 034911 (2010)  

 \bibitem{EXODUS}
  Description of the EXODUS software: \\
  \url{https://www.phenix.bnl.gov/WWW/offline/wikioffline/index.php/EXODUS}
  
 \bibitem{CGC}
   F.\ Gelis, E.\ Iancu, J.\ Jalilian-Marian, R.\ Venugopalan, \\
   \url{arXiv:1002.0333v1} [hep-ph]

 \bibitem{Adcox:2004mh}
   K.~Adcox {\it et al.}  [PHENIX Coll.],  Nucl.\ Phys.\  A {\bf 757}, 184  (2005)

 \bibitem{Adams:2005dq}
   J.~Adams {\it et al.}  [STAR Coll.],  Nucl.\ Phys.\  A {\bf 757}, 102  (2005)

 \bibitem{Back:2004je}
   B.~B.~Back {\it et al.} [PHOBOS Coll.],  Nucl.\ Phys.\  A {\bf 757}, 28  (2005)

 \bibitem{Arsene:2004fa}
   I.~Arsene {\it et al.}  [BRAHMS Coll.],  Nucl.\ Phys.\  A {\bf 757}, 1 (2005)

 \bibitem{PHENIXcentrality}
  S.\ S.\ Adler {\it et al.} [PHENIX Coll.], Phys.\ Rew.\ C {\bf 69}, 034909 (2004)

 \bibitem{HIJING}
  M.\ Gyulassy, Xin-Nian Wang, Phys.\ Rev.\ D {\bf 44}, 3501 (1991) \\
  Detailed documentation and users-guide on the website: \\
  \url{http://www-nsdth.lbl.gov/~xnwang/hijing/index.html}
 \bibitem{PISA}
  Description of the PISA software:\\
  \url{https://www.phenix.bnl.gov/WWW/offline/wikioffline/index.php/Simulations#Tutorials}
  
 \bibitem{PPG026}
   S.\ S.\ Adler {\it et al.} [PHENIX Coll.], Phys.\ Rev.\ C {\bf 69}, 034909 (2004)    
 
 \bibitem{PPG101}
   A.\ Adare {\it et al.} [PHENIX Coll.], Phys.\ Rev.\ C {\bf 83}, 064903 (2011) 
 
 \bibitem{PDGnum}
   Review of Particle Properties, Particle Data Group, \\ Phys.\ Lett.\ B {\bf 204} (1988), 113-114. Available in most Scientific Libraries.

 \bibitem{ALCOR}
   T.\ S.\ Bíró, P. L\'evai, and J.\ Zim\'anyi, Phys.\ Lett.\ B {\bf 347}, 6 (1995)

 \bibitem{Kaneta}
   M.\ Kaneta and N.\ Xu, \url{arXiv:nucl-th/0405068}

 \bibitem{Letessier}
   J.\ Letessier and J.\ Rafelski, Eur.\ Phys.\ J.\ A {\bf 35}, 221 (2008)

 \bibitem{UrQMD}
   M.\ Bleicher et al., J.\ Phys.\ G {\bf 25}, 1859 (1999)

 \bibitem{Stachel}
   S.\ A.\ Bass et al., Nucl.\ Phys.\ A {\bf 661}, 205 (1999)

 \bibitem{Vertesi:2010si}
   R.~V\'ertesi, T.~Cs\"org\H o, J.~Sziklai, Phys.\ Rev.\ C {\bf 83}, 054903 (2011)
   
 \bibitem{Rafelski}
  J.\ Letessier, J.\ Rafelski, Eur.\ Phys.\ J.\ A {\bf 35}, 221
(2008)

 \bibitem{PRC75}
   S.\ S.\ Adler {\it et al.} [PHENIX Coll.], Phys.\ Rev.\ C {\bf 775},
024909 (2007).

  
 \bibitem{chiral2}
 S.\ E.\  Vance, T.\ Csörgő, D.\ Kharzeev, Phys.\ Rev.\ Lett.\ {\bf 81} (1998) 2205-2208 
 
 \bibitem{Csorgo:2001xm}
  T.~Cs\"org\H{o}, S.~V.~Akkelin, Y.~Hama, B.~Luk\'acs and Y.~.M.~Sinyukov,
  Phys.\ Rev.\ C {\bf 67} (2003) 034904
  [hep-ph/0108067].

\end{thebibliography}
\end{document}